\documentclass[12pt,notitlepage]{article}
\usepackage{eurosym}
\usepackage{amsfonts}
\usepackage{amsmath}
\usepackage{geometry}
\usepackage[utf8]{inputenc}
\usepackage{graphicx}
\usepackage{multirow}
\usepackage{subfig}
\usepackage[onehalfspacing]{setspace}
\usepackage{tikz}
\usepackage{xcolor}

\usetikzlibrary{positioning}
\newtheorem{theorem}{Theorem}

\newtheorem{corollary}{Corollary}

\newtheorem{definition}{Definition}

\newtheorem{lemma}{Lemma}

\newtheorem{proposition}{Proposition}

\newenvironment{proof}[1][Proof]{\textbf{#1} }{\ \rule{0.5em}{0.5em}}
\input{tcilatex}

\begin{document}

\title{\textbf{Multidimensional Signaling }\\
\textbf{with a Resource Constraint}\thanks{
We are grateful to Maxim Ivanov for his useful comments. Han gratefully
acknowledges financial support from Social Sciences and Humanities Research
Council of Canada for this research.}}
\author{Seungjin Han and Alex Sam\thanks{
Department of Economics, McMaster University, 1280 Main Street Hamilton, ON,
Canada L8S 4M4. Phone: (905) 525-9140. Han's email: hansj@mcmaster.ca, and
Sam's email: sama1@mcmaster.ca.}}
\maketitle

\begin{abstract}
We study multidimensional signaling (cognitive/non-cognitive) as a sender's
portfolio choice with a resource constraint. We establish the existence of a
unique monotone D1 equilibrium where the cognitive (non-cognitive) signal
increases (decreases) in sender type and the sum of the two increases in
sender type. The equilibrium is characterized by two threshold sender types.
The low threshold is one where a kink occurs in signaling. The constraint is
binding only for sender types above it. The high threshold is the other one,
above which all types spend all the resources in cognitive signal with
pooling and discontinuity on the top.

\medskip


\end{abstract}


\section{Introduction}

It is often observed in practice that a privately-informed individual
(sender) chooses multiple actions to signal their type. For example, in the
labor market, a potential worker may signal their ability through a
combination of a degree of theoretical knowledge and vocational training.
While multidimensional signaling is a fascinating research topic, it has
been underdeveloped. Technical challenges associated with potential pooling
on discontinuities or kinks in the equilibrium outcomes have posed
difficulties in characterizing equilibrium multidimensional signaling.
Perhaps this is why the early literature on multidimensional signaling
focuses on separating equilibrium and its existence (Milgrom and Roberts
(1986), Wilson (1985), Quinzii and Rochet (1985), Engers (1987), and Chen
(1997)).

As assumed in the one-dimensional signaling model, one of common assumptions
in the multidimensional signaling models is that there is no constraint on
the total signaling resource. However, it would be really surprising to see
that a sender has unlimited resources for signaling. For example, a
potential worker has only a limited amount of time or resources to spend
investing in signaling. A sender's multidimensional signaling choice given a
resource constraint is then very similar to the financial portfolio choice
between multiple assets given a constraint on the total amount of money,
i.e., wealth.

In their classical papers (Arrow (1964), Pratt (1965)), they provide the
well-known results: An increase in wealth leads to an increase in the amount
of money invested (or in the fraction of the portfolio held) in a risky
asset if the individual's absolute (relative) risk aversion is decreasing in
their wealth. Our paper provides a similar but unique characterization of a
sender's equilibrium multidimensional signaling as their portfolio choice
given a resource constraint in a competitive signaling market.

In this competitive market, there is a continuum of heterogeneous senders
and homogeneous receivers. Senders invest in multidimensional signals
(cognitive and non-cognitive), facing a resource constraint. In the
classical portfolio model (Arrow (1964), Pratt (1965)), investors differ in
the financial constraint but they face the same market price for each asset.
In our model, senders have the same resource constraint, but different
sender types face different costs of acquiring cognitive signal, while
facing the same cost of acquiring non-cognitive signal. In spirit with
Spence (1973), we assume that the cost function of a cognitive signal
satisfies the single crossing property.

In signaling models, multiplicity of equilibrium is prevalent due to the
multiplicity of off-the-path beliefs and ours is not an exception. To
sharpen equilibrium characterization, our focus is the equilibrium with
\emph{reasonable D1 beliefs}, that is, reasonable beliefs that satisfy
Criterion D1. A belief is \emph{reasonable} if given any pair of signals
that makes the resource constraint slack, a receiver's belief (on-path and
off-path) is unaffected by a change in the non-cognitive signal that keeps
the resource constraint slack. We show that when the belief is reasonable,
an equilibrium is monotone in the sense that a sender's equilibrium
cognitive signal is non-decreasing in her type, the sender's equilibrium
non-cognitive signal is non-increasing in her type, and the sum of her
equilibrium signals is non-decreasing in her type. We then show that
Criterion D1 (Banks and Sobel (1987) and Cho and Sobel (1990)) can select a
unique equilibrium among monotone equilibria even with multi-dimensional
signaling.\footnote{%
With multidimensional signaling, Criterion D1 does not ensure a unique
equilibrium among all equilibria, nor does it imply the monotonicity of the
sender's signaling choice.} We call it a \emph{monotone D1 equilibrium}.

Separating monotone D1 equilibrium is no longer guaranteed because upward
deviations are not always possible given a resource constraint.
Discontinuity, pooling and a kink are together a unique feature of a
monotone D1 equilibrium of multidimensional signaling with a resource
constraint, compared to the equilibrium portfolio in Arrow (1964) and Pratt
(1965). Not only do we establish the existence of a unique monotone D1
equilibrium by painstakingly handling them, but we provide a sharp
characterization of them.

Given any amount of resources, a unique monotone D1 equilibrium is
characterized by (at most) two threshold sender types. Any sender type below
the low threshold chooses the efficient level of non-cognitive signal but
their cognitive signal is not high enough to make their resource constraint
binding. This is because the cost associated with acquiring cognitive signal
is too high for them. The resource constraint is binding for only those
sender types above the low threshold, creating a kink at the low threshold.
Interestingly, those senders above the low threshold reduce resources spent
in acquiring non-cognitive signal to increase their cognitive signal. As the
sender type increases, the resource used in cognitive (non-cognitive) signal
increases (decreases), which makes cognitive (non-cognitive) signal increase
(decrease) in the sender type up until the high threshold. All sender types
above the high threshold spends all the resources in cognitive signal, which
causes discontinuity and pooling.

\subsection*{Related Work}

Our paper is closely related to Milgrom and Roberts (1986) where the quality
of a firm's product is its private information and the price of the product
and advertising both play a role of signaling the product quality. Given the
binary quality of the product, they established the existence of a unique
separating equilibrium that satisfies the Intuitive Criterion (Kreps
(1984)). Allowing for continuous quality, Wilson (1985) extends Milgrom and
Roberts (1986) and shows the existence of a separating equilibrium.

Our paper also considers a model where there is one dimensional type and two
dimensional signals (cognitive and non-cognitive). However, the key
difference is that the sender has the constraint of resources that she can
use in selecting signals, whereas there is no resource constraint in Milgrom
and Roberts (1986) and Wilson (1985). While our notion of refinement,
reasonable Criterion D1 ensures a unique monotone equilibrium in terms of a
sender's signaling choice, it no longer guarantees a separating equilibrium.
Given an amount of resources, kinks and discontinuities are pervasive.
Despite that, we fully characterize a monotone D1 equilibrium given any
amount of resources.

Quinzii and Rochet (1985), Engers (1987), and Chen (1997) extend Milgrom and
Roberts (1986) allowing for multidimensional type but there is also no
resource constraint for selecting signaling. They also focus on only a
separating equilibrium in order to avoid technical difficulties encountered
in pooling on discontinuity or kinks in the wage schedule, whereas our paper
fully address kinks and discontinuities in characterizing every possible
type of a monotone D1 equilibrium.

In multidimensional signaling, Bagwell (2007) introduces multidimensional
binary types (patience and cost) of the incumbent facing a potential entrant
to admit pooling in (multiple) intuitive equilibria (Cho and Kreps (1987)).
Frankel and Kartik (2019) show that the market faces muddled information
(i.e., pooling) when the agent care about a market's belief on their quality
on one of the two dimensions in their type. Our paper suggests that muddled
information is part of the signaling portfolio choice with a resource
constraint.

Recent developments in monotone signaling equilibrium are based on one
dimensional signaling. With the monotone-supermodular property of the
sender's utility, the monotonicity of the sender's signaling choice is
established in a two-period signaling game between one sender and one
receiver (Liu and Pei (2020)) and a dynamic incomplete information games
with multiple players (Mensch (2020)). Han, Sam and Shin (2024) extend the
signaling model to matching markets with two-sided heterogeneity and provide
a full monotone equilibrium analysis: Suppose that the sender utility is
monotone-supermodular and the receiver's utility is weakly
monotone-supermodular. Then, the sender's action function, the receiver's
reaction function, the matching function and the belief in a signaling
equilibrium are all stronger monotone if and only if the belief passes
Criterion D1. This implies the equivalence between the stronger monotone
belief and the belief that satisfies Criterion D1 with one dimensional
signaling.\footnote{%
It is not clear under what conditions such an equivalence holds with
multidimensional signaling.} While the monotone-supermodular property of the
sender's utility can alone ensure the monotonicity of the sender's
one-dimensional signaling choice, we show that it is crucial to impose the
reasonable belief along with the monotone-single crossing property of the
sender's utility for the monotonicity of the sender's multidimensional
signaling choice.

\section{The Model}

The model consists of a continuum of heterogenous senders and homogenous
receivers. For example, in the labour market, senders are workers, while
receivers are the firms a worker works for. Senders are characterized by
their privately observed types which a receiver cannot observe. Let $T= %
\left[\underline{t}, \overline{t}\right]$ with $0<\underline{t}<\overline{t}
<\infty $ be the type space of the sender. Each sender's type $t\in T$ is
drawn from a probability distribution $G$ with full support. $G$ is a public
information.

Let $M>0$ denote the amount of resources that senders can invest. Given any $%
M>0$, a sender of type $t$ can invest in two types of signals: $m_{1}$ and $%
m_{2}$, and incur the cost
\begin{equation*}
h(m_{2})+c(m_{1},t)
\end{equation*}
while facing a resource constraint,
\begin{equation*}
\mathcal{C}:=\{(m_{1},m_{2})\in
\mathbb{R}
_{+}^{2}:m_{1}+m_{2}\leq M\}.
\end{equation*}
When a sender of type $t$ chooses the signals $m_{1}$ and $m_{2}$, the
receiver values the sender according to $\alpha m_{2}+f(m_{1},t)$, where $%
\alpha $ is some nonnegative constant. The payoffs of a sender and a
receiver after a sender chooses $m_{1}$ and $m_{2}$ and receives a transfer $%
w\in \mathbb{R}_{+}$ from the receiver are $%
u(t,m_{1},m_{2},w)=w-h(m_{2})-c(m_{1},t)$, and $v(t,m_{1},m_{2},w)=\alpha
m_{2}+f(m_{1},t)-w$ respectively.

We view a sender's type as her inherent intellectual ability. A sender's
type $t$ only affects the cost of acquiring $m_{1}$ and the output produced
by $m_{1}$. Therefore, one can view $m_{1}$ as a \textit{cognitive signal},
whereas $m_{2}$ is a \textit{non-cognitive signal}.

The timing of events is summarized as follows: Nature draws a sender's
private type $t$ from the prior distribution $G$. Upon privately learning
their own type $t$, given $M>0$, each sender selects their public signal
pair $(m_{1},m_{2})\in \mathcal{C}$. After observing the distribution of the
signal pairs $(m_{1},m_{2})\in \mathcal{C}$, a receiver updates his beliefs
about a sender's type $t$, and takes an action $w\in \mathbb{R} _{+} $.

We begin with the following assumptions.

\begin{description}
\item[Assumption 1] (a) $c$ is differentiable everywhere, (b) $c$ is
strictly increasing in $m_{1}$ and strictly decreasing in $t$, (c) $f$ is
differentiable everywhere, (d) $f$ is non-decreasing in $m_{1}$ and strictly
increasing in $t$.

\item[Assumption 2] $c$ satisfies the single crossing property: For all $%
m_{1}>m_{1}^{\prime }$ and $t>t^{\prime }$, $c(m_{1},t)+c(m_{1}^{\prime
},t^{\prime })<c(m_{1},t^{\prime })+c(m_{1}^{\prime },t).$

\item[Assumption 3] (a) $f-c$ is strictly concave in $m_{1}$, (b) $h$ is
strictly convex in $m_{2}$, (c) $h$ is differentiable everywhere.

\item[Assumption 4] $0<m_{2}^{\circ }<M,$ where $m_{2}^{\circ }$ is a unique
solution for $\max_{m_{2}\in
\mathbb{R}
_{+}}\alpha m_{2}-h(m_{2}).$
\end{description}

Assumptions 1.(b) and 2 are together referred to as the monotone-single
crossing property of the sender's utility function. Let $\mu :\mathcal{C}
\rightarrow \Delta (T)$ be the belief function that specifies a probability
distribution over a sender's possible types conditional on her signals. Let $%
\mathbb{E}_{\mu (m_{1},m_{2})}f(m_{1},t)$ be the expected value of $%
f(m_{1},t)$ given the probability distribution $\mu (m_{1},m_{2})$ over $T$
conditional on the signal pair $(m_{1},m_{2})$.

We define a competitive signaling equilibrium as follows. The consistency of
$\mu$ follows the definition in Ramey (1996).

\begin{definition}[Competitive Signaling Equilibrium]
$\{\hat{m}_{1},\hat{m}_{2},\hat{w},\hat{\mu}\}$ is a competitive signaling
equilibrium (henceforth an equilibrium) if it satisfies

\begin{enumerate}
\item Sender's optimal signaling decision: for all $t\in T,$ $(\hat{m}
_{1}(t),\hat{m}_{2}(t))$ solves
\begin{equation}  \label{eq: competitive}
\underset{\left( m_{1},m_{2}\right) \in \mathcal{C}}{\max }\hat{w}
(m_{1},m_{2})-h(m_{2})-c(m_{1},t)\text{,}
\end{equation}
where $\hat{w}(m_{1},m_{2})=\alpha m_{2}+\mathbb{E}_{\hat{\mu}
(m_{1},m_{2})}f(m_{1},t)$ $\forall (m_{1},m_{2})\in \mathcal{C}$.

\item Consistency of $\hat{\mu}$:

\begin{enumerate}
\item If $(m_{1},m_{2})\in $ \emph{range}$(\hat{m}_{1},\hat{m}_{2})$
satisfies $\hat{\mu} \left( \left\{ t|\left( \hat{m}_{1}\left( t\right) ,
\hat{m} _{2}\left( t\right) \right) =\left( m_{1},m_{2}\right) \right\}
\right) >0,$ then $\hat{\mu} (m_{1},m_{2})$ is determined by $G$ and $\left(
\hat{m}_{1},\hat{m} _{2}\right) $ using Bayes' rule.

\item If $(m_{1},m_{2})\in $ \emph{range}$(\hat{m}_{1},\hat{m}_{2})$
satisfies $\hat{\mu} \left( \left\{ t|\left( \hat{m}_{1}\left( t\right) ,
\hat{m} _{2}\left( t\right) \right) =\left( m_{1},m_{2}\right) \right\}
\right) =0,$ then $\hat{\mu} (m_{1},m_{2})$ is any probability distribution
satisfying \emph{\ supp} $\hat{\mu} (m_{1},m_{2})=$ cl $\left\{ t|\left(
\hat{m}_{1}\left( t\right) , \hat{m}_{2}\left( t\right) \right) =\left(
m_{1},m_{2}\right) \right\} $.

\item If $(m_{1},m_{2})\notin $ \emph{range}$(\hat{m}_{1},\hat{m}_{2})$,
then $\hat{\mu} (m_{1},m_{2})$ is unrestricted.
\end{enumerate}
\end{enumerate}
\end{definition}

Condition 2(a) and 2(b) imposes the consistency of the belief in
equilibrium. However, Condition 2(c) leaves the indeterminacy of the belief
conditional on any off-path message.

All proofs are relegated to the appendix, except for the proofs of
Proposition \ref{weakmonotone}, Lemma \ref{lemma1}, and Theorem \ref{prop2}.

\section{Monotone equilibrium}

We focus our attention to monotone equilibria. The main goal of this section
is to provide a sufficient condition for an equilibrium to be monotone. The
latter part of the section introduces the Criterion D1 (an equilibrium
refinement technique) to select a unique monotone equilibrium.

We begin by defining a monotone equilibrium.

\begin{definition}
\label{monotone} An equilibrium $\{\hat{m}_{1},\hat{m}_{2},\hat{w},\hat{\mu}
\}$ is monotone if $\hat{m}_{1}\left( t\right) $ is non-decreasing, $\hat{m}
_{2}\left( t\right) $ is non-increasing and $\hat{m}_{1}\left( t\right) +
\hat{m}_{2}\left( t\right) $ is non-decreasing.
\end{definition}

Note that the belief $\mu (m_{1},m_{2})$ on the sender's type generally
depends on her choice of both signals, $m_{1}$ and $m_{2}$. For any $%
m_{1},m_{2},$ let $r(m_{1},m_{2})$ take one of the two values, $\{b,s\}$
such that

\begin{equation*}
r(m_{1},m_{2})=\left\{
\begin{array}{cc}
b & \text{if }m_{1}+m_{2}=M \\
s & \text{if }m_{1}+m_{2}<M%
\end{array}
\right. .
\end{equation*}
We define a reasonable belief as follows.

\begin{definition}
A belief $\mu :\mathcal{C}\rightarrow \Delta (T)$ is \emph{reasonable} if
\begin{equation*}
r(m_{1},m_{2})=r(m_{1},m_{2}^{\prime })\Rightarrow \mu (m_{1},m_{2})=\mu
(m_{1},m_{2}^{\prime })\text{ }\forall (m_{1},m_{2}),(m_{1},m_{2}^{\prime
})\in \mathcal{C}.
\end{equation*}
Let $\Psi $ be the set of all reasonable beliefs.
\end{definition}

An implication for reasonable beliefs is that for any $\mu \in \Psi $ and
any $(m_{1},m_{2}),(m_{1},m_{2}^{\prime })\in \mathcal{C},$
\begin{multline*}
\hat{w}(m_{1},m_{2})-\hat{w}(m_{1},m_{2}^{\prime })= \\
\left\{
\begin{array}{cc}
\alpha m_{2}-\alpha m_{2}^{\prime } & \text{if }
r(m_{1},m_{2})=r(m_{1},m_{2}^{\prime }) \\
\alpha m_{2}+\mathbb{E}_{\mu (m_{1},m_{2})}f(m_{1},t)-\left( \alpha
m_{2}^{\prime }+\mathbb{E}_{\mu (m_{1},m_{2}^{\prime })}f(m_{1},t)\right) &
\text{if }r(m_{1},m_{2})\neq r(m_{1},m_{2}^{\prime })%
\end{array}
\right. \text{ }.
\end{multline*}

Intuitively, by focusing on reasonable beliefs, we mean that, given a pair
of signals $(m_{1},m_{2})$ that makes the constraint slack, a receiver's
belief (on-path and off-path) is unaffected by a change in the non-cognitive
signal $(m_{2}\rightarrow m_{2}^{\prime })$ when the constraint is still
slack with $(m_{1},m_{2}^{\prime })$.

The proposition below establishes the monotonicity of $\hat{m} _{1}$ and $%
\hat{m}_{2}$.

\begin{proposition}
\label{weakmonotone} In any equilibrium $\{\hat{m}_{1},\hat{m}_{2},\hat{w},
\hat{\mu}\}$, $\hat{m}_{1}$ is non-decreasing and $\hat{m}_{2}$ is
non-increasing if $\hat{\mu}\in \Psi $,
\end{proposition}

\begin{proof}
The non-decreasing property of $\hat{m}_{1}$ directly comes from Assumption
2 (single crossing property of $c$). The proof is standard, so it is omitted.

Let us prove the non-increasing property of $\hat{m}_{2}$. Given Assumption
3.(b) (strict convexity of $h$) and Assumption 4, $\hat{m}_{2}(t)\leq
m_{2}^{\circ }$ for all $t\in T$. Further, $\hat{m}_{2}(t)\leq M-\hat{m}
_{1}(t)$ for all $t\in T$ because of the resource constraint. Therefore,
\begin{equation}
\hat{m}_{2}(t)\leq \min \left[ m_{2}^{\circ },M-\hat{m}_{1}(t)\right] \text{
}\forall t\in T.  \label{prop1_e1}
\end{equation}
We prove the non-increasing property of $\hat{m}_{2}$ by contradiction.
Towards a contradiction, suppose that there is a pair of $t$ and $t^{\prime
} $ such that $t<t^{\prime }$ and $\hat{m}_{2}(t)<\hat{m}_{2}(t^{\prime }).$
Because the expression on the right hand side of (\ref{prop1_e1}) is
non-increasing in $t$ given the non-decreasing property of $\hat{m}_{1}$,
this claim holds only when
\begin{equation*}
\hat{m}_{2}(t)<\min \left[ m_{2}^{\circ },M-\hat{m}_{1}(t)\right] ,
\end{equation*}
which implies that
\begin{equation}
\hat{m}_{1}(t)+\hat{m}_{2}(t)<M.  \label{prop1_e2}
\end{equation}
On the other hand, $\hat{m}_{2}(t)<\hat{m}_{2}(t^{\prime })$ implies
\begin{equation}
\hat{m}_{2}(t)<m_{2}^{\circ }  \label{prop1_e3}
\end{equation}
because $\hat{m}_{2}(t^{\prime })\leq m_{2}^{\circ }$ given the strict
convexity of $h$ (Assumption 3.(b)) and Assumption 4.

Given (\ref{prop1_e2}) and (\ref{prop1_e3}), there exists $m_{2}^{\prime }$
such that (i) $\hat{m}_{2}(t)<m_{2}^{\prime }<m_{2}^{\circ }$ and (ii) $r(
\hat{m}_{1}(t),m_{2}^{\prime })=s.$ Because $r(\hat{m}_{1}(t),\hat{m}
_{2}(t))=r(\hat{m}_{1}(t),m_{2}^{\prime })=s$, we have $\hat{\mu}(\hat{m}
_{1}(t),\hat{m}_{2}(t))=\hat{\mu}(\hat{m}_{1}(t),m_{2}^{\prime })$ given $%
\hat{\mu}\in \Psi $. Accordingly, the change in the utility for type $t$ by
deviation to $\left( \hat{m}_{1}(t),m_{2}^{\prime }\right) $ is
\begin{gather*}
\hat{w}(\hat{m}_{1}(t),m_{2}^{\prime })-h(m_{2}^{\prime })-c(\hat{m}
_{1}(t),t)-\left[ \hat{w}(\hat{m}_{1}(t),\hat{m}_{2}(t))-h(\hat{m}
_{2}(t))-c( \hat{m}_{1}(t),t)\right] = \\
\alpha m_{2}^{\prime }-h\left( m_{2}^{\prime }\right) -[\alpha \hat{m}
_{2}(t)-h\left( \hat{m}_{2}(t)\right) ],
\end{gather*}
where the equality holds because $\hat{w}(\hat{m}_{1}(t),m_{2}^{\prime })-
\hat{w}(\hat{m}_{1}(t),\hat{m}_{2}(t))=$ $\alpha m_{2}^{\prime }-\alpha \hat{%
			m}_{2}(t)$ given $r(\hat{m}_{1}(t),\hat{m}_{2}(t))=r(\hat{m}
_{1}(t),m_{2}^{\prime })$.

Because $\hat{m}_{2}(t)<m_{2}^{\prime }<m_{2}^{\circ },$ the strict
convexity of $h$ implies that $\alpha m_{2}^{\prime }-h\left( m_{2}^{\prime
}\right) >\alpha \hat{m}_{2}(t)-h\left( \hat{m}_{2}(t)\right) $. Therefore,
the utility difference is strictly positive. This contradicts that $\hat{m}
_{2}(t)$ is an equilibrium choice of a non-cognitive signal for type $t$. We
conclude that $\hat{m}_{2}$ must be non-increasing.
\end{proof}

\bigskip

The proof of Proposition \ref{weakmonotone} relies on the single crossing
condition (Assumption 1(b)), and the condition that a belief is reasonable.
It is important to mention that only the monotonicity of $\hat{m}_{2}$
requires that a belief be reasonable.

The following lemma is crucial in establishing the monotonicity of $\hat{m}
_{1} +\hat{m}_{2}$.

\begin{lemma}
\label{lemma1}In any equilibrium $\{\hat{m}_{1},\hat{m}_{2},\hat{w},\hat{\mu}
\}$ with $\hat{\mu}\in \Psi$, the following properties are satisfied:

\begin{enumerate}
\item[(a)] If $\hat{m}_{1}\left( t\right) +\hat{m}_{2}\left( t\right) =M$
for some $t<\bar{t},$ then $\hat{m}_{1}\left( t^{\prime }\right) +\hat{m}
_{2}\left( t^{\prime }\right) =M$ for all $t^{\prime }>t$.

\item[(b)] For any pair of two types, $t$ and $t^{\prime }$ such that $\hat{%
				m			}_{1}\left( t^{\prime }\right) +\hat{m}_{2}\left( t^{\prime }\right)
<M$ and $\hat{m}_{1}\left( t\right) +\hat{m}_{2}\left( t\right) <M$, we have
that $\hat{m}_{1}\left( t\right) +\hat{m}_{2}\left( t\right) \leq \hat{m}
_{1}\left( t^{\prime }\right) +\hat{m}_{2}\left( t^{\prime }\right) $ if $%
t<t^{\prime }$.
\end{enumerate}
\end{lemma}

\begin{proof}
We prove the first statement. Suppose that $\hat{m}_{1}\left( t\right) +\hat{%
			m}_{2}\left( t\right) =M$ for some $t<\bar{t}$. Let us show that no type $%
t^{\prime }>t$ with $\left( \hat{m}_{1},\hat{m}_{2}\right) $ such that $\hat{%
			m}_{1}\left( t^{\prime }\right) +\hat{m}_{2}\left( t^{\prime }\right) <M$%
. Suppose towards a contradiction that there is a type $t^{\prime }>t$ such
that $\hat{m}_{1}\left( t^{\prime }\right) +\hat{m}_{2}\left( t^{\prime
}\right) <M.$ Because $\hat{m}_{2}$ is non-increasing in type, we have that $%
\hat{m}_{2}\left( t\right) \geq \hat{m}_{2}\left( t^{\prime }\right) $ given
$t^{\prime }>t.$

If $\hat{m}_{2}\left( t\right) =\hat{m}_{2}\left( t^{\prime }\right) $, then
$\hat{m}_{1}\left( t^{\prime }\right) <\hat{m}_{1}\left( t\right) $ given $%
\hat{m}_{1}\left( t\right) +\hat{m}_{2}\left( t\right) =M$ and $\hat{m}
_{1}\left( t^{\prime }\right) +\hat{m}_{2}\left( t^{\prime }\right) <M.$
Because $t^{\prime }>t$, $\hat{m}_{1}\left( t^{\prime }\right) <\hat{m}
_{1}\left( t\right) $ violates the non-decreasing property of $\hat{m}_{1}.$

Consider the case where $\hat{m}_{2}\left( t\right) >\hat{m}_{2}\left(
t^{\prime }\right) .$ Since $\hat{m}_{2}\left( t\right) $ is an equilibrium
choice for type $t,$ we have that $\hat{m}_{2}\left( t\right) \leq
m_{2}^{\circ }$. Therefore, $\hat{m}_{2}\left( t^{\prime }\right)
<m_{2}^{\circ }$. Because $r(\hat{m}_{1}\left( t^{\prime }\right) ,\hat{m}
_{2}\left( t^{\prime }\right) )=s$ and $\hat{m}_{2}\left( t^{\prime }\right)
<m_{2}^{\circ }$, there exists $m_{2}^{\prime }$ for type $t^{\prime }$ such
that (i) $\hat{m}_{2}(t^{\prime })<m_{2}^{\prime }<m_{2}^{\circ }$ and (ii) $%
r(\hat{m}_{1}(t^{\prime }),m_{2}^{\prime })=s.$

Because $r(\hat{m}_{1}(t^{\prime }),m_{2}^{\prime })=r(\hat{m}_{1}(t^{\prime
}),\hat{m}_{2}\left( t^{\prime }\right) )=s$, we have $\hat{\mu}(\hat{m}
_{1}(t^{\prime }),m_{2}^{\prime })=\hat{\mu}(\hat{m}_{1}(t^{\prime }),\hat{m}
_{2}(t^{\prime }))$ given $\hat{\mu}\in \Psi $. Accordingly, the utility
difference for type $t^{\prime }$ by deviation to $(\hat{m}_{1}(t^{\prime
}),m_{2}^{\prime })$ is
\begin{gather*}
\hat{w}(\hat{m}_{1}(t^{\prime }),m_{2}^{\prime })-h(m_{2}^{\prime })-c(\hat{
m }_{1}(t^{\prime }),t^{\prime })-\left[ \hat{w}(\hat{m}_{1}(t^{\prime }),
\hat{ m}_{2}(t^{\prime }))-h(\hat{m}_{2}(t^{\prime }))-c(\hat{m}
_{1}(t^{\prime }),t^{\prime })\right] = \\
\alpha m_{2}^{\prime }-h\left( m_{2}^{\prime }\right) -[\alpha \hat{m}
_{2}(t^{\prime })-h(\hat{m}_{2}(t^{\prime }))],
\end{gather*}
where the equality holds because $\hat{w}(\hat{m}_{1}(t^{\prime
}),m_{2}^{\prime })-\hat{w}(\hat{m}_{1}(t^{\prime }),\hat{m}_{2}(t^{\prime
}))=$ $\alpha m_{2}^{\prime }-\alpha \hat{m}_{2}(t^{\prime })$ given $r(\hat{%
			m}_{1}(t^{\prime }),m_{2}^{\prime })=r(\hat{m}_{1}(t^{\prime }),\hat{m}
_{2}(t^{\prime }))$.

Because $\hat{m}_{2}(t^{\prime })<m_{2}^{\prime }<m_{2}^{\circ },$ the
strict convexity of $h$ implies that $\alpha m_{2}^{\prime }-h\left(
m_{2}^{\prime }\right) >\alpha \hat{m}_{2}(t^{\prime })-h\left( \hat{m}
_{2}(t^{\prime })\right) $. Therefore, the utility difference is strictly
positive. This contradicts that $\hat{m}_{2}(t^{\prime })$ is an equilibrium
choice of a non-cognitive signal for type $t^{\prime }$. We conclude that
the first statement is true. The second statement can be proved analogously,
so we omit its proof.
\end{proof}

\bigskip

Proposition \ref{weakmonotone} and Lemma \ref{lemma1} together establish the
monotonicity of $\hat{m}_{1},$ $\hat{m}_{2}$, and $\hat{m}_{1}+\hat{m}_{2}$.
We summarize the results in the theorem below.

\begin{theorem}
\label{prop2} An equilibrium $\{\hat{m}_{1},\hat{m}_{2},\hat{w},\hat{\mu} \}
$ is monotone if $\hat{\mu}\in \Psi$.
\end{theorem}

\begin{proof}
According to Proposition \ref{weakmonotone}, $\hat{m}_{1}$ is non-decreasing
and $\hat{m}_{2}$ is non-increasing when $\hat{\mu}\in \Psi$. It remains to
show that $\hat{m}_{1}+\hat{m}_{2}$ is non-decreasing. Because of Lemma \ref%
{lemma1}, it is sufficient to show the following statement: If $\hat{m}
_{1}\left( t\right) +\hat{m}_{2}\left( t\right) <M$ and $\hat{m}_{1}\left(
t^{\prime }\right) +\hat{m}_{2}\left( t^{\prime }\right) =M$ for a pair of
two different $t$ and $t^{\prime },$ then $t<t^{\prime }.$ Suppose that $%
\hat{m}_{1}\left( t\right) +\hat{m}_{2}\left( t\right) <M$ and $\hat{m}
_{1}\left( t^{\prime }\right) +\hat{m}_{2}\left( t^{\prime }\right) =M$ for
a pair of two different $t$ and $t^{\prime }$, with $t>t^{\prime }.$ Because
$\hat{m}_{1}\left( t^{\prime }\right) + \hat{m}_{2}\left( t^{\prime }\right)
=M,$ we must have $\hat{m}_{1}\left( t\right) +\hat{m}_{2}\left( t\right) =M$
as well for $t>t^{\prime }$, according to Lemma \ref{lemma1}(a). This
contradicts $\hat{m}_{1}\left( t\right) +\hat{m}_{2}\left( t\right) <M.$
\end{proof}

\bigskip

It is important to note that Criterion D1 (Cho \& Sobel (1990), and Banks \&
Sobel (1987)) is not needed to ensure the monotonicity of a signaling
equilibrium $\{\hat{m}_{1},\hat{m}_{2},\hat{w},\hat{\mu}\}$ nor does it
imply that the belief $\hat{\mu}$ is reasonable. To see this point, given an
equilibrium $\{\hat{m}_{1},\hat{m}_{2},\hat{w},\hat{\mu}\}$, we define type $%
t^{\prime }s$ equilibrium utility $U(t)$ by $U(t):=\hat{w}(\hat{m_{1}}(t),
\hat{m_{2}}(t))-h(\hat{m_{2}}(t))-c(\hat{m_{1}}(t),t)$. Then, Criterion D1
is defined as follows.

\begin{definition}
(\textbf{Criterion D1}) Fix any $(m_{1},m_{2})\notin $ \emph{range}$(\hat{m}
_{1},\hat{m}_{2})$ and any transfer $w\in \mathbb{R} _{+}$. Suppose that
there is a non-empty set $T^{\prime}\subset \left[\underline{t}, \overline{t}
\right]$ such that the following is true: for each $t\notin T^{\prime},$
there exists $t^{\prime}$ such that

\begin{equation}  \label{D1}
w-h(m_{2})-c(m_{1},t)\geq U(t) \implies w-h(m_{2})-c(m_{1},t^{\prime})>
U(t^{\prime}).
\end{equation}
Then, the equilibrium is said to violate Criterion D1 unless it is the case
that $supp$ $\mu(m_{1},m_{2})$ $\subset T^{\prime}$.
\end{definition}

The intuition for the definition above is as follows: Upon observing a
signal pair that is not chosen in equilibrium, the receiver places all the
weight on a sender type who finds it strictly profitable to deviate from her
equilibrium action pair whenever another sender type finds it weakly
profitable to deviate from their equilibrium actions.

In Criterion D1, the reference utility for a sender is her equilibrium
utility, but a reasonable belief is defined without reference to equilibrium
utility. We adopt Criterion D1 to further refine the reasonable belief,
leading a unique monotone equilibrium given any amount of resources $M>0.$
When a reasonable belief $\mu \in \Psi $ satisfies Criterion D1, we call it
a reasonable D1 belief.

\section{Unique Monotone D1 Equilibrium}

\label{characterization}

In this section, we precisely characterize all unique monotone equilibria
that satisfy the Criterion D1 given any amount of resources $M>0.$

We focus on reasonable D1 beliefs. By focusing on reasonable beliefs, we are
restricting our attention to monotone equilibria according to Theorem \ref%
{prop2}. 

\subsection{Description of a Unique Monotone D1 Equilibrium}

We begin by describing a unique monotone D1 equilibrium. Given any amount of
resources $M>0$, a \textit{(unique) monotone D1 equilibrium} is precisely
characterized by two sender threshold types $t_{\ell}$ and $t_{h}$: The low
threshold type $t_{\ell}$, is one where a kink occurs for the sender's
signaling choice and the resource constraint is binding for any type of a
sender above it. The high threshold type $t_{h}$, is the other above which
all senders choose a pooled cognitive signal with no non-cognitive signal
and the resource constraint is binding. The discontinuity of the signals can
happen only at the high threshold type.

If $t_{h}$ does not exist (i.e. no discontinuity in the signals), then there
is no pooling part in equilibrium. There are three sub-cases to examine in
this situation: First, suppose that $M$ is large enough such that $%
\underline{t}<t_{\ell }=\overline{t}$. Then every sender type has a
nonbinding resource constraint in equilibrium, and a unique monotone D1
equilibrium is simply the baseline separating equilibrium. If $M$ is not too
large such that $\underline{t}<t_{\ell }<\overline{t}$, then there is a
positive measure of sender types with a binding resource constraint in
equilibrium---the corresponding unique monotone D1 equilibrium has two
distinct separating parts with no pooling, and a sender whose type belongs
to the interval $\left[ t_{\ell },\overline{t}\right] $ has binding resource
constraint in equilibrium. Finally, suppose that $M$ is sufficiently low
such that $\underline{t}=t_{\ell }<\overline{t}$. Here, a unique monotone D1
equilibrium has only one separating part with no pooling, and every sender
(including the lowest sender type) has binding resource constraint in
equilibrium.

If $t_{h}$ exists (i.e. discontinuity in the signals), then one part of a
unique monotone D1 equilibrium is pooling. There are three sub-cases to
consider here: If $M$ is such that $\underline{t}<t_{\ell }<t_{h}<\overline{t%
}$, then a unique monotone D1 equilibrium has two separating parts for
senders in $\left[ \underline{t},t_{h}\right) $, and one pooling part for
senders in $\left[ t_{h},\overline{t}\right] $. Further, every sender type
in $\left[ t_{\ell },\overline{t}\right] $ has binding resource constraint
in equilibrium. If $M$ is such that $\underline{t}<t_{\ell }=t_{h}<\overline{%
t}$, then a unique monotone D1 equilibrium has only one separating part, and
one pooling part for senders in $\left[ t_{\ell },\overline{t}\right] $, and
every sender type in $\left[ t_{\ell },\overline{t}\right] $ has binding
resource constraint in equilibrium. If $M$ is sufficiently low such that $%
\underline{t}=t_{\ell }<t_{h}<\overline{t}$, then a unique monotone D1
equilibrium has only one separating part and a pooling part for senders in $%
\left[ t_{h},\overline{t}\right] $, and every sender type in $\left[
\underline{t},\overline{t}\right] $ has binding resource constraint in
equilibrium.

\subsection{Characterization of a Unique Monotone D1 Equilibrium}

Given any amount of resources $M$, we provide a sharp characterization of
any unique monotone D1 equilibria that is induced by such $M$.

\subsubsection{Unique Monotone D1 Equilibrium with no pooling part}

Let us start with the case where $M$ is such that $\underline{t}<t_{\ell }<%
\overline{t}$. All other cases would naturally be followed from this case.
Consider some $M$ that induces $\underline{t}<t_{\ell }<\overline{t}$. This
means that, all sender types in the interval $\left[ t_{\ell },\overline{t}%
\right] $ have binding resource constraint in equilibrium. We denote a
unique monotone D1 equilibrium which is generated by such $M$ by $%
\{m_{1}^{\ast },m_{2}^{\ast },\mu ^{\ast },w^{\ast }\}$. Because $\underline{%
t}$ has no information rent, her equilibrium action pair $(m_{1}^{\ast }(%
\underline{t}),m_{2}^{\ast }(\underline{t}))$ that solves Problem \ref{eq:
competitive} is efficient, and moreover, because $\underline{t}<t_{\ell }$,
we have $m_{1}^{\ast }(\underline{t})+m_{2}^{\ast }(\underline{t})<M$.
Therefore, the following first order necessary equations for Problem \ref%
{eq: competitive} must be satisfied at $(m_{1}^{\ast }(\underline{t}%
),m_{2}^{\ast }(\underline{t}))$:

\begin{equation}
f_{m_{1}}(m_{1},\underline{t})-c_{m_{1}}(m_{1},\underline{t})=0
\label{eq:foc_lowest1}
\end{equation}
\begin{equation}
\alpha -h^{\prime }(m_{2})=0.  \label{eq:foc_lowest2}
\end{equation}
By applying Assumption 3.(a) and 3.(b) on equations (\ref{eq:foc_lowest1})
and (\ref{eq:foc_lowest2}) respectively, we see that the pair $(m_{1}^{\ast
}(\underline{t}),m_{2}^{\ast }(\underline{t}))$ is unique. In particular,
Assumption 4 implies that $m_{2}^{\ast }(\underline{t})=m_{2}^{\circ }$.

We now derive the equilibrium actions for senders in $\left(\underline{t}
,t_{\ell}\right)$. Because $m^{\ast}_{1}(t)+m^{\ast}_{2}(t)<M$ given $t\in
\left(\underline{t},t_{\ell}\right)$, we have that, for each $t\in \left(
\underline{t},t_{\ell}\right)$, $m^{\ast}_{2}(t)$ must also satisfy (\ref%
{eq:foc_lowest2}). Assumption 4 further implies that $m^{
\ast}_{2}(t)=m_{2}^{\circ}$ for all $t\in \left(\underline{t}
,t_{\ell}\right) $. It follows that, because $m^{\ast}_{2}(t)$ is a constant
on the interval $\left(\underline{t},t_{\ell}\right)$, the equilibrium
belief $\mu^{\ast}$ can only be affected by changes in $m^{\ast}_{1}$.
Therefore, for each sender type $t\in \left(\underline{t},t_{\ell}\right)$,
given that $m^{\ast}_{1}(t)+m^{\ast}_{2}(t)<M$, the equilibrium signal pair $%
\left(m^{\ast}_{1}(t), m^{\ast}_{2}(t)\right)$ that solves Problem \ref{eq:
competitive} must satisfy the following first order necessary condition at $%
m_{1}=m_{1}^{\ast}(t)$:
\begin{equation}  \label{eq:foc_lowest3}
f_{m_1}\left(m_{1}, \mu(m_{1})\right) +f_t\left(m_{1},
\mu(m_{1})\right)\mu^{\prime}(m_{1}) -c_{m_1}\left(m_{1}, t\right)=0.
\end{equation}
We can rewrite Equation (\ref{eq:foc_lowest3}) as
\begin{equation}  \label{eq:foc_lowest4}
\mu^{\prime}(m_{1})= \dfrac{-\left[f_{m_1}\left(m_{1}, \mu(m_{1})\right)-
c_{m_1}\left(m_{1}, t\right)\right]}{f_t\left(m_{1}, \mu(m_{1})\right)}.
\end{equation}
The denominator in (\ref{eq:foc_lowest4}) is strictly positive because of
Assumption 1.(d). We obtain $m^{\ast}_{1}$ by taking the inverse of $\mu$ in
(\ref{eq:foc_lowest4}).\footnote{%
We show in Theorem \ref{theorem_1} below that $\mu$ is increasing and
continuous in $m_{1}$, making it possible to obtain its inverse.} The
uniqueness of $m^{\ast}_{2}(t)$ for $t\in \left(\underline{t}
,t_{\ell}\right) $ is straightforward (Assumption 4). It remains to show
that for every $t\in \left(\underline{t},t_{\ell}\right)$, $m^{\ast}_{1}(t) $
is unique. Lemma \ref{lemma_2} below establishes the uniqueness of $%
m^{\ast}_{1}(t)$ when $t\in \left(\underline{t},t_{\ell}\right)$.

Now consider the case where $t\in\left[t_{\ell},\overline{t}\right]$. For
each $t\in\left[t_{\ell},\overline{t}\right]$, because $m_{1}^{
\ast}(t)+m_{2}^{\ast}(t)=M$, we have that $\left(m_{1}^{\ast}(t),m_{2}^{
\ast}(t)\right)$ that solves Problem \ref{eq: competitive} must satisfy the
following first order necessary condition at $m_{1}=m_{1}^{\ast}(t)$:
\begin{equation}  \label{eq:foc_lowest5}
f_{m_1}\left(m_{1}, \mu(m_{1})\right) +f_t\left(m_{1},
\mu(m_{1})\right)\mu^{\prime}(m_{1}) -c_{m_1}\left(m_{1}, t\right) -
\alpha+h^{\prime}(M-m_{1}) =0.
\end{equation}
We can rewrite (\ref{eq:foc_lowest5}) as
\begin{equation}  \label{eq:foc_lowest6}
\mu^{\prime}(m_{1})= \dfrac{-\left[f_{m_1}\left(m_{1},
\mu(m_{1})\right)-c_{m_1}\left(m_{1}, t\right) -\alpha+h^{\prime}(M-m_{1}) %
\right]}{f_t\left(m_{1}, \mu(m_{1})\right)}
\end{equation}
for all $t\in\left[t_{\ell},\overline{t}\right]$. Again, (\ref%
{eq:foc_lowest6}) is defined because of Assumption 1.(d). For each $t\in %
\left[t_{\ell},\overline{t}\right]$, $m_{1}^{\ast}(t)$ is obtained by taking
the inverse of $\mu$ which is derived from (\ref{eq:foc_lowest6}).\footnote{%
Because $m_{1}^{\ast}(t)+m_{2}^{\ast}(t)=M$, when $t\in\left[t_{\ell},
\overline{t}\right]$, we have $m_{2}^{\ast}(t)=M-m_{1}^{\ast}(t)$ when $t\in %
\left[t_{\ell},\overline{t}\right]$. We are able to take the inverse of $\mu$
to obtain $m_{1}^{\ast}$ because of Theorem \ref{theorem_1}.}

For any $t\in(\underline{t}, t_{\ell})$, and for any $m_{1}\in
\left(m_{1}^{\ast}(\underline{t}),m_{1}^{\ast}(t_{\ell})\right)$, we define
a function $\phi_{s} (m_{1},t)$ as
\begin{equation}  \label{eq:foc_lowest7}
\phi_{s} (m_{1},t):= \dfrac{-\left[f_{m_1}\left(m_{1}, \mu(m_{1})\right)-
c_{m_1}\left(m_{1}, t\right)\right]}{f_t\left(m_{1}, \mu(m_{1})\right)}.
\end{equation}
For any $t\in(t_{\ell},\overline{t})$, and for any $m_{1}\in
\left(m_{1}^{\ast}(t_{\ell}),m_{1}^{\ast}(\overline{t})\right)$, define a
function $\phi_{b} (m_{1},t)$ as
\begin{equation}  \label{eq:foc_lowest8}
\phi_{b} (m_{1},t):= \dfrac{-\left[f_{m_1}\left(m_{1},
\mu(m_{1})\right)-c_{m_1}\left(m_{1}, t\right) -\alpha+h^{\prime}(M-m_{1}) %
\right]}{f_t\left(m_{1}, \mu(m_{1})\right)}.
\end{equation}

Consider the first order nonlinear differential equation $%
\mu^{\prime}(m_{1})=\phi (m_{1},t)$ where
\begin{equation}  \label{eq:foc_lowest8c}
\phi (m_{1},t):=\left\{
\begin{array}{cc}
\phi_{s} (m_{1},t) & \text{if }t\in(\underline{t}, t_{\ell}), \text{and}~
m_{1}\in \left(m_{1}^{\ast} (\underline{t}),m_{1}^{\ast}(t_{\ell})\right) \\
\phi_{b} (m_{1},t) & \text{if }t\in(t_{\ell},\overline{t}), \text{and}~m_{1}
\in \left(m_{1}^{\ast}(t_{\ell}),m_{1}^{\ast}(\overline{t})\right)%
\end{array}
\right. .
\end{equation}

The lemma below establishes the solution for the above first order nonlinear
differential equation.

\begin{lemma}
\label{lemma_2} Let $f$, $c$, and $h$ be such that $\phi$ which is defined
in (\ref{eq:foc_lowest8c}) is a uniformly Lipschitz continuous function. The
solution for the first order differential equation $\mu^{\prime}(m_{1})=\phi
(m_{1},t)$ exists and it is unique.
\end{lemma}

The proof of Lemma \ref{lemma_2} follows from Picard-Lindelof's existence
and uniqueness theorem. Let $\mu^{\ast}$ denote the unique solution for the
differential equation in Lemma \ref{lemma_2}. Given the unique solution $%
\mu^{\ast}$, we obtain the unique equilibrium level of a cognitive signal $%
m_{1}^{\ast}$ by $m_{1}^{\ast}=\mu^{\ast-1}$. Given $M$, for every $%
t\in(t_{\ell},\overline{t})$, we have $m_{2}^{\ast}(t)=M-m_{1}^{\ast}(t)$,
which is unique because the uniqueness of $m_{1}^{\ast}$.

In Theorem \ref{theorem_1} below, the compactness of the range of $%
m_{1}^{\ast}$, and the range of $m_{2}^{\ast}$ follows directly from the
continuity of $m_{1}^{\ast}$ and $m_{2}^{\ast}$ on the compact type space $%
\left[\underline{t},\overline{t}\right]$, which we prove in the Appendix. In
Item 2 of Theorem \ref{theorem_1}, the differentiability of the belief $\mu$
comes from the continuity of $m_{1}^{\ast}$ and $m_{2}^{\ast}$, and
Assumptions 1.(a), 1.(c), and 3.(c).

An important observation is that the equilibrium belief $\mu$ forms a kink
at $m_{1}^{\ast}(t_{\ell})$. This happens because of the differences in the
right hand sides of (\ref{eq:foc_lowest4}) and (\ref{eq:foc_lowest6}). See
Lemma \ref{continuous} in the Appendix for the continuity of $m_{1}^{\ast}$
and $m_{2}^{\ast}$.

\begin{theorem}
\label{theorem_1} In any separating unique monotone D1 equilibrium with $%
\underline{t}<t_{\ell }<\overline{t}$, the following hold:

\begin{enumerate}
\item \emph{Range}$(m_{1}^{\ast})$ and \emph{Range}$(m_{2}^{\ast})$ are
compact intervals.

\item $\mu:\mathbb{R}_{+}\rightarrow \Delta(T)$ is strictly increasing and
continuous on \emph{Range}$(m_{1}^{\ast})$ and it has a continuous
derivative $\mu^{\prime}$ on $\left( m_{1}^{\ast}(\underline{t}
),m_{1}^{\ast}(t_{\ell}) \right) \cup
\left(m_{1}^{\ast}(t_{\ell}),m_{1}^{\ast}(\overline{t})\right)$.
\end{enumerate}
\end{theorem}

Because $\mu$ is the inverse of $m_{1}^{\ast}$, Item 2 of Theorem \ref%
{theorem_1} implies that $m_{1}^{\ast}$ is differentiable. Also, because $%
\mu $ is kinked at $m_{1}^{\ast}(t_{\ell})$, the equilibrium action $%
m_{1}^{\ast} $ also forms a kink at $t_{\ell}$.

We show in Proposition \ref{prop_5} that the respective first order
conditions are both necessary and sufficient for $(m_{1}^{\ast},m_{2}^{
\ast}) $ to be optimal for each respective sender type.

\begin{proposition}
\label{prop_5}Consider the $M$ that induces a unique separating monotone D1
equilibrium $\{m_{1}^{\ast },m_{2}^{\ast },\mu ^{\ast },w^{\ast }\}$ with $%
\underline{t}<t_{\ell }<\overline{t}$. Then the following are the necessary
and sufficient conditions for such an equilibrium:

\begin{enumerate}
\item $f_{m_{1}}(m_{1}^{\ast}(\underline{t}),\underline{t})-
c_{m_{1}}(m_{1}^{\ast}(\underline{t}),\underline{t}) =0$ and $\alpha-
h^{\prime}(m_{2}^{\ast}(\underline{t}))=0$ such that $m_{1}^{\ast}(
\underline{t})+m_{2}^{\ast}(\underline{t})<M$.

\item $(m_{1}^{\ast}(t),m_{2}^{\ast}(t))$ satisfies $f_{m_1}\left(m_{1},
\mu(m_{1})\right) +f_t\big(m_{1}, \mu(m_{1})\big)\mu^{\prime}(m_{1})
-c_{m_1} \big(m_{1}, t\big)=0$ and $\alpha-h^{\prime}(m_{2})=0$ at $%
m_{1}=m_{1}^{\ast}(t)$, $m_{2}=m_{2}^{\ast}(t)$ such that $m_{1}+m_{2}<M$
for all $t\in (\underline{t},t_{\ell})$.

\item $(m_{1}^{\ast}(t),m_{2}^{\ast}(t))$ satisfies $f_{m_1}\left(m_{1},
\mu(m_{1})\right) +f_t\big(m_{1}, \mu(m_{1})\big)\mu^{\prime}(m_{1})
-c_{m_1} \big(m_{1}, t\big)- \alpha+h^{\prime}(M-m_{1})=0$ at $%
m_{1}=m_{1}^{\ast}(t)$ , $m_{2}=m_{2}^{\ast}(t)$ such that $m_{1}+m_{2}=M$
for all $t\in \left(t_{\ell},\overline{t}\right)$.
\end{enumerate}
\end{proposition}

Following the characterization of the unique separating monotone D1
equilibrium in Proposition \ref{prop_5}, all other unique separating
monotone D1 equilibrium follow naturally: Suppose that $M$ is sufficiently
large so that every sender's resource constraint is not binding in
equilibrium, i.e., $\underline{t}<t_{\ell }=\overline{t}$. Here, the induced
unique monotone D1 equilibrium is the \textit{baseline separating
equilibrium } which we denote by $\{\bar{m}_{1},\bar{m}_{2},\bar{\mu},\bar{w}%
\}$. Because $t_{\ell }=\overline{t}$, we have $\bar{m}(t)+\bar{m}_{2}(t)<M$
for all $t\in \left[ \underline{t},\overline{t}\right) $. Therefore, the
equilibrium behavior of a sender with type in $\left[ \underline{t},%
\overline{t}\right) $ coincides with the equilibrium behavior of a sender
with type $t\in \left[ \underline{t},t_{\ell }\right) $ in a unique monotone
D1 equilibrium with $\underline{t}<t_{\ell }<\overline{t}$, which we have
previously analyzed (Proposition \ref{prop_5}). In summary, suppose that $M$
is sufficiently large so that every sender's resource constraint is not
binding in equilibrium, i.e., $\underline{t}<t_{\ell }=\overline{t}$, then
for each $t\in \left[ \underline{t},t_{\ell }\right) $, we have $\bar{m}%
_{1}(t)=m_{1}^{\ast }(t)$ and $\bar{m}_{2}(t)=m_{2}^{\ast }(t)=m_{2}^{\circ
} $.

On the other hand, if $M$ is extremely low such that $\underline{t}=t_{\ell
}<\overline{t}$, then the induced unique monotone D1 equilibrium which we
denote by $\{\check{m}_{1},\check{m}_{2},\check{\mu},\check{w}\}$, has only
one separating part with no pooling, and every sender has binding resource
constraint in equilibrium. That is, for each $t\in \left[ \underline{t},%
\overline{t}\right] $, $\check{m}_{1}(t)+\check{m}_{2}(t)=M$. Therefore, the
equilibrium behavior of a sender with type in $\left[ \underline{t},%
\overline{t}\right] $ coincides with the equilibrium behavior of a sender
with type $t\in \left[ t_{\ell },\overline{t}\right] $ in the unique
monotone D1 equilibrium with $\underline{t}<t_{\ell }<\overline{t}$ which we
earlier studied (Proposition \ref{prop_5}).

\subsubsection{Unique Monotone D1 Equilibrium with a pooling part}

Consider an amount of resources $M$ that induces an equilibrium in which the
resource constraint is binding for a positive measure of sender types in $%
\left[t_{\ell}, \overline{t}\right]$, but there is discontinuity in the
signals. This is the case where a unique monotone equilibrium has a pooling
part i.e. $t_{h}\in T$ exists. In this case, we let $Z(m_{1},m_{2}) $ denote
the set of types of senders who choose the same signal pair $(m_{1},m_{2})$
in equilibrium. We begin our analysis by considering the case in which such $%
M$ induces an equilibrium with $\underline{t}<t_{\ell}< t_{h}<\overline{t}$.
All other unique monotone equilibria with a pooling part follow naturally.

In Theorem \ref{lemma40}, we show that resource constraint must be binding
for senders who choose a pooled pair of cognitive and non-cognitive signal
in equilibrium.

\begin{theorem}
\label{lemma40} $m_{1}+m_{2}=M$ if $Z(m_{1},m_{2})$ has a positive measure
in equilibrium.
\end{theorem}

In the lemma below, we show that if a positive measure of sender types
choose the same pair of cognitive and non-cognitive actions in equilibrium,
and the highest among them is an interior type, then there is an interval of
discontinuity created by such type.

\begin{lemma}
\label{lemma4} Suppose that $Z(m_{1},m_{2})$ has a positive measure in
equilibrium. Then $\displaystyle{\hat{m_{1}}(t_{\circ}) <\lim_{t\searrow
t_{\circ}} \hat{m_{1}}(t)}$ and $\displaystyle{\hat{m_{2}}(t_{\circ}) >
\lim_{t\searrow t_{\circ}} \hat{m_{2}}(t)}$ when $\overline{t}> t_{\circ}:=
\text{max~} Z(m_{1},m_{2})$.
\end{lemma}

When a positive measure of sender types choose the same pair of cognitive
and non-cognitive actions in equilibrium, this is referred to as \textit{\
bunching}. In Theorem \ref{prop3} below, we show that bunching must occur
only on the top if it happens. Moreover, all sender types who choose a
pooled pair of action pair in equilibrium must allocate all their resources
to the cognitive signal with no level of a non-cognitive signal.

\begin{theorem}
\label{prop3} Suppose that $Z(m_{1},m_{2})$ has a positive measure in
equilibrium.

\begin{enumerate}
\item[(a)] $Z(m_{1},m_{2})$ is a connected interval with $\text{max~}
Z(m_{1},m_{2})=\overline{t}$.

\item[(b)] $m_{1}=M$.
\end{enumerate}
\end{theorem}

\begin{corollary}
\label{prop4} There is no bunching on the bottom if Criterion D1 is
satisfied in equilibrium.
\end{corollary}

According to Theorem \ref{prop3}, when bunching happens it must be only on
the top, and all sender types who choose a pooled pair of action pair in
equilibrium must allocate all their resources to the cognitive signal with
no level of a non-cognitive signal. Therefore for bunching to occur, it is
necessary for some interior sender type to spend all her resources on
cognitive signal investment. This is captured in Condition A.

\begin{description}
\item[Condition A] : \label{eq:cond_A}There exists $t^{\prime}\in$ int($T$)
such that $m_{1}^{\ast}(t^{\prime})=M$ where $m_{1}^{\ast}(t^{\prime})$
solves equation (\ref{eq:foc_lowest5}) at $t=t^{\prime}$.
\end{description}

Condition A guarantees that a positive measure of sender types will choose
the pooled action pair $(M,0)$ in equilibrium according to Theorem \ref%
{prop3}.

Given such $M$, we denote a unique monotone D1 equilibrium with $\underline{
t }<t_{\ell}< t_{h}<\overline{t}$ by $\{\tilde{m_{1}},\tilde{m_{2}},\tilde{
\mu} ,\tilde{w}\}$ such that
\begin{equation}  \label{eq:nonbinding51}
\left(\tilde{m_{1}}(t),\tilde{m_{2}}(t)\right) =\left\{
\begin{array}{cc}
\left(m_{1}^{\ast}(t),m_{2}^{\ast}(t)\right) & \text{if }t\in[\underline{t}
,t_{h}) \\
\left(M,0\right) & \text{if }t\in\big[t_{h},\overline{t}\big]%
\end{array}
\right.
\end{equation}
for some $t_{h}$ $\in(\underline{t},t^{\prime}]$ with $t^{\prime}\in$ int($T$
) where $t_{h}$ solves the indifference condition in (\ref{eq:nonbinding52})
at $t=t_{h}$:
\begin{equation}  \label{eq:nonbinding52}
\alpha m_{2}^{\ast}(t)+f(m_{1}^{\ast}(t),t)-c(m_{1}^{\ast}(t),t)-h(m_{2}^{
\ast}(t))= \mathbb{E}f(M,z|z\geq t)- c(M,t).
\end{equation}
The following proposition establishes the existence and uniqueness of such $%
t_{h}$.

\begin{proposition}
\label{nonbinding53}A unique solution $t_{h}$ $\in (\underline{t},
t^{\prime} ]$ for equation (\ref{eq:nonbinding52}) exists if
\begin{equation}  \label{eq:nonbinding54}
\mathbb{E}f(M,z|z\geq \underline{t})- c(M,\underline{t})<\alpha
m_{2}^{\ast}( \underline{t})+ f(m_{1}^{\ast}(\underline{t}),\underline{t})-
c(m_{1}^{\ast}( \underline{t}),\underline{t})- h(m_{2}^{\ast}(\underline{t}
)).
\end{equation}
\end{proposition}

Condition (\ref{eq:nonbinding54}) states that the lowest sender type cannot
choose the pooled action pair $(M,0)$ in equilibrium.

Given $M$, the induced unique monotone D1 equilibrium $\{\tilde{m_{1}},
\tilde{m_{2}},\tilde{\mu},\tilde{w}\}$ with $\underline{t}<t_{\ell}< t_{h}<
\overline{t}$ can be graphically described in Figure \ref{fig:figure1}
below. The blue colored parts represent the equilibrium actions for senders
in a unique monotone D1 equilibrium with $\underline{t}<t_{\ell}< t_{h}<
\overline{t}$.\footnote{%
Note that $\lim_{t\nearrow t_{h}}\tilde{m}_{1}(t)=m^{\ast}_{1}(t_{h})$ and $%
\lim_{t\nearrow t_{h}}\tilde{m}_{2}(t)=m^{\ast}_{2}(t_{h})$ in Figure \ref%
{fig:figure1}}
\begin{figure}[tph]
\centering
\subfloat{{\includegraphics[width=8cm]{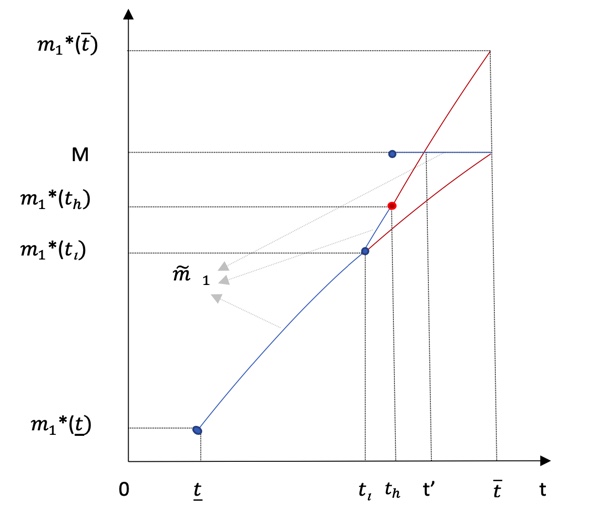} }} \subfloat{{								%
\includegraphics[height=8cm, width=8.2cm]{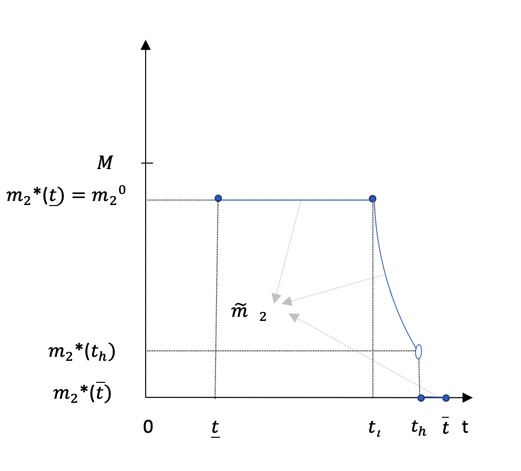} }} \vspace{-5mm}
\caption{Senders' equilibrium action pairs in a unique monotone D1
equilibrium with $\protect\underline{t}<t_{\ell}< t_{h}<\overline{t}$.}
\label{fig:figure1}
\end{figure}

The theorem below provides the full characterization of a unique monotone D1
equilibrium $\{\tilde{m_{1}},\tilde{m_{2}},\tilde{\mu},\tilde{w}\}$ which is
induced by some $M$ that gives $\underline{t}<t_{\ell }<t_{h}<\overline{t}$.

\begin{theorem}
\label{theorem_2} Fix the $M>0$ that induces $\underline{t}<t_{\ell}< t_{h}<
\overline{t}$. Then, there exists a unique monotone D1 equilibrium $\{\tilde{
m_{1}},\tilde{m_{2}},\tilde{\mu},\tilde{w}\}$. We characterize it as follow:

\begin{enumerate}
\item $\left(\tilde{m_{1}},\tilde{m_{2}}\right)$ follows (i) $\left(\tilde{
m_{1}}(t),\tilde{m_{1}}(t)\right)=\left(m_{1}^{\ast}(t),m_{2}^{\ast}(t)
\right)$ such that $m_{1}^{\ast}(t)+m_{2}^{\ast}(t)<M$ for all $t\in[
\underline{t}, t_{\ell})$; (ii) $\left(\tilde{m_{1}}(t),\tilde{m_{1}}
(t)\right)=\left(m_{1}^{\ast}(t),m_{2}^{\ast}(t)\right)$ such that $%
m_{1}^{\ast}(t)+m_{2}^{\ast}(t)=M$ for all $t\in[t_{\ell},t_{h})$; (iii) $%
\tilde{m_{1}}(t)=M$ if $t\in \big[t_{h}, \overline{t}\big]$.

\item $\tilde{\mu}$ follows\newline
(i) $\text{supp}~\tilde{\mu}(m_{1}^{\prime},m_{2}^{\prime})=\{\underline{t}
\} $ if $m_{1}^{\prime}\in (0,m_{1}^{\ast}(\underline{t}) )$ such that $%
m_{1}^{\prime}+ m_{2}^{\prime}\leq M$.

(ii) $\text{supp}~\tilde{\mu}(m_{1}^{\prime},m_{2}^{\prime})=\{t_{h}\}$ if $%
m_{1}^{\prime}\in [m_{1}^{\ast}(t_{h}),M)$ such that $m_{1}^{\prime}+
m_{2}^{\prime}= M$ where $m_{1}^{\ast}(t_{h}):=\lim_{t\searrow t_{h}} \tilde{
m_{1}}(t)$.

(iii) $\tilde{\mu}(m_{1}, m_{2}^{\prime})=t^{\ast}\in \left[\underline{t}
,t_{\ell}\right)$ such that $m_{1}^{\ast}(t^{\ast})=m_{1}$ with $m_{1}\in %
\left[m_{1}^{\ast}(\underline{t}),m_{1}^{\ast}(t_{\ell})\right)$ such that $%
m_{1}+m_{2}^{\prime}\leq M$.

(iv) $\tilde{\mu}(m_{1}, m_{2}^{\prime})=t^{\ast}\in \left[
t_{\ell},t_{h}\right)$ such that $m_{1}^{\ast}(t^{\ast})=m_{1}$ with $%
m_{1}\in \left[m_{1}^{\ast}(t_{\ell}),m_{1}^{\ast}(t_{h})\right)$ such that $%
m_{1}+m_{2}^{\prime}\leq M$.

(v) $\tilde{\mu}(m_{1},m_{2}^{\circ})=m_{1}^{*-1}(m_{1})$ for $m_{1}\in %
\left[m_{1}^{\ast}(\underline{t}),m_{1}^{\ast}(t_{\ell})\right)$ such that $%
m_{1}+m_{2}^{\circ}<M$.

(vi) $\tilde{\mu}(m_{1},m_{2})=m_{1}^{*-1}(m_{1})$ for all $m_{1}\in \left[
m_{1}^{\ast}(t_{\ell}),m_{1}^{\ast}(t_{h})\right)$ such that $m_{1}+m_{2}=M$.

(vii) $\text{supp}~\tilde{\mu}(m_{1},m_{2})=\big[t_{h}, \overline{t}\big]$
if $m_{1}=M$ and $m_{2}=0$.
\end{enumerate}
\end{theorem}

Note that Theorem \ref{theorem_2} allows for the possibility of all other
unique monotone D1 equilibria. For example, if $t_{\ell }\rightarrow t_{h}$
and $t_{h}\rightarrow \overline{t}$, then a unique monotone D1 equilibrium
is simply the baseline separating equilibrium. In a special case where $M$
induces $\underline{t}<t_{\ell }=t_{h}<\overline{t}$, the corresponding
unique monotone D1 equilibrium has only one separating part, with a pooling
part for senders in $\left[ t_{\ell },\overline{t}\right] $, and every
sender type in $\left[ t_{\ell },\overline{t}\right] $ has binding resource
constraint in equilibrium.

\section{Conclusion}

Our paper provides a sharp characterization of a unique monotone D1
equilibria given any resource constraint. Given any amount of resources $M>0$
, a \textit{(unique) monotone D1 equilibrium} is precisely characterized by
two sender threshold types $t_{\ell }$ and $t_{h}$: The low threshold type $%
t_{\ell }$, is one where a kink occurs for the sender's signaling choice and
the resource constraint is binding for any type of a sender above it. The
high threshold type $t_{h}$, is the other above which all senders choose a
pooled cognitive signal with no non-cognitive signal and the resource
constraint is binding. The discontinuity of the signals can happen only at
the high threshold type.

The sharp monotone equilibrium characterization opens the way to study a
decision maker's optimal choice of the amount of signaling resources that
senders can invest in their multidimensional signaling to maximize the
decision maker's aggregate net surplus. A follow-up paper of ours (Han and
Sam (2024)) explores such an optimal design problem. In that paper, we
identify a trade-off in the decision maker's optimal choice that is not
present in a one-dimensional signaling model. Compared to the baseline
separating equilibrium with no restrictions on the amount of resources, a
binding resource constraint due to a lower amount of feasible resources
leads to an equilibrium where high types reduce investment in non-cognitive
signal (efficiency loss) to finance the investment in the cognitive signal.
Nonetheless, the investment in the cognitive signal is lower than that in
the baseline separating equilibrium and it may leads to efficiency gain.

\addcontentsline{toc}{section}{Appendices}

\renewcommand{\thesection}{\Alph{section}} \setcounter{section}{0} %
\setcounter{equation}{0} \renewcommand{\theequation}{A\arabic{equation}}




\section*{Proof of Lemma \protect\ref{lemma_2} and Theorem \protect\ref%
{theorem_1}}

We break the proofs in lemmas. Lemma \ref{lemma_Off2} below is crucial in
establishing the continuity of $m_{1}^{\ast}$ and $m_{2}^{\ast}$ (Lemma \ref%
{continuous} below), which are useful for the proof of Theorem \ref%
{theorem_1}.

\begin{lemma}
\label{lemma_Off2} Let $\{\hat{m}_{1},\hat{m}_{2},\hat{w},\hat{\mu}\}$ with $%
\hat{\mu}\in \Psi$ be any equilibrium such that $\hat{m}_{1}+\hat{m}_{2}=M$.
Suppose $\hat{\mu}(m_{1}^{\prime},m_{2}^{\prime})$ conditional on any $%
(m_{1}^{\prime},m_{2}^{\prime})\notin $ \emph{range}$(\hat{m}_{1},\hat{m}
_{2})$ passes Criterion D1. Then $\text{supp~}\hat{\mu}(m^{\prime}_{1},m^{
\prime}_{2}) =\{z\}$ if $m_{1}^{\prime }$ and $m_{2}^{\prime }$ belongs to
the interval of off-path actions induced by the discontinuity of $\hat{m}
_{1} $ and $\hat{m}_{2}$ at $z$ such that $m_{1}^{\prime }+m_{2}^{\prime
}\leq M$.
\end{lemma}

\begin{proof}
. Because $\hat{m}_{1}+\hat{m}_{2}=M$, $\hat{m}_{2}$ is discontinuous at $z$
when $\hat{m}_{1}$ is discontinuous at $z$. Let us assume that $\hat{m}_{1} $
is right continuous at $z$, so that the product interval of off-path actions
induced by the discontinuity of $\hat{m}_{1}$ at $z$ is $[\hat{m}
_{1}(z_{-}), \hat{m}_{1}(z)) \times (\hat{m}_{2}(z),\hat{m}_{2}(z_{-})]$,
where $\hat{m} _{1}(z_{-})=\lim_{t\nearrow z} \hat{m}_{1}(t)$, and $\hat{m}
_{2}(z_{-})=\lim_{t\nearrow z} \hat{m}_{2}(t)$. Therefore, the intervals $[
\hat{m}_{1}(z_{-}),\hat{m}_{1}(z))$ and $(\hat{m}_{2}(z),\hat{m}_{2}(z_{-})]
$ are nonempty. Because $\hat{\mu}\in \Psi$, $\hat{m}_{1}$ and $\hat{m}_{2} $
are monotone according to Theorem \ref{prop2}. We show that $\text{supp~}
\hat{\mu}(m^{\prime}_{1},m^{\prime}_{2}) =\{z\}$ for any $%
(m_{1}^{\prime},m_{2}^{\prime})\notin $ \emph{range}$(\hat{m}_{1},\hat{m}
_{2})$ with $m_{1}^{\prime}\in[\hat{m}_{1}(z_{-}),\hat{m}_{1}(z))$ and $%
m_{2}^{\prime}\in (\hat{m}_{2}(z),\hat{m}_{2}(z_{-})]$ such that $%
m_{1}^{\prime }+m_{2}^{\prime }\leq M$, if Criterion D1 is satisfied.

We first show that $t>z$ cannot be in the support of $\hat{\mu}
(m^{\prime}_{1},m^{\prime}_{2})$ for any $(m_{1}^{\prime},m_{2}^{\prime})
\notin $ \emph{range}$(\hat{m}_{1},\hat{m}_{2})$ with $m_{1}^{\prime}\in[
\hat{m}_{1}(z_{-}),\hat{m}_{1}(z))$ and $m_{2}^{\prime}\in (\hat{m}_{2}(z),
\hat{m}_{2}(z_{-})]$ such that $m_{1}^{\prime }+m_{2}^{\prime }< M$. On the
contrary, suppose that $t\in \text{supp}~\hat{\mu}(m^{\prime}_{1},m^{
\prime}_{2})$ for some $(m_{1}^{\prime},m_{2}^{\prime})\notin $ \emph{range}$%
(\hat{m}_{1},\hat{m}_{2})$ with $m_{1}^{\prime}\in[\hat{m}_{1}(z_{-}),\hat{m}
_{1}(z))$ and $m_{2}^{\prime}\in (\hat{m}_{2}(z),\hat{m}_{2}(z_{-})]$ such
that $m_{1}^{\prime }+m_{2}^{\prime }\leq M$, when $t>z$. Given the wage $%
w^{\prime}$ chosen by the receiver after observing such off-path message,
let
\begin{equation}  \label{eq:bunchinge1}
w^{\prime}-c(m^{\prime}_{1},z)-h(m^{\prime}_{2})\leq \hat{w}(\hat{m_{1}}(z),
\hat{m_{2}}(z))-c(\hat{m_{1}}(z),z)-h(\hat{m_{2}}(z)),
\end{equation}
which can be rewritten as
\begin{equation}  \label{eq:bunchinge2}
w^{\prime}-h(m^{\prime}_{2})- \left[\hat{w}(\hat{m_{1}}(z),\hat{m_{2}}
(z))-h( \hat{m_{2}}(z))\right] \leq c(m^{\prime}_{1},z)-c(\hat{m_{1}}(z),z).
\end{equation}
Because $m_{1}^{\prime}\in[\hat{m}_{1}(z_{-}),\hat{m}_{1}(z))$, we have that
$m_{1}^{\prime}<\hat{m}_{1}(z)$. Therefore, we can use the strict single
crossing condition of $c$ together with (\ref{eq:bunchinge2}) to show that
for $t>z$,
\begin{equation}  \label{eq:bunching30}
w^{\prime}-c(m^{\prime}_{1},t)-h(m^{\prime}_{2})< \hat{w}(\hat{m_{1}}(z),
\hat{m_{2}}(z))-c(\hat{m_{1}}(z),t)-h(\hat{m_{2}}(z)).
\end{equation}
Also, in equilibrium, we must have
\begin{equation}  \label{eq:bunching40}
\hat{w}(\hat{m_{1}}(z),\hat{m_{2}}(z))-c(\hat{m_{1}}(z),t)-h(\hat{m_{2}}
(z))\leq \hat{w}(\hat{m_{1}}(t),\hat{m_{2}}(t))-c(\hat{m_{1}}(t),t)-h(\hat{
m_{2}}(t)).
\end{equation}
Combining (\ref{eq:bunching30}) and (\ref{eq:bunching40}) leads to
\begin{equation}  \label{eq:bunching50}
w^{\prime}-c(m^{\prime}_{1},t)-h(m^{\prime}_{2})< \hat{w}(\hat{m_{1}}(t),
\hat{m_{2}}(t))-c(\hat{m_{1}}(t),t)-h(\hat{m_{2}}(t))
\end{equation}
which shows that $t>z$ cannot be in the support of $\hat{\mu}
(m^{\prime}_{1},m^{\prime}_{2})$ for any $(m_{1}^{\prime},m_{2}^{\prime})
\notin $ \emph{range}$(\hat{m}_{1},\hat{m}_{2})$ with $m_{1}^{\prime}\in[
\hat{m}_{1}(z_{-}),\hat{m}_{1}(z))$ such that $m_{1}^{\prime }+m_{2}^{\prime
}\leq M$. It follows from the contrapositive of (\ref{D1}) in the definition
of Criterion D1 that any $t>z$ cannot be in the support of $\hat{\mu}
(m^{\prime}_{1},m^{\prime}_{2})$ $(m_{1}^{\prime},m_{2}^{\prime})\notin $
\emph{range}$(\hat{m}_{1},\hat{m}_{2})$ with $m_{1}^{\prime}\in[\hat{m}
_{1}(z_{-}),\hat{m}_{1}(z))$ and $m_{2}^{\prime}\in (\hat{m}_{2}(z),\hat{m}
_{2}(z_{-})]$ such that $m_{1}^{\prime }+m_{2}^{\prime }\leq M$.

Finally, we show that $t<z$ cannot be in the support of $\hat{\mu}
(m^{\prime}_{1},m^{\prime}_{2})$ for any $(m_{1}^{\prime},m_{2}^{\prime})
\notin $ \emph{range}$(\hat{m}_{1},\hat{m}_{2})$ with $m_{1}^{\prime}\in[
\hat{m}_{1}(z_{-}),\hat{m}_{1}(z))$ such that $m_{1}^{\prime }+m_{2}^{\prime
} \leq M$. On the contrary, suppose that $t\in \text{supp}~\hat{\mu}
(m^{\prime}_{1},m^{\prime}_{2})$ for some $(m_{1}^{\prime},m_{2}^{\prime})
\notin $ \emph{range}$(\hat{m}_{1},\hat{m}_{2})$ with $m_{1}^{\prime}\in[
\hat{m}_{1}(z_{-}),\hat{m}_{1}(z))$ such that $m_{1}^{\prime }+m_{2}^{\prime
}\leq M$, when $t<z$. If $t<z$ then we have $t^{\prime}$ such that $%
t<t^{\prime}<z$. Given the wage $w^{\prime}$ chosen by the receiver after
observing such off-path message, let
\begin{equation}  \label{eq:bunching60}
w^{\prime}-c(m^{\prime}_{1},t)-h(m^{\prime}_{2})\geq \hat{w}(\hat{m_{1}}(t),
\hat{m_{2}}(t))-c(\hat{m_{1}}(t),t)-h(\hat{m_{2}}(t)).
\end{equation}
Because
\begin{equation}  \label{eq:bunching70}
\hat{w}(\hat{m_{1}}(t),\hat{m_{2}}(t))-c(\hat{m_{1}}(t),t)-h(\hat{m_{2}}
(t))\geq \hat{w}(\hat{m_{1}}(t^{\prime}),m_{2}(t^{\prime}))-c(\hat{m_{1}}
(t^{\prime}),t)-h(m_{2}(t^{\prime}))
\end{equation}
in equilibrium, it follows from (\ref{eq:bunching50}) that
\begin{equation}  \label{eq:bunching80}
w^{\prime}-c(m^{\prime}_{1},t)-h(m^{\prime}_{2})\geq \hat{w}(\hat{m_{1}}
(t^{\prime}),m_{2}(t^{\prime}))-c(\hat{m_{1}}(t^{\prime}),t)-h(m_{2}(t^{
\prime}))
\end{equation}
which can be rewritten as
\begin{equation}  \label{eq:bunching90}
w^{\prime}-\hat{w}(\hat{m_{1}}(t^{\prime}),m_{2}(t^{\prime}))+h(m_{2}(t^{
\prime}))-h(m^{\prime}_{2})\geq c(m^{\prime}_{1},t)-c(\hat{m_{1}}
(t^{\prime}),t).
\end{equation}
Because $m^{\prime}_{1}>\hat{m_{1}}(t^{\prime})$, we can use the single
crossing condition to show that (\ref{eq:bunching90}) implies that for $%
t^{\prime}>t$,
\begin{equation}  \label{eq:bunching100}
w^{\prime}-c(m^{\prime}_{1},t^{\prime})-h(m^{\prime}_{2})> \hat{w}(\hat{
m_{1} }(t^{\prime}),m_{2}(t^{\prime}))-c(\hat{m_{1}}(t^{\prime}),t^{
\prime})-h(m_{2}(t^{\prime}))
\end{equation}
which shows that the sender of type $z$ is strictly better off with the same
deviation. Criterion D1 implies that any $t<z$ cannot be in the support of $%
\hat{\mu}(m^{\prime}_{1},m^{\prime}_{2})$ for any $(m_{1}^{\prime},m_{2}^{
\prime})\notin $ \emph{range}$(\hat{m}_{1},\hat{m}_{2})$ with $%
m_{1}^{\prime} \in[\hat{m}_{1}(z_{-}),\hat{m}_{1}(z))$ and $%
m_{2}^{\prime}\in (\hat{m} _{2}(z),\hat{m}_{2}(z_{-})]$ such that $%
m_{1}^{\prime }+m_{2}^{\prime } \leq M$. We can similarly prove this lemma
when $\hat{m}_{1}$ is left continuous at $z$.
\end{proof}

\begin{lemma}
\label{continuous} Let $\{m_{1}^{\ast },m_{2}^{\ast },\mu ^{\ast }\}$ with $%
\mu ^{\ast }\in \Psi $ be any separating D1 equilibrium such that $%
\underline{t}<t_{\ell }<\overline{t}$. Then $m_{1}^{\ast }$ and $m_{2}^{\ast
}$ are continuous on $\left[ \underline{t},\overline{t}\right] $.
\end{lemma}

\begin{proof}
\textbf{of Lemma \ref{continuous}.} Let us split the interval $\left[
\underline{t},\overline{t}\right]$ into two parts: $\left[\underline{t}
,t_{\ell}\right)$ and $\left[t_{\ell},\overline{t}\right]$. We show that $%
m_{1}^{\ast}$ and $m_{2}^{\ast}$ are continuous on each of the given
subintervals.

Consider the subinterval $\left[t_{\ell},\overline{t}\right]$. We prove by
contradiction. Suppose that at least one of the equilibrium action pair is
discontinuous at some $t\in\left[t_{\ell},\overline{t}\right]$. Because $%
m_{1}^{\ast}(t)+m_{2}^{\ast}(t)=M$ when $t\in \left[t_{\ell},\overline{t} %
\right]$, if $m_{1}^{\ast}$ is discontinuous at some $t\in\left[t_{\ell},
\overline{t}\right]$, then $m_{2}^{\ast}$ is also discontinuous at such $%
t\in \left[t_{\ell},\overline{t}\right]$, and vice versa. Because $%
\mu^{\ast}\in \Psi$, we have that $m_{1}^{\ast}$ and $m_{2}^{\ast}$ are
monotone functions (Theorem \ref{weakmonotone}). Therefore, the
discontinuity of $m_{1}^{\ast}$ and $m_{2}^{\ast}$ at $t\in $$\big[
\underline{t}, \overline{t}\big]$ is a jump discontinuity according to
Froda's Theorem\footnote{\textbf{Froda's Theorem:}\label{frodas} Let $f$ be
a monotone function on the interval $[a,b] $. Then $f$ has at most a
countably infinite number of jump discontinuities.}. Let $m_{1}^{\ast}$ be
left continuous at some $t\in\left[ t_{\ell},\overline{t}\right]$ so that
the discontinuity of $m_{1}^{\ast}$ will only be to the right. It implies
that $\displaystyle{\ m_{1}^{\ast}(t)<\lim_{k\searrow t} m_{1}^{\ast}(k)}$
and $\displaystyle{\ m_{2}^{\ast}(t)>\lim_{k\searrow t} m_{2}^{\ast}(k)}$.
Hence, the discontinuity of $m_{1}^{\ast}$ and $m_{2}^{\ast}$ are
discontinuous at some $t\in\left[t_{\ell},\overline{t}\right]$ creates a
product interval of off-path messages $D$ defined as follows:
\begin{equation*}
D=\Big\{(s_1,s_2)\in \mathcal{C}: (s_1,s_2)\in \displaystyle{\big( %
m_{1}^{\ast}(t),\lim_{k\searrow t} m_{1}^{\ast}(k) \big)\times \big( %
\lim_{k\searrow t} m_{2}^{\ast}(k),m_{2}^{\ast}(t)\big)}\Big\}.
\end{equation*}
Criterion D1 places all the weight on such $t\in\left[t_{\ell},\overline{t} %
\right]$ for any off-path deviation in $D$ according to Lemma \ref%
{lemma_Off2}. We show that such type $t\in\left[t_{\ell},\overline{t}\right]$
has a profitable deviation to any off-path deviation in $D$. Let $%
m_{1}^{\prime}:=m_{1}^{\ast}(t)+\varepsilon$, and any $m_{2}^{
\prime}:=m_{2}^{\ast}(t)-\varepsilon$ for some $\varepsilon >0$ small enough
such that $(m_{1}^{\prime},m_{2}^{\prime})\in D$. We show that
\begin{align}  \label{eq:nonbinding07}
\alpha m_{2}^{\ast}(t)-\alpha \varepsilon +\mathbb{E}_{\mu^{\ast}(m_{1}^{
\prime},m_{2}^{\prime})}f(m_{1}^{\ast}(t)+\varepsilon,t)-c(m_{1}^{\ast}(t)+
\varepsilon,t)-h(m_{2}^{\ast}(t)-\varepsilon)>&  \notag \\
\alpha
m_{2}^{\ast}(t)+f(m_{1}^{\ast}(t),t)-c(m_{1}^{\ast}(t),t)-h(m_{2}^{\ast}(t)).
\end{align}
Because
\begin{equation}  \label{eq:nonbinding08}
\max~supp~\mu^{\ast}(m_{1}^{\ast}(t),m_{2}^{\ast}(t))=\lim_{k\nearrow t}inf
~supp~\mu^{\ast} (m_{1}^{\ast}(k),m_{2}^{\ast}(k))=t,
\end{equation}
we have that
\begin{equation}  \label{eq:nonbinding09}
\mathbb{E}_{\mu(m_{1}^{\prime},m_{2}^{\prime})}f(m_{1}^{\prime},t)>f(m_{1}^{
\ast}(t),t)
\end{equation}
Because $f$, $c$ and $h$ are continuous in the sender's actions, and
inequality (\ref{eq:nonbinding09}), we have that
\begin{align}  \label{eq:nonbinding10}
\lim_{\varepsilon\searrow 0}\Big[\alpha m_{2}^{\ast}(t)-\alpha \varepsilon +
\mathbb{E}_{\mu(m_{1}^{\prime},m_{2}^{\prime})}f(m_{1}^{\ast}(t)+
\varepsilon,t)-c(m_{1}^{\ast}(t)+
\varepsilon,t)-h(m_{2}^{\ast}(t)-\varepsilon)\Big] >&  \notag \\
\alpha
m_{2}^{\ast}(t)+f(m_{1}^{\ast}(t),t)-c(m_{1}^{\ast}(t),t)-h(m_{2}^{\ast}(t)).
\end{align}
Inequality (\ref{eq:nonbinding10}) implies that there exists $\varepsilon>0$
such that (\ref{eq:nonbinding07}) holds. This is a contradiction to the
definition of a equilibrium. Therefore, both $m_{1}^{\ast}$ and $%
m_{2}^{\ast} $ are continuous on $\left[t_{\ell},\overline{t}\right]$.

Consider the subinterval $\left[ \underline{t},t_{\ell }\right) $. Again, we
prove by contradiction. Suppose that at least one of the equilibrium action
pair is discontinuous at some $t\in \left[ \underline{t},t_{\ell }\right) $.
Because $m_{2}^{\ast }(t)=m_{2}^{\circ }$ when $t\in \left[ \underline{t}%
,t_{\ell }\right) $, we have that $m_{2}^{\ast }(t)$ is constant on the
subinterval $\left[ \underline{t},t_{\ell }\right) $. This implies that $%
m_{2}^{\ast }$ is always continuous on $\left[ \underline{t},t_{\ell
}\right) $. Therefore we consider only the case where $m_{1}^{\ast }$ is
discontinuous at some $t\in \left[ \underline{t},t_{\ell }\right) $. Suppose
that $m_{1}^{\ast }$ is discontinuous at some $t\in \left[ \underline{t}%
,t_{\ell }\right) $. Criterion D1 places all the weight on such $t\in \left[
\underline{t},t_{\ell }\right) $. We follow the proof in the previous case
to show that type $t\in \left[ \underline{t},t_{\ell }\right) $ has an
incentive to profitably deviate from her equilibrium actions. This leads to
a contradiction of the equilibrium choice of type $t\in \left[ \underline{t}%
,t_{\ell }\right) $. Hence, $m_{1}^{\ast }$ must be continuous on $\in \left[
\underline{t},t_{\ell }\right) $. In conclusion, $(m_{1}^{\ast },m_{2}^{\ast
})$ is continuous on $\left[ \underline{t},\overline{t}\right] $ in any
separating monotone D1 equilibrium.
\end{proof}


\begin{proof}
\textbf{of Lemma \ref{lemma_2}.} The proof is standard. For all $t\in(
\underline{t}, t_{\ell})$ and for all $m_{1}\in \left(m_{1}^{\ast}(
\underline{t}),m_{1}^{\ast}(t_{\ell})\right)$, define a function $\Phi_{s}
(m_{1},t)$ as
\begin{equation}  \label{eq:100}
\Phi_{s} (m_{1},t):= \dfrac{-\big[f_{m_1}\left(m_{1}, \mu(m_{1})\right)-
c_{m_1}\big(m_{1}, t\big)\big]}{f_t\big(m_{1}, \mu(m_{1})\big)}.
\end{equation}
For all $t\in(t_{\ell},\overline{t})$ and for all $m_{1}\in
\left(m_{1}^{\ast}(t_{\ell}),m_{1}^{\ast}(\overline{t})\right)$, define a
function $\Phi_{b} (m_{1},t)$ as
\begin{equation}  \label{eq:101}
\Phi_{b} (m_{1},t):= \dfrac{-\big[f_{m_1}\left(m_{1},
\mu(m_{1})\right)-c_{m_1}\big(m_{1}, t\big) -\alpha+h^{\prime}(M-m_{1})\big]
}{f_t\big(m_{1}, \mu(m_{1})\big)}.
\end{equation}
Combining (\ref{eq:foc_lowest4}) and (\ref{eq:100}), we generate a first
order differential equation $\mu^{\prime}(m_{1})=\Phi_{s} (m_{1},t)$ for all
$t\in(\underline{t}, t_{\ell})$ and for all $m_{1}\in \left(m_{1}^{\ast}(
\underline{t}),m_{1}^{\ast}(t_{\ell})\right)$. Also, combining (\ref%
{eq:foc_lowest6}) and (\ref{eq:101}), we generate a first order differential
equation $\mu^{\prime}(m_{1})=\Phi_{b} (m_{1},t)$ for all $t\in(t_{\ell},
\overline{t})$ and for all $m_{1}\in
\left(m_{1}^{\ast}(t_{\ell}),m_{1}^{\ast}(\overline{t})\right)$. Given the
initial condition $\mu(m_{1}^{\ast}(\underline{t}))=\underline{t}$ for (\ref%
{eq:100}), and the initial condition $\mu(m_{1}^{\ast}(t_{\ell}))=t_{\ell}$
for (\ref{eq:101}), implies that there exists a unique solution $\mu^{\ast} $
for each of the differential equations generated by (\ref{eq:100}) and (\ref%
{eq:101}) when $\Phi_{s} (m_{1},t)$ and $\Phi_{b} (m_{1},t)$ are uniformly
Lipschitz continuous according to Picard-Lindelof's Existence and Uniqueness
Theorem\footnote{\textbf{Picard-Lindelof's Existence and Uniqueness Theorem:}
Consider the initial value problem
\begin{equation*}
y^{\prime}(t)=f(t,y(t)),\qquad y(t_{\circ})=y_{0}.
\end{equation*}
Suppose $f$ is uniformly Lipschitz continuous in $y$ (meaning the Lipschitz
constant can be taken independent of $t$) and continuous in $t$, then for
some value $\varepsilon > 0$, there exists a unique solution $y(t)$ to the
initial value problem on the interval $[t_{\circ}-\varepsilon
,t_{\circ}+\varepsilon ]$.}. Given the unique solution $\mu^{\ast}$, we
obtain equilibrium level of cognitive signal $m_{1}^{\ast}=\mu^{*-1}$.
\end{proof}



\begin{proof}
\textbf{of Theorem \ref{theorem_1}.} The statement in item (1) is a
corollary to Lemma \ref{continuous}. Because $m_{1}^{\ast}$ and $%
m_{2}^{\ast} $ are both continuous on a compact interval $\big[\underline{t}
, \overline{t}\big]$, it follows from the definition of a continuous
function that Range$(m_{1}^{\ast})$ and Range$(m_{2}^{\ast})$ are compact
intervals.

The first part of item (2) can be easily shown as follows. Fix $t\in\big[
\underline{t}, \overline{t}\big]$. Because $m_{1}^{\ast}$ is increasing and
continuous over the compact interval $\big[\underline{t}, \overline{t}\big]
$, it implies that for all $m_{1}\in$ Range$(m_{1}^{\ast})$, the belief
function $\mu^{\ast}$ defined by $\mu^{\ast}(m_{1})=m_{1}^{*-1}(m_{1})$ is
also increasing and continuous over Range$(m_{1}^{\ast})$.

Next, we prove the second statement in item (2): $\mu^{\ast}$ has continuous
derivative $\mu^{\ast\prime}$ on $\left( m_{1}^{\ast}(\underline{t}
),m_{1}^{\ast}(t_{\ell}) \right)\cup
\left(m_{1}^{\ast}(t_{\ell}),m_{1}^{\ast}(\overline{t})\right)$. We apply
Theorem 7.21\footnote{\textbf{Theorem 7.21 in Wheeden and Zygmund (1977):}
Let $f(x)$, for $x\in \mathbb{R},$ be monotone increasing and finite on
interval $(a, b)$. Then $f$ has a measurable nonenegative derivative $%
f^{\prime}$ almost everywhere in $(a, b)$. Moreover,
\begin{equation*}
0 \leq \int_{a}^{b}f^{\prime} \leq f(b_{-})-f(a_{+}).
\end{equation*}%
} in Wheeden and Zygmund (1977) to show that $\mu^{\ast}$ is differentiable
almost everywhere on $\left( m_{1}^{\ast}(\underline{t}),m_{1}^{\ast}(t_{
\ell}) \right)\cup \left(m_{1}^{\ast}(t_{\ell}),m_{1}^{\ast}(\overline{t}
)\right)$. Because $\mu^{\ast}:=m_{1}^{*-1}$ is strictly increasing on Range$%
(m_{1}^{\ast})$ it is monotone increasing on Range$(m_{1}^{\ast})$. Also,
because $\mu^{\ast}\in \Delta(T)$, it is finite.

We partition Range$(m_{1}^{\ast})$ into two sub-intervals: $\left[
m_{1}^{\ast}(\underline{t}),m_{1}^{\ast}(t_{\ell}) \right)$ and $\left[
m_{1}^{\ast}(t_{\ell}),m_{1}^{\ast}(\overline{t})\right]$. First, suppose
towards a contradiction that $\mu^{\ast}$ is not differentiable at some $%
m_{1} \in \left( m_{1}^{\ast}(\underline{t}),m_{1}^{\ast}(t_{\ell}) \right)$
. Because $\mu^{\ast}$ is not differentiable at only finitely many points,
there exists $s_{1}^{\prime}$,$s_{1}^{\prime\prime}$ $\in \left(
m_{1}^{\ast}(\underline{t}),m_{1}^{\ast}(t_{\ell}) \right)$ such that $%
\mu^{\ast}$ is differentiable everywhere on $(s_{1}^{\prime},m_{1})$ and $%
(m_{1},s_{1}^{\prime\prime})$. Fix $t\in\left(\underline{t},t_{\ell}\right)$
and $m_{1}^{\ast}(t)\in $ $(s_{1}^{\prime},m_{1})$. For any given $%
\varepsilon >0,$ let $\mu^{\ast\prime}(m_{1}^{-}):=
\lim_{\varepsilon\downarrow 0}\mu^{\ast\prime}(m_{1}(t)-\varepsilon)$ and $%
\displaystyle{\mu^{\ast\prime}(m_{1}^{+}):=\lim_{\varepsilon\downarrow
0}\mu^{\ast\prime}(m_{1}(t)+\varepsilon)}$. Given $f_t>0$ (Assumption 1(d))
and in addition to the assumption that $f_{m_{1}}$, $c_{m_{1}}$, and $f_{t}$
are all continuous over Range$(m_{1}^{\ast})$, it follows from equation (\ref%
{eq:foc_lowest4}) that
\begin{equation}  \label{eq:nonbinding12}
\mu^{\ast\prime}(m_{1}^{-})= \dfrac{-\big[f_{m_1}\left(m_{1},
\mu^{\ast}(m_{1})\right) -c_{m_1}\big(m_{1}, t\big)\big]}{f_t\big(m_{1},
\mu^{\ast}(m_{1})\big)}=\mu^{\ast\prime}(m_{1}^{+}).
\end{equation}
Equation (\ref{eq:nonbinding12}) implies that $\mu^{\ast\prime}(m_{1}^{-})=
\mu^{\ast\prime}(m_{1})= \mu^{\ast\prime}(m_{1}^{+})$, which further implies
that $\mu^{\ast}$ is differentiable at $m_{1}$. This is a contradiction to
the hypothesis that $\mu^{\ast}$ is not differentiable at $m_{1}$. Further,
since $f_{m_{1}}$, $c_{m_{1}}$, and $f_{t}$ are all continuous over Range$%
(m_{1}^{\ast})$, $\mu^{\ast\prime}$ is continuous over $\left( m_{1}^{\ast}(
\underline{t}),m_{1}^{\ast}(t_{\ell}) \right)$. Thus, $\mu^{\ast\prime}$ is
continuously differentiable on the $\left( m_{1}^{\ast}(\underline{t}
),m_{1}^{\ast}(t_{\ell}) \right)$.

Second, suppose towards a contradiction that $\mu^{\ast}$ is not
differentiable at some $m_{1} \in\left(m_{1}^{\ast}(t_{\ell}),m_{1}^{\ast}(
\overline{t})\right)$. Because $\mu^{\ast}$ is not differentiable at only
finitely many points, there exists $s_{1}^{\prime}$,$s_{1}^{\prime\prime}$ $%
\in \left(m_{1}^{\ast}(t_{\ell}),m_{1}^{\ast}(\overline{t})\right)$ such
that $\mu^{\ast}$ is differentiable everywhere on $(s_{1}^{\prime},m_{1})$
and $(m_{1},s_{1}^{\prime\prime})$. Fix $t\in\left(t_{\ell}, \overline{t}
\right)$ and $m_{1}^{\ast}(t)\in $ $(m_{1},s_{1}^{\prime\prime})$. For any
given $\varepsilon >0,$ let $\displaystyle{\mu^{\ast\prime}(m_{1}^{-}):
=\lim_{\varepsilon\downarrow 0}\mu^{\ast\prime}(m_{1}(t)-\varepsilon)}$ and $%
\displaystyle{\mu^{\ast\prime}(m_{1}^{+}):= \lim_{\varepsilon\downarrow
0}\mu^{\ast\prime}(m_{1}(t)+\varepsilon)}$. Given $f_t>0$ (Assumption 1(d))
and in addition to the assumption that $f_{m_{1}}$, $c_{m_{1}}$, $h^{\prime}$
and $f_{t}$ are all continuous over Range$(m_{1}^{\ast})$, it follows from
equation (\ref{eq:foc_lowest6}) that
\begin{equation}  \label{eq:nonbinding13}
\mu^{\ast\prime}(m_{1}^{-})= \dfrac{-\big[f_{m_1}\left(m_{1},
\mu^{\ast}(m_{1})\right) -c_{m_1}\big(m_{1}, t\big) -\alpha+h^{
\prime}(M-m_{1})\big]}{f_t\big(m_{1}, \mu^{\ast}(m_{1})\big)}
=\mu^{\ast\prime}(m_{1}^{+}).
\end{equation}
Equation (\ref{eq:nonbinding13}) implies that $\mu^{\ast\prime}(m_{1}^{-})=
\mu^{\ast\prime}(m_{1})= \mu^{\ast\prime}(m_{1}^{+})$, which further implies
that $\mu^{\ast}$ is differentiable at $m_{1}$. This is a contradiction to
the hypothesis that $\mu^{\ast}$ is not differentiable at $m_{1}$. Further,
since $f_{m_{1}}$, $c_{m_{1}}$, $h^{\prime}$ and $f_{t}$ are all continuous
over $\left(m_{1}^{\ast}(t_{\ell}),m_{1}^{\ast}(\overline{t})\right)$. Thus $%
\mu^{\ast\prime}$ is continuously differentiable on $\left(m_{1}^{\ast}(t_{
\ell}),m_{1}^{\ast}(\overline{t})\right)$.

It is clear that $\mu ^{\ast }$ is not differentiable at $m_{1}^{\ast
}(t_{\ell })$. For all $m_{1}\in \left( m_{1}^{\ast }(\underline{t}%
),m_{1}^{\ast }(t_{\ell })\right) $, the belief function follows the
differentiable equation in (\ref{eq:foc_lowest4}) and for all $m_{1}\in
\left( m_{1}^{\ast }(t_{\ell }),m_{1}^{\ast }(\overline{t})\right) $, the
belief function follows the differentiable equation in (\ref{eq:foc_lowest6}%
). Because the right hand side of (\ref{eq:foc_lowest4}) and the right hand
side of (\ref{eq:foc_lowest6}) are not equal, we conclude that $\mu ^{\ast }$
is not differentiable at $m_{1}^{\ast }(t_{\ell })$.

Therefore, $\mu^{\ast} $ is continuously differentiable on $%
\left(m_{1}^{\ast}(\underline{t}),m_{1}^{\ast}(t_{\ell})\right)\cup
\left(m_{1}^{\ast}(t_{\ell}),m_{1}^{\ast}(\overline{t})\right)$. The proof
is complete.
\end{proof}

\section*{Proof of Proposition \protect\ref{prop_5}}

We break the proof of Proposition \ref{prop_5} in lemmas.

\begin{lemma}
\label{lemma_Off} Let $\{\hat{m}_{1},\hat{m}_{2},\hat{w},\hat{\mu}\}$ with $%
\hat{\mu}\in \Psi$ be any equilibrium. Suppose $\hat{\mu}(m_{1}^{
\prime},m_{2}^{\prime})$ conditional on any $(m_{1}^{\prime},m_{2}^{\prime})
\notin $ \emph{range}$(\hat{m}_{1},\hat{m}_{2})$ passes Criterion D1. Then $%
\text{supp~}\hat{\mu}(m^{\prime}_{1},m^{\prime}_{2}) =\{\underline{t}\}$ for
$m_{1}^{\prime }\in (0,\hat{m}_{1}(\underline{t}))$ such that $m_{1}^{\prime
}+m_{2}^{\prime }\leq M$.
\end{lemma}

\begin{proof}
. We show that any $t>\underline{t}$ cannot be in the support of $\hat{\mu}
(m^{\prime}_{1},m^{\prime}_{2})$ for any $(m_{1}^{\prime},m_{2}^{\prime})
\notin $ \emph{range}$(\hat{m}_{1},\hat{m}_{2})$ with $m_{1}^{\prime }\in
(0, \hat{m}_{1}(\underline{t}))$ such that $m_{1}^{\prime }+m_{2}^{\prime
}\leq M $. On the contrary, suppose that $t\in \text{supp}~\hat{\mu}
(m^{\prime}_{1},m^{\prime}_{2})$ for some $(m_{1}^{\prime},m_{2}^{\prime})
\notin $ \emph{range}$(\hat{m}_{1},\hat{m}_{2})$ with $m_{1}^{\prime }\in
(0, \hat{m}_{1}(\underline{t}))$ such that $m_{1}^{\prime }+m_{2}^{\prime
}\leq M $ when $t>\underline{t}$. It is enough to show that if $\underline{t}
$ is weakly worse off by deviating to such $(m^{\prime}_{1},m^{\prime}_{2}) $
, then type $t$ is strictly worse off with the same deviation. Given the
wage $w^{\prime}$ chosen by the receiver after observing such $%
(m^{\prime}_{1},m^{ \prime}_{2})$, let
\begin{equation}
w^{\prime }-c\big(m_{1}^{\prime },\underline{t}\big)-h(m_{2}^{\prime })\leq
\alpha \hat{m}_{2}(\underline{t})+f\big(\hat{m}_{1}(\underline{t}),
\underline{t}\big)-c\big(\hat{m}_{1}(\underline{t}),\underline{t}\big) -h(
\hat{m}_{2}(\underline{t})).  \label{eq:nonbinding100}
\end{equation}
Because $m_{1}^{\prime }<\hat{m}_{1}(\underline{t})$, using the single
crossing property of $c$ for $t>\underline{t}$, we have that
\begin{equation}
c\big(\hat{m}_{1}(\underline{t}),\underline{t}\big)-c\big(m_{1}^{\prime },
\underline{t}\big)>c\big(\hat{m}_{1}(\underline{t}),t\big)-c\big( %
m_{1}^{\prime },t\big).  \label{eq:nonbinding101}
\end{equation}
By combining equation (\ref{eq:nonbinding100}) and (\ref{eq:nonbinding101}),
we have that
\begin{equation}
w^{\prime }-c\big(m_{1}^{\prime },t\big)-h(m_{2}^{\prime })<\alpha \hat{m}
_{2}(\underline{t})+f\big(\hat{m}_{1}(\underline{t}),\underline{t} \big)-c %
\big(\hat{m}_{1}(\underline{t}),t\big)-h(\hat{m}_{2}(\underline{t} )).
\label{eq:nonbinding102}
\end{equation}
On the other hand, we have
\begin{equation}
\alpha \hat{m}_{2}(\underline{t})+f\big(\hat{m}_{1}(\underline{t}),
\underline{t}\big)-c\big(\hat{m}_{1}(\underline{t}),t\big)-h(\hat{m}_{2}(
\underline{t}))\leq \alpha \hat{m}_{2}(t)+f\big(\hat{m}_{1}(t),t\big)-c\big(
\hat{m}_{1}(t),t\big)-h(\hat{m}_{2}(t))  \label{eq:nonbinding103}
\end{equation}
in equilibrium. By combining equation (\ref{eq:nonbinding102}) and (\ref%
{eq:nonbinding103}), we have that
\begin{equation}
w^{\prime }-c\big(m_{1}^{\prime },t\big)-h(m_{2}^{\prime })<\alpha \hat{m}
_{2}(t)+f\big(\hat{m}_{1}(t),t\big)-c\big(\hat{m}_{1}(t),t\big) -h(\hat{m}
_{2}(t))  \label{eq:nonbinding105}
\end{equation}
which shows that type $t>\underline{t}$ is strictly worse off by deviating
to an off-path action pair $(m_{1}^{\prime },m_{2}^{\prime })$ with $%
m_{1}^{\prime }\in (0,\hat{m}_{1}(\underline{t}))$ and $m_{1}^{\prime
}+m_{2}^{\prime }\leq M$. The contrapositive of (\ref{D1}) in the definition
of Criterion D1 implies that any $t>\underline{t}$ cannot be in the support
of $\hat{\mu}(m^{\prime}_{1},m^{\prime}_{2})$ for any $(m_{1}^{
\prime},m_{2}^{\prime})\notin $ \emph{range}$(\hat{m}_{1},\hat{m}_{2})$ with
$m_{1}^{\prime }\in (0,\hat{m}_{1}(\underline{t}))$ such that $m_{1}^{\prime
}+m_{2}^{\prime }\leq M$.
\end{proof}

\begin{lemma}
\label{lemma_Off3} Let $\{\hat{m}_{1},\hat{m}_{2},\hat{w},\hat{\mu}\}$ with $%
\hat{\mu}\in \Psi$ be any equilibrium. Suppose $\hat{\mu}(m_{1}^{
\prime},m_{2}^{\prime})$ conditional on any $(m_{1}^{\prime},m_{2}^{\prime})
\notin $ \emph{range}$(\hat{m}_{1},\hat{m}_{2})$ passes Criterion D1. Then $%
\text{supp~}\hat{\mu}(m^{\prime}_{1},m^{\prime}_{2}) =\{\overline{t}\}$ for $%
m_{1}^{\prime}\in \left(\hat{m}_{1}(\overline{t}),M\right]$ such that $%
m_{1}^{\prime }+m_{2}^{\prime }\leq M$.
\end{lemma}

\begin{proof}
. We show that any $t<\overline{t}$ cannot be in the support of $\hat{\mu}
(m^{\prime}_{1},m^{\prime}_{2})$ for any $(m_{1}^{\prime},m_{2}^{\prime})
\notin $ \emph{range}$(\hat{m}_{1},\hat{m}_{2})$ with $m_{1}^{\prime}\in
\left(\hat{m}_{1}(\overline{t}),M\right]$ such that $m_{1}^{\prime
}+m_{2}^{\prime }\leq M$. On the contrary, suppose that $t\in \text{supp}~
\hat{\mu}(m^{\prime}_{1},m^{\prime}_{2})$ for some $m_{1}^{\prime}\in \left(
\hat{m}_{1}(\overline{t}),M\right]$ such that $m_{1}^{\prime }+m_{2}^{\prime
}\leq M$ when $t>\underline{t}$. It is enough to show that if $\overline{t}$
is weakly worse off by deviating to such $(m^{\prime}_{1},m^{\prime}_{2})$,
then type $t$ is strictly worse off with the same deviation. Given the wage $%
w^{\prime}$ chosen by the receiver after observing such $(m^{\prime}_{1},m^{
\prime}_{2})$, let
\begin{equation}  \label{eq:nonbinding22}
w^{\prime}-c\big(m_{1}^{\prime},\overline{t}\big) -h(m_{2}^{\prime}) \leq
\alpha \hat{m}_{2}(\overline{t}) +f\big(\hat{m}_{1}(\overline{t}), \overline{
t}\big)- c\big(\hat{m}_{1}(\overline{t}),\overline{t}\big) -h(\hat{m}_{2}(
\overline{t})).
\end{equation}
Because $m_{1}^{\prime}>\hat{m}_{1}(\overline{t})$, using the single
crossing property of $c$ for $t<\overline{t}$, we have that
\begin{equation}  \label{eq:nonbinding23}
c\big(m_{1}^{\prime},\overline{t}\big)- c\big(\hat{m}_{1}(\overline{t}),
\overline{t}\big) < c\big(m_{1}^{\prime},t\big)-c\big(\hat{m}_{1}(\overline{
t }),t\big).
\end{equation}
By combining equation (\ref{eq:nonbinding22}) and (\ref{eq:nonbinding23}),
we have that
\begin{equation}  \label{eq:nonbinding24}
w^{\prime}-c\big(m_{1}^{\prime},t\big)-h(m_{2}^{\prime}) < \alpha \hat{m}
_{2}(\overline{t}) +f\big(\hat{m}_{1}(\overline{t}), \overline{t}\big)-c %
\big( \hat{m}_{1}(\overline{t}),t\big)-h(\hat{m}_{2}(\overline{t})).
\end{equation}
Because $((\hat{m}_{1}(t),\hat{m}_{2}(t))$ is incentive compatible for type $%
t<\overline{t}$, we have that
\begin{equation}  \label{eq:nonbinding25}
\alpha \hat{m}_{2}(\overline{t}) +f\big(\hat{m}_{1}(\overline{t}), \overline{
t}\big)-c\big(\hat{m}_{1}(\overline{t}),t\big)- h(\hat{m}_{2}(\overline{t}
))< \alpha \hat{m}_{2}(t) +f\big(\hat{m}_{1}(t), t\big)-c\big(\hat{m}
_{1}(t),t\big)-h(\hat{m}_{2}(t)).
\end{equation}
By combining equation (\ref{eq:nonbinding24}) and (\ref{eq:nonbinding25}),
we have that
\begin{equation}  \label{eq:nonbinding26}
w^{\prime}-c\big(m_{1}^{\prime},t\big)-h(m_{2}^{\prime}) < \alpha \hat{m}
_{2}(t) +f\big(\hat{m}_{1}(t), t\big)- c\big(\hat{m}_{1}(t),t\big)-h(\hat{m}
_{2}(t))
\end{equation}
which shows that type $t<\overline{t}$ is strictly worse off by deviating to
an off-path action pair $(m_{1}^{\prime },m_{2}^{\prime })$ with $%
m_{1}^{\prime}\in \left(\hat{m}_{1}(\overline{t}),M\right]$ such that $%
m_{1}^{\prime }+m_{2}^{\prime }\leq M$. The contrapositive of (\ref{D1}) in
the definition of Criterion D1 implies that any $t<\overline{t}$ cannot be
in the support of $\hat{\mu}(m^{\prime}_{1},m^{\prime}_{2})$ for any $%
(m_{1}^{\prime},m_{2}^{\prime})\notin $ \emph{range}$(\hat{m}_{1},\hat{m}
_{2})$ with $m_{1}^{\prime}\in \left(\hat{m}_{1}(\overline{t}),M\right]$
such that $m_{1}^{\prime }+m_{2}^{\prime }\leq M$.
\end{proof}

\begin{proof}
\textbf{of Proposition \ref{prop_5}.} Because of Assumption 3.(a) (strict
concavity of $f-c$) and Assumption 3.(b) (strict convexity of $h$), we
conclude that Condition 1 in Proposition \ref{prop_5} is sufficient for $%
m_{1}^{\ast}(\underline{t})$ and $m_{2}^{\ast}(\underline{t})$ to be the
equilibrium signaling actions for type $\underline{t}$. Because type $%
\underline{t}$ has no information rent, it is clear that Condition 1 is her
first order necessary condition.

Given that $c$ and $h$ are differentiable, and the second item of Theorem %
\ref{theorem_1}, it is clear that Conditions 2 and 3 in Proposition \ref%
{prop_5} are the first order necessary conditions for $(m_{1}^{\ast
}(t),m_{2}^{\ast }(t))$ to be an optimal pair of action for types $t\in (%
\underline{t},t_{\ell })$ and $t\in \left( t_{\ell },\overline{t}\right) $
respectively. We now show that both conditions are also sufficient
conditions for the respective types in a separating monotone D1 equilibrium
with $\underline{t}<t_{\ell }<\overline{t}$. \newline
\textbf{Incentive Compatibility:}\newline
(a) We show that $(m_{1}^{\ast }(t),m_{2}^{\ast }(t))$ is incentive
compatible for type $t\in \left( t_{\ell },\overline{t}\right) $ when
Condition 2 holds. That is, we show that for any $t,t^{\prime }\in \left(
t_{\ell },\overline{t}\right) $ with $t\neq t^{\prime }$, $\alpha
m_{2}^{\ast }(t)+f(m_{1},\mu (m_{1}(t)))-c(m_{1}^{\ast }(t),t)-h(m_{2}^{\ast
}(t))>\alpha m_{2}^{\ast }(t^{\prime })+f(m_{1}^{\ast }(t^{\prime }),\mu
(m_{1}(t^{\prime })))-c(m_{1}^{\ast }(t^{\prime }),t)-h(m_{2}^{\ast
}(t^{\prime })).$ For $t,t^{\prime }\in \left( t_{\ell },\overline{t}\right)
$ with $t^{\prime }<t$, the strict single crossing property of $c$ implies
that $c_{m_{1}}\big(m_{1}^{\ast }(t^{\prime }),t)\big)<c_{m_{1}}\big(%
m_{1}^{\ast }(t^{\prime }),t^{\prime }\big)$. This follows that $%
f_{m_{1}}\left( m_{1}^{\ast }(t^{\prime }),\mu (m_{1}^{\ast }(t^{\prime
}))\right) +f_{t}\big(m_{1}^{\ast }(t^{\prime }),\mu (m_{1}^{\ast
}(t^{\prime }))\big)\mu ^{\prime }(m_{1}^{\ast }(t^{\prime }))-c_{m_{1}}\big(%
m_{1}^{\ast }(t^{\prime }),t\big)-\alpha +h^{\prime }(M-m_{1}^{\ast
}(t^{\prime }))>$$f_{m_{1}}\left( m_{1}^{\ast }(t^{\prime }),\mu
(m_{1}^{\ast }(t^{\prime }))\right) +f_{t}\big(m_{1}^{\ast }(t^{\prime
}),\mu (m_{1}^{\ast }(t^{\prime }))\big)\mu ^{\prime }(m_{1}^{\ast
}(t^{\prime }))-c_{m_{1}}\big(m_{1}^{\ast }(t^{\prime }),t^{\prime }\big)%
-\alpha +h^{\prime }(M-m_{1}^{\ast }(t^{\prime }))=0$. The equality here is
because of Condition 3. This leads to
\begin{align}
& f_{m_{1}}\left( m_{1}^{\ast }(t^{\prime }),\mu (m_{1}^{\ast }(t^{\prime
}))\right) +f_{t}\big(m_{1}^{\ast }(t^{\prime }),\mu (m_{1}^{\ast
}(t^{\prime }))\big)\mu ^{\prime }(m_{1}^{\ast }(t^{\prime }))-c_{m_{1}}\big(%
m_{1}^{\ast }(t^{\prime }),t\big)-\alpha  \notag  \label{eq:nonbinding14} \\
& +h^{\prime }(M-m_{1}^{\ast }(t^{\prime }))>0
\end{align}%
for $t,t^{\prime }\in \left( t_{\ell },\overline{t}\right) $ with $t^{\prime
}<t$.

Similarly, for $t,t^{\prime}\in \left(t_{\ell},\overline{t}\right)$ with $%
t^{\prime}>t$, the strict single crossing property of $c$ implies that $%
c_{m_{1}}\big(m_{1}^{\ast}(t^{\prime}), t\big)> c_{m_{1}}\big(
m_{1}^{\ast}(t^{\prime}), t^{\prime})\big)$. This follows immediately: $%
f_{m_{1}}\left(m_{1}^{\ast}(t^{\prime}),
\mu(m_{1}^{\ast}(t^{\prime}))\right) +f_{t}\big(m_{1}^{\ast}(t^{\prime}),
\mu(m_{1}^{\ast} (t^{\prime}))\big)\mu^{\prime}(m_{1}^{\ast}(t^{\prime}))
-c_{m_{1}}\big(m_{1}^{\ast}(t^{\prime}), t\big)-
\alpha+h^{\prime}(M-m_{1}^{\ast}(t^{\prime}))<f_{m_{1}}\left(m_{1}^{
\ast}(t^{\prime}), \mu(m_{1}^{\ast}(t^{\prime}))\right) +f_{t}\big(
m_{1}^{\ast}(t^{\prime}), \mu(m_{1}^{\ast}(t^{\prime})) \big)
\mu^{\prime}(m_{1}^{\ast}(t^{\prime})) -c_{m_{1}}\big(m_{1}^{\ast}(t^{
\prime}), t^{\prime}\big)- \alpha+h^{\prime}(M-m_{1}^{\ast}(t^{\prime}))=0$.
The equality here is because of Condition 3. This leads to
\begin{align}  \label{eq:nonbinding15}
&f_{m_{1}}\left(m_{1}^{\ast}(t^{\prime}),
\mu(m_{1}^{\ast}(t^{\prime}))\right) +f_{t}\big(m_{1}^{\ast}(t^{\prime}),
\mu(m_{1}^{\ast} (t^{\prime}))\big)\mu^{\prime}(m_{1}^{\ast}(t^{\prime}))
-c_{m_{1}}\big(m_{1}^{\ast}(t^{\prime}), t\big)- \alpha  \notag \\
&+h^{\prime}(M-m_{1}^{\ast}(t^{\prime}))<0
\end{align}
for $t,t^{\prime}\in \left(t_{\ell},\overline{t}\right)$ with $t^{\prime}>t$
. Given Assumption 3.(a) and 3.(b), the inequalities (\ref{eq:nonbinding14})
and (\ref{eq:nonbinding15}) imply that $m_{1}^{\ast}(t)$ is the maximizer of
$w^{\ast}\left(m_{1}^{\ast}(t),m_{2}^{\ast}(t)\right)-c(m_{1},t)-h(m_{2})$
for type $t\in \left(t_{\ell},\overline{t}\right)$. That is,
\begin{equation}  \label{eq:nonbinding16}
w^{\ast}\left(m_{1}^{\ast}(t),m_{2}^{\ast}(t)\right)
-c(m_{1}^{\ast}(t),t)-h(m_{2}^{\ast}(t))>w^{\ast}\left(m_{1}^{\ast}(t^{
\prime}),m_{2}^{\ast}(t^{\prime})\right)-
c(m_{1}^{\ast}(t^{\prime}),t)-h(m_{2}^{\ast}(t^{\prime}))
\end{equation}
for all $t,t^{\prime}\in \left(t_{\ell},\overline{t}\right)$ with $t\neq
t^{\prime}$, which implies $\left(m_{1}^{\ast}, m_{2}^{\ast}\right)$ is
incentive compatible for type $t\in \left(t_{\ell},\overline{t}\right)$.

(b) We show that $(m_{1}^{\ast}(t),m_{2}^{\ast}(t))$ is incentive compatible
for all $t\in \left(\underline{t},t_{\ell}\right)$.
The proof is similar part (a), so we omit it.

(c) We show that sender types in $\left(\underline{t},t_{\ell}\right)$ do
not have incentive to deviate to some on-path message $(m_{1}, m_{2})$ with $%
m_{1}\in \left [m_{1}^{\ast}(t_{\ell}),m_{1}^{\ast}(\overline{t})\right]$,
and a corresponding $m_{2}$ such that $m_{1}+m_{2}= M$.

Fix any $t^{\prime}$ with $t^{\prime}<t_{\ell}$. Because of (a) and (b) in
this proof, $(m_{1}^{\ast}(t^{\prime}),m_{2}^{\ast}(t^{\prime}))$ is
incentive compatibility for $t^{\prime}$ and moreover, because $m_{1}^{\ast}
$, $m_{2}^{\ast}$, $f$, $c$, $h$, are all continuous, we have that
\begin{align}  \label{eq:1}
&\alpha m_{2}^{\ast}(t^{\prime})+f(m_{1}^{\ast}(t^{\prime}),t^{\prime})
-c(m_{1}^{\ast}(t^{\prime}),t^{\prime})-h(m_{2}^{\ast}(t^{\prime}))  \notag
\\
&>\lim_{t\searrow t_{\ell}}\left\{\alpha
m_{2}^{\ast}(t)+f(m_{1}^{\ast}(t),t)-c(m_{1}^{\ast}(t),t^{\prime})-h(m_{2}^{
\ast}(t))\right\}
\end{align}
for $t^{\prime}<t_{\ell}$. Inequality (\ref{eq:1}) implies that there exists
some $\varepsilon >0$ such that
\begin{align}  \label{eq:2}
&\alpha m_{2}^{\ast}(t^{\prime})+f(m_{1}^{\ast}(t^{\prime}),t^{\prime})
-c(m_{1}^{\ast}(t^{\prime}),t^{\prime})-h(m_{2}^{\ast}(t^{\prime}))  \notag
\\
&>\alpha
m_{2}^{\ast}(t_{\ell}+\varepsilon)+f(m_{1}^{\ast}(t_{\ell}+\varepsilon),t_{
\ell} +\varepsilon)-c(m_{1}^{\ast}(t_{\ell}+\varepsilon),t^{\prime})
-h(m_{2}^{\ast}(t_{\ell}+\varepsilon))
\end{align}
for $t^{\prime}<t_{\ell}$.

Incentive compatibility for $t_{\ell}+\varepsilon$ implies that
\begin{align}  \label{eq:3}
&\alpha m_{2}^{\ast}(t_{\ell}+\varepsilon)+f(m_{1}^{\ast}(t_{\ell}+
\varepsilon),t_{\ell}+\varepsilon)-c(m_{1}^{\ast}(t_{\ell}+
\varepsilon),t_{\ell}+\varepsilon)-h(m_{2}^{\ast}(t_{\ell}+\varepsilon))
\notag \\
&>\alpha m_{2}^{\ast}(t)+f(m_{1}^{\ast}(t),t)-c(m_{1}^{\ast}(t),t_{\ell}+
\varepsilon)-h(m_{2}^{\ast}(t))
\end{align}
for all $t>t_{\ell}+\varepsilon$.

Because $m_{1}^{\ast}(t)>m_{1}^{\ast}(t_{\ell}+\varepsilon)$, we can apply
the single crossing condition for $t^{\prime}<t_{\ell}+\varepsilon$.
\begin{align}  \label{eq:4}
c(m_{1}^{\ast}(t_{\ell}+\varepsilon),t_{\ell}+\varepsilon)-
c(m_{1}^{\ast}(t),t_{\ell}+\varepsilon)>c(m_{1}^{\ast}(t_{\ell}+
\varepsilon),t^{\prime})-c(m_{1}^{\ast}(t),t^{\prime}).
\end{align}
By combining (\ref{eq:3}) and (\ref{eq:4}), we have that for all $%
t^{\prime}<t_{\ell}$,
\begin{align}  \label{eq:5}
&\alpha m_{2}^{\ast}(t_{\ell}+\varepsilon)+f(m_{1}^{\ast}(t_{\ell}+
\varepsilon),t_{\ell}+\varepsilon)-c(m_{1}^{\ast}(t_{\ell}+
\varepsilon),t^{\prime})-h(m_{2}^{\ast}(t_{\ell}+\varepsilon))  \notag \\
&>\alpha
m_{2}^{\ast}(t)+f(m_{1}^{\ast}(t),t)-c(m_{1}^{\ast}(t),t^{\prime})-h(m_{2}^{
\ast}(t))
\end{align}
for all $t>t_{\ell}+\varepsilon$. By combining (\ref{eq:2}) and (\ref{eq:5}
), we have that for all $t^{\prime}<t_{\ell}$,
\begin{align}  \label{eq:6}
&\alpha m_{2}^{\ast}(t^{\prime})+f(m_{1}^{\ast}(t^{\prime}),t^{\prime})-
c(m_{1}^{\ast}(t^{\prime}),t^{\prime})-h(m_{2}^{\ast}(t^{\prime}))  \notag \\
&>\alpha
m_{2}^{\ast}(t)+f(m_{1}^{\ast}(t),t)-c(m_{1}^{\ast}(t),t^{\prime})-h(m_{2}^{
\ast}(t))
\end{align}
for all $t>t_{\ell}+\varepsilon$ which establishes that sender types $%
t^{\prime}\in \left(\underline{t},t_{\ell}\right)$ do not have incentive to
deviate to the on-path message pair $(m_{1}, m_{2})$ such that $m_{1}\in
\left [m_{1}^{\ast}(t_{\ell}),m_{1}^{\ast}(\overline{t})\right]$ and $%
m_{1}+m_{2}= M$.

(d) sender types in $\left[t_{\ell}, \overline{t}\right]$ do not have
incentive to deviate to on-path message pair $(m_{1}, m_{2}^{\circ})$ such
that $m_{1}\in \left [m_{1}^{\ast}(\underline{t}),m_{1}^{\ast}(t_{\ell})
\right)$, and $m_{1}+m_{2}^{\circ}< M$.

Fix any $t^{\prime}$ with $t^{\prime}>t_{\ell}$. Because of (a) and (b) in
this proof, $(m_{1}^{\ast}(t^{\prime}),m_{2}^{\ast}(t^{\prime}))$ is
incentive compatibility for $t^{\prime}$ and moreover, because $m_{1}^{\ast}
$, $m_{2}^{\ast}$, $f$, $c$, $h$, are all continuous, we have that
\begin{align}  \label{eq:11}
&\alpha m_{2}^{\ast}(t^{\prime})+f(m_{1}^{\ast}(t^{\prime}),t^{\prime})-
c(m_{1}^{\ast}(t^{\prime}),t^{\prime})-h(m_{2}^{\ast}(t^{\prime}))  \notag \\
&>\lim_{t\nearrow t_{\ell}}\left\{\alpha
m_{2}^{\ast}(t)+f(m_{1}^{\ast}(t),t)-c(m_{1}^{\ast}(t),t^{\prime})-h(m_{2}^{
\ast}(t))\right\}
\end{align}
for $t^{\prime}>t_{\ell}$. Inequality (\ref{eq:11}) implies that there
exists some $\varepsilon >0$ such that
\begin{align}  \label{eq:12}
&\alpha m_{2}^{\ast}(t^{\prime})+f(m_{1}^{\ast}(t^{\prime}),t^{\prime})-
c(m_{1}^{\ast}(t^{\prime}),t^{\prime})-h(m_{2}^{\ast}(t^{\prime}))  \notag \\
&>\alpha
m_{2}^{\ast}(t_{\ell}-\varepsilon)+f(m_{1}^{\ast}(t_{\ell}-\varepsilon),t_{
\ell}-
\varepsilon)-c(m_{1}^{\ast}(t_{\ell}-\varepsilon),t^{\prime})-h(m_{2}^{
\ast}(t_{\ell}-\varepsilon))
\end{align}
for $t^{\prime}>t_{\ell}$.

Incentive compatibility for $t_{\ell}-\varepsilon$ implies that
\begin{align}  \label{eq:13}
&\alpha
m_{2}^{\ast}(t_{\ell}-\varepsilon)+f(m_{1}^{\ast}(t_{\ell}-\varepsilon),t_{
\ell}- \varepsilon)-c(m_{1}^{\ast}(t_{\ell}-\varepsilon),t_{\ell}-
\varepsilon)-h(m_{2}^{\ast}(t_{\ell}-\varepsilon))  \notag \\
&>\alpha m_{2}^{\ast}(t)+f(m_{1}^{\ast}(t),t)-c(m_{1}^{\ast}(t),t_{\ell}-
\varepsilon)-h(m_{2}^{\ast}(t))
\end{align}
for all $t<t_{\ell}-\varepsilon$.

Because $m_{1}^{\ast}(t_{\ell}-\varepsilon)>m_{1}^{\ast}(t)$, we can apply
the single crossing condition for $t^{\prime}>t_{\ell}-\varepsilon$.
\begin{align}  \label{eq:14}
c(m_{1}^{\ast}(t_{\ell}-\varepsilon),t_{\ell}-\varepsilon)-c(m_{1}^{
\ast}(t),t_{\ell}-
\varepsilon)>c(m_{1}^{\ast}(t_{\ell}-\varepsilon),t^{\prime})-c(m_{1}^{
\ast}(t),t^{\prime}).
\end{align}
By combining (\ref{eq:13}) and (\ref{eq:14}), we have that for all $%
t^{\prime}>t_{\ell}$,
\begin{align}  \label{eq:15}
&\alpha
m_{2}^{\ast}(t_{\ell}-\varepsilon)+f(m_{1}^{\ast}(t_{\ell}-\varepsilon),t_{
\ell}-
\varepsilon)-c(m_{1}^{\ast}(t_{\ell}-\varepsilon),t^{\prime})-h(m_{2}^{
\ast}(t_{\ell} -\varepsilon))  \notag \\
&>\alpha
m_{2}^{\ast}(t)+f(m_{1}^{\ast}(t),t)-c(m_{1}^{\ast}(t),t^{\prime})-h(m_{2}^{
\ast}(t))
\end{align}
for all $t<t_{\ell}-\varepsilon$. By combining (\ref{eq:12}) and (\ref{eq:15}
), we have that for all $t^{\prime}>t_{\ell}$,
\begin{align}  \label{eq:16}
&\alpha m_{2}^{\ast}(t^{\prime})+f(m_{1}^{\ast}(t^{\prime}),t^{\prime})-
c(m_{1}^{\ast}(t^{\prime}),t^{\prime})-h(m_{2}^{\ast}(t^{\prime}))  \notag \\
&>\alpha
m_{2}^{\ast}(t)+f(m_{1}^{\ast}(t),t)-c(m_{1}^{\ast}(t),t^{\prime})-h(m_{2}^{
\ast}(t))
\end{align}
for all $t<t_{\ell}-\varepsilon$ which establishes that sender types $%
t^{\prime}\in\left[t_{\ell}, \overline{t}\right]$ do not have incentive to
deviate to the on-path message pair $(m_{1}, m_{2})$ such that $m_{1}\in
\left [m_{1}^{\ast}(\underline{t}),m_{1}^{\ast}(t_{\ell})\right)$ and $%
m_{1}+m_{2}< M$.\newline
\textbf{Individual Rationality:}\newline
Let $u_{b}^{\ast}(t):=\alpha m_{2}^{\ast}(t)+f\big(m_{1}^{\ast}(t),
\mu(m_{1}^{\ast}(t))\big)-c\big(m_{1}^{\ast}(t),t\big)-h(m_{2}^{\ast}(t))$
for $t\in \left(t_{\ell},\overline{t}\right)$. We apply the envelope Theorem
to that show $u_{b}^{\ast\prime}(t)>0$ for all $t\in \left(t_{\ell},
\overline{t}\right)$. Because $m_{1}^{\ast}(t)+m_{2}^{\ast}(t)=M$ for all $%
t\in \left(t_{\ell},\overline{t}\right)$, it follows that $%
u_{b}^{\ast\prime}(t)=\big[f_{m_1}\left(m_{1}^{\ast}(t),
\mu(m_{1}^{\ast}(t))\right) +f_t\big(m_{1}^{\ast}(t), \mu(m_{1}^{\ast}(t)) %
\big)\mu^{\prime}(m_{1}^{\ast}(t)) -c_{m_1}\big(m_{1}^{\ast}(t), t\big)-
\alpha+h^{\prime}(M-m_{1}^{\ast}(t))\big]m_{1}^{\ast\prime}(t)-c_{t}(m_{1}^{
\ast}(t),t) =-c_{t}(m_{1}^{\ast}(t),t)>0$. The squared bracket term in the
middle expression of $u_{b}^{\ast\prime}(t)$ is zero because of Condition 3.
The strict inequality is because of the single crossing property of $c$.
Therefore, $u_{b}^{\ast\prime}(t)>0$ for all $t\in \left(t_{\ell},\overline{
t }\right)$.

Also, from part (d), we have that for all $t^{\prime}\in \left(t_{\ell},
\overline{t}\right)$ and for all $t\in \left(\underline{t},t_{\ell}\right)$,
\begin{align}  \label{eq:17}
&\alpha m_{2}^{\ast}(t^{\prime})+f(m_{1}^{\ast}(t^{\prime}),t^{\prime})-
c(m_{1}^{\ast}(t^{\prime}),t^{\prime})-h(m_{2}^{\ast}(t^{\prime}))  \notag \\
&>\alpha
m_{2}^{\ast}(t)+f(m_{1}^{\ast}(t),t)-c(m_{1}^{\ast}(t),t^{\prime})-h(m_{2}^{
\ast}(t)).
\end{align}
Because $t<t^{\prime}$, by combining (\ref{eq:17}) and $c(m_{1}^{
\ast}(t),t^{\prime})<c(m_{1}^{\ast}(t),t)$ (Assumption 1(b)), we have that
for all $t^{\prime}\in \left(t_{\ell},\overline{t}\right)$ and for all $t\in
\left(\underline{t},t_{\ell}\right)$,
\begin{align}  \label{eq:18}
&\alpha m_{2}^{\ast}(t^{\prime})+f(m_{1}^{\ast}(t^{\prime}),t^{\prime})-
c(m_{1}^{\ast}(t^{\prime}),t^{\prime})-h(m_{2}^{\ast}(t^{\prime}))  \notag \\
&>\alpha
m_{2}^{\ast}(t)+f(m_{1}^{\ast}(t),t)-c(m_{1}^{\ast}(t),t)-h(m_{2}^{\ast}(t)).
\end{align}

Similarly, we show that $u^{\ast \prime }(t)>0$ for all $t\in \left(
\underline{t},t_{\ell }\right) $. Let $u_{s}^{\ast }(t):=\alpha m_{2}^{\ast
}(t)+f\big(m_{1}^{\ast }(t),\mu (m_{1}^{\ast }(t))\big)-c\big(m_{1}^{\ast
}(t),t\big)-h(m_{2}^{\ast }(t))$ for all $t\in \left( \underline{t},t_{\ell
}\right) $. Because $m_{2}^{\ast }(t)=m_{2}^{\circ }$ is constant on $\left(
\underline{t},t_{\ell }\right) $, by applying the envelope Theorem we have
that $u_{s}^{\ast \prime }(t)=\big[f_{m_{1}}\left( m_{1}^{\ast }(t),\mu
(m_{1}^{\ast }(t))\right) +f_{t}\big(m_{1}^{\ast }(t),\mu (m_{1}^{\ast }(t))%
\big)\mu ^{\prime }(m_{1}^{\ast }(t))-c_{m_{1}}\big(m_{1}^{\ast }(t),t\big)%
\big]m_{1}^{\ast \prime }(t)-c_{t}(m_{1}^{\ast }(t),t)=-c_{t}(m_{1}^{\ast
}(t),t)>0$ . The squared bracket term in the middle expression of $%
u_{s}^{\ast \prime }(t)$ is zero because of the first part of Condition 2.
The strict inequality is because of the single crossing property of $c$.
Therefore, $u_{s}^{\ast \prime }(t)>0$ for $t\in \left( \underline{t}%
,t_{\ell }\right) $. Let $u^{\ast }(\underline{t}):=\alpha m_{2}^{\ast }(%
\underline{t})+f\big(m_{1}^{\ast }(\underline{t}),\underline{t}\big)-c\big(%
m_{1}^{\ast }(\underline{t}),\underline{t}\big)-h(m_{2}^{\ast }(\underline{t}%
))$. Because $(m_{1}^{\ast }(\underline{t}),m_{2}^{\ast }(\underline{t}))$
is efficient, Assumption (3) implies that $u^{\ast }(\underline{t})\geq 0$.

In summary, $u_{b}^{\ast }(t^{\prime })>u_{s}^{\ast }(t)>u^{\ast }(%
\underline{t})\geq 0$ for any $t^{\prime }\in \left[ t_{\ell },\overline{t}%
\right) $ and for any $t\in \left( \underline{t},t_{\ell }\right) $. This
shows the individual rationality.\newline
\textbf{Off-path deviations:}\newline
(a) We show that no sender type has a profitable deviation to off-path
message pair $(m_{1}^{\prime },m_{2}^{\prime })$ with $m_{1}^{\prime }\in
(0,m_{1}^{\ast }(\underline{t}))$, such that $m_{1}^{\prime }+m_{2}^{\prime
}\leq M$. Lemma \ref{lemma_Off} shows that Criterion D1 places all the
weight on type $\underline{t}$ for such deviations. Therefore if $\underline{%
t}$ has no profitable deviation to such off-path message, then no other type
does. Because the equilibrium choice of type $\underline{t}$ is efficient
and unique, she does not deviate from her equilibrium action pair. Therefore
no sender type profitably deviates to such off-path message pair.\newline
(b) We show that no sender type has a profitable deviation to off-path
message $(m_{1}^{\prime },m_{2}^{\prime })$ with $m_{1}^{\prime
}>m_{1}^{\ast }(\overline{t})$ such that $m_{1}^{\prime }+m_{2}^{\prime
}\leq M$. According to Lemma \ref{lemma_Off3}, Criterion D1 places all the
weight on type $\overline{t}$ for such deviations. Therefore if $\overline{t}
$ has no profitable deviation to such off-path message, then no other type
does. We now show that type $\overline{t}$ is strictly worse off by
deviating to off-path message pair $(m_{1}^{\prime },m_{2}^{\prime })$ with $%
m_{1}^{\prime }>m_{1}^{\ast }(\overline{t})$ such that $m_{1}^{\prime
}+m_{2}^{\prime }\leq M$. Because $m_{1}^{\ast }(\overline{t})$ is
inefficiently high, given that $m_{1}^{\prime }>m_{1}^{\ast }(\overline{t})$%
, by using the strict concavity of $f-c$, we have that
\begin{equation}
f\big(m_{1}^{\ast }(\overline{t}),\overline{t}\big)-c\big(m_{1}^{\ast }(%
\overline{t}),\overline{t}\big)>f\big(m_{1}^{\prime },\overline{t}\big)-c%
\big(m_{1}^{\prime },\overline{t}\big).  \label{eq:nonbinding28}
\end{equation}%
See that $m_{1}^{\prime }+m_{2}^{\prime }\leq M$ implies $m_{2}^{\prime
}\leq M-m_{1}^{\prime }<M-m_{1}^{\ast }(\overline{t})=m_{2}^{\ast }(%
\overline{t})$ when $m_{1}^{\prime }>m_{1}^{\ast }(\overline{t})$. Therefore
$m_{2}^{\prime }<m_{2}^{\ast }(\overline{t})$ when $m_{1}^{\prime
}>m_{1}^{\ast }(\overline{t})$. Because $h$ is strictly convex (Assumption
3.(b)), $m_{2}^{\prime }<m_{2}^{\ast }(\overline{t})<m_{2}^{\circ }$,
implies that
\begin{equation}
\alpha m_{2}^{\prime }-h(m_{2}^{\prime })<\alpha m_{2}^{\ast }(\overline{t}%
)-h(m_{2}^{\ast }(\overline{t})).  \label{eq:nonbinding27}
\end{equation}%
By combining equation (\ref{eq:nonbinding27}) and (\ref{eq:nonbinding28}),
we have that
\begin{equation}
\alpha m_{2}^{\ast }(\overline{t})+f\big(m_{1}^{\ast }(\overline{t}),%
\overline{t}\big)-c\big(m_{1}^{\ast }(\overline{t}),\overline{t}\big)%
-h(m_{2}^{\ast }(\overline{t}))>\alpha m_{2}^{\prime }+f\big(m_{1}^{\prime },%
\overline{t}\big)-c\big(m_{1}^{\prime },\overline{t}\big)-h(m_{2}^{\prime })
\label{eq:nonbinding29}
\end{equation}%
which shows that any sender type $\overline{t}$ is strictly worse off by
deviating to off-path message pair $(m_{1}^{\prime },m_{2}^{\prime })$ with $%
m_{1}^{\prime }>m_{1}^{\ast }(\overline{t})$ such that $m_{1}^{\prime
}+m_{2}^{\prime }\leq M$.

(c) No sender type in $\left[\underline{t},t_{\ell}\right)$ will choose the
off-path message pair $(m_{1}, m_{2}^{\prime})$ with $m_{1}\in \left[
m_{1}^{\ast}(\underline{t}),m_{1}^{\ast}(t_{\ell})\right)$, and $%
m_{2}^{\prime}<m_{2}^{\ast}(\overline{t})$ such that $m_{1}+m_{2}^{\prime}<
M $.

Define $t^{\ast}$ as the type such that $m_{1}^{\ast}(t^{\ast})=m_{1}$. Then
$\mathbb{E}_{\mu(m_{1}, m_{2}^{\prime})}f(m_{1},z)=f(m_{1},t^{\ast})$ for
some generic type $z$. Let $w^{\prime}:=\alpha m_{2}^{\prime}+f\big(m_{1},
t^{\ast}\big)$ denote the wage a sender receives upon choosing the action
pair $(m_{1}, m_{2}^{\prime})$ with $m_{1}\in \left[m_{1}^{\ast}(\underline{
t }),m_{1}^{\ast}(t_{\ell})\right)$, and $m_{2}^{\prime}<m_{2}^{\ast}(
\overline{t})$ such that $m_{1}+m_{2}^{\prime}< M$. For each $m_{1}\in \left[
m_{1}^{\ast}(\underline{t}),m_{1}^{\ast}(t_{\ell})\right]$, let $m_{2}$ be
the non-cognitive signal chosen by a sender along with $m_{1} $ on the
equilibrium path. Because $h$ is strictly convex, $m_{2}^{\prime}<m_{2}^{
\ast}(\overline{t})\leq m_{2}= m_{2}^{\circ}$ implies that
\begin{equation}  \label{eq:nonbinding30}
\alpha m_{2}^{\prime}-h(m_{2}^{\prime})<\alpha m_{2}-h(m_{2}).
\end{equation}
Because $((m_{1}^{\ast}(t),m_{2}^{\ast}(t))$ is incentive compatible for
type $t\in \left[\underline{t},t_{\ell}\right)$, we have that for each $t\in %
\left[\underline{t},t_{\ell}\right)$,
\begin{equation}  \label{eq:nonbinding31}
\alpha m_{2}^{\ast}(t) +f\big(m_{1}^{\ast}(t), t\big)-c\big( %
m_{1}^{\ast}(t),t \big) -h(m_{2}^{\ast}(t))\geq \alpha m_{2}+f\big(m_{1},
t^{\ast}\big)- c\big( m_{1},t\big)-h(m_{2}).
\end{equation}
By combining equation (\ref{eq:nonbinding30}) and (\ref{eq:nonbinding31}),
we have that for each $t\in \left[\underline{t},t_{\ell}\right)$,
\begin{equation}  \label{eq:nonbinding32}
\alpha m_{2}^{\prime}+f\big(m_{1}, t^{\ast}\big)-c\big(m_{1},t\big)-
h(m_{2}^{\prime}) < \alpha m_{2}^{\ast}(t) +f\big(m_{1}^{\ast}(t), t\big) -c %
\big(m_{1}^{\ast}(t),t\big)-h(m_{2}^{\ast}(t))
\end{equation}
which shows that each sender type $t\in \left[\underline{t},t_{\ell}\right)$
is strictly worse off by deviating to off-path message pair $(m_{1},
m_{2}^{\prime})$ with $m_{1}\in \left[m_{1}^{\ast}(\underline{t}
),m_{1}^{\ast}(t_{\ell})\right)$, and $m_{2}^{\prime}<m_{2}^{\ast}(\overline{
t})$ such that $m_{1}+m_{2}^{\prime}< M$.

(d) No sender type $t\in \left[t_{\ell},\overline{t}\right]$ will choose the
off-path pair $(m_{1}, m_{2}^{\prime})$ with $m_{1}\in \left[
m_{1}^{\ast}(t_{\ell}),m_{1}^{\ast}(\overline{t})\right]$ and $%
m_{2}^{\prime}<m_{2}^{\ast}(\overline{t})$ such that $m_{1}+m_{2}^{\prime}<
M $.

Define $t^{\ast}$ as the type such that $m_{1}^{\ast}(t^{\ast})=m_{1}$ for $%
m_{1}\in \left[m_{1}^{\ast}(t_{\ell}),m_{1}^{\ast}(\overline{t})\right]$.
Then $\mathbb{E}_{\mu(m_{1}, m_{2}^{\prime})}f(m_{1},z)=f(m_{1},t^{\ast})$
for some generic type $z$. Let $w^{\prime}:=\alpha
m_{2}^{\prime}+f\left(m_{1}, t^{\ast}\right)$ denote the wage a sender
receives upon choosing the action pair $(m_{1}, m_{2}^{\prime})$ such that $%
m_{1}+ m_{2}^{\prime}<M$. For each $m_{1}\in \left[m_{1}^{\ast}(t_{
\ell}),m_{1}^{\ast}(\overline{t})\right]$, let $m_{2}$ be the non-cognitive
signal chosen by a sender along with $m_{1}$ on the equilibrium path.
Because $m_{1}+ m_{2}=M$, it is clear that $m_{2}^{\prime}<m_{2}$ when $%
m_{1}+ m_{2}^{\prime}<M$. Because $h$ is strictly convex, $%
m_{2}^{\prime}<m_{2}<m_{2}^{\circ}$ implies that
\begin{equation}  \label{eq:nonbinding43}
\alpha m_{2}^{\prime}-h(m_{2}^{\prime})<\alpha m_{2}-h(m_{2}).
\end{equation}
Because $((m_{1}^{\ast}(t),m_{2}^{\ast}(t))$ is incentive compatible for
type $t\in \left[t_{\ell},\overline{t}\right]$, we have that for each $t\in %
\left[t_{\ell},\overline{t}\right]$,
\begin{equation}  \label{eq:nonbinding44}
\alpha m_{2}^{\ast}(t) +f\left(m_{1}^{\ast}(t),
t\right)-c\left(m_{1}^{\ast}(t),t\right) -h(m_{2}^{\ast}(t))\geq \alpha
m_{2}+f\left(m_{1}, t^{\ast}\right)-c\left(m_{1},t\right)-h(m_{2}).
\end{equation}
By combining equation (\ref{eq:nonbinding30}) and (\ref{eq:nonbinding31}),
we have that for each $t\in \left[t_{\ell},\overline{t}\right]$,
\begin{equation}  \label{eq:nonbinding45}
\alpha m_{2}^{\ast}(t) +f\left(m_{1}^{\ast}(t),
t\right)-c\left(m_{1}^{\ast}(t),t\right)-h(m_{2}^{\ast}(t)) >
w^{\prime}-c\left(m_{1},t\right)-h(m_{2}^{\prime})
\end{equation}
where $w^{\prime}=\alpha m_{2}^{\prime}+f\left(m_{1}, t^{\ast}\right)$. (\ref%
{eq:nonbinding45}) shows that each sender type $t\in \left[t_{\ell},
\overline{t}\right]$ is strictly worse off by deviating to $(m_{1},
m_{2}^{\prime})$ with $m_{1}\in \left[m_{1}^{\ast}(t_{\ell}),m_{1}^{\ast}(
\overline{t})\right]$, and $m_{2}^{\prime}<m_{2}^{\ast}(\overline{t})$ such
that $m_{1}+m_{2}^{\prime}< M$.

(e) No sender type in $\left[\underline{t},t_{\ell}\right)$ will choose the
off-path message pair $(m_{1}, m_{2}^{\prime})$ with $m_{1}\in
\left
		[m_{1}^{\ast}(\underline{t}),m_{1}^{\ast}(t_{\ell})\right)$ and $%
m_{2}^{\prime}>m_{2}^{\circ}$ such that $m_{1}+m_{2}^{\prime}\leq M$.

Define $t^{\ast}$ as the type such that $m_{1}^{\ast}(t^{\ast})=m_{1}$. Then
$\mathbb{E}_{\mu(m_{1}, m_{2}^{\prime})}f(m_{1},z)=f(m_{1},t^{\ast})$ for
some generic type $z$. Let $w^{\prime}:=\alpha m_{2}^{\prime}+f\big(m_{1},
t^{\ast}\big)$ denote the wage a sender receives upon choosing the action
pair $(m_{1}, m_{2}^{\prime})$ with $m_{1}\in \left [m_{1}^{\ast}(\underline{
t}),m_{1}^{\ast}(t_{\ell})\right)$ and $m_{2}^{\prime}>m_{2}^{\circ}$ such
that $m_{1}+m_{2}^{\prime}\leq M$. Because $h$ is strictly convex, $%
m_{2}^{\circ}<m_{2}^{\prime}$ we have that
\begin{equation}  \label{eq:nonbinding39}
\alpha m_{2}^{\prime}-h(m_{2}^{\prime})<\alpha
m_{2}^{\circ}-h(m_{2}^{\circ}).
\end{equation}

For each $m_{1}\in \left [m_{1}^{\ast}(\underline{t}),m_{1}^{\ast}(t_{\ell})
\right)$, let $m_{2}$ be the non-cognitive signal chosen by a sender along
with $m_{1}$ on the equilibrium path. Because $((m_{1}^{\ast}(t),m_{2}^{
\ast}(t))$ is incentive compatible for type $t\in \left[\underline{t}
,t_{\ell}\right)$, we have that for each $t\in \left[\underline{t}
,t_{\ell}\right)$,
\begin{equation}  \label{eq:nonbinding40}
\alpha m_{2}^{\ast}(t) +f\big(m_{1}^{\ast}(t), t\big)- c\big( %
m_{1}^{\ast}(t),t\big) -h(m_{2}^{\ast}(t))\geq \alpha m_{2}+f\big(m_{1},
t^{\ast}\big) -c\big(m_{1},t\big)-h(m_{2}).
\end{equation}
Because $m_{2}^{\ast}(t)=m_{2}=m_{2}^{\circ}$ for all $t\in \left[\underline{
t},t_{\ell}\right)$, (\ref{eq:nonbinding40}) becomes
\begin{equation}  \label{eq:nonbinding41}
\alpha m_{2}^{\circ} +f\big(m_{1}^{\ast}(t), t\big) -c\big(m_{1}^{\ast}(t),t %
\big) -h(m_{2}^{\circ})\geq \alpha m_{2}^{\circ}+ f\big(m_{1}, t^{\ast}\big) %
-c\big(m_{1},t\big)-h(m_{2}^{\circ}).
\end{equation}
By combining equation (\ref{eq:nonbinding39}) and (\ref{eq:nonbinding41}),
we have that for each $t\in \left[\underline{t},t_{\ell}\right)$,
\begin{equation}  \label{eq:nonbinding42}
\alpha m_{2}^{\prime}+f\big(m_{1}, t^{\ast}\big)- c\big(m_{1},t\big) %
-h(m_{2}^{\prime}) < \alpha m_{2}^{\ast}(t) +f\big(m_{1}^{\ast}(t), t\big)-
c \big(m_{1}^{\ast}(t),t\big)-h(m_{2}^{\ast}(t))
\end{equation}
which shows that each sender type $t\in \left[\underline{t},t_{\ell}\right)$
is strictly worse off by deviating to off-path message pair$(m_{1},
m_{2}^{\prime})$ with $m_{1}\in \left [m_{1}^{\ast}(\underline{t}
),m_{1}^{\ast}(t_{\ell})\right)$ and $m_{2}^{\prime}>m_{2}^{\circ}$ such
that $m_{1}+m_{2}^{\prime}\leq M$.

(f) No sender type $t\in \left[t_{\ell},\overline{t}\right]$ will choose the
off-path pair $(m_{1}, m_{2}^{\prime})$ with $m_{1}\in \left[
m_{1}^{\ast}(t_{\ell}),m_{1}^{\ast}(\overline{t})\right]$, and $%
m_{2}^{\prime}>m_{2}^{\circ}$ such that $m_{1}+m_{2}^{\prime}\leq M$.

Define $t^{\ast}$ as the type such that $m_{1}^{\ast}(t^{\ast})=m_{1}$. Then
$\mathbb{E}_{\mu(m_{1}, m_{2}^{\prime})}f(m_{1},z)=f(m_{1},t^{\ast})$ for
some generic type $z$. Let $w^{\prime}:=\alpha m_{2}^{\prime}+f\big(m_{1},
t^{\ast}\big)$ denote the wage a sender receives upon choosing the action
pair $(m_{1}, m_{2}^{\prime})$ with $m_{1}\in \left[m_{1}^{\ast}(t_{
\ell}),m_{1}^{\ast}(\overline{t})\right]$, and $m_{2}^{\prime}>m_{2}^{\circ}$
such that $m_{1}+m_{2}^{\prime}\leq M$. For each $m_{1}\in \left[
m_{1}^{\ast}(t_{\ell}),m_{1}^{\ast}(\overline{t})\right]$, let $m_{2}$ be
the non-cognitive signal chosen by a sender along with $m_{1}$ on the
equilibrium path. Because $m_{2}^{\prime}>m_{2}^{\circ} $, it is clear that $%
m_{2}> m_{2}^{\prime}>m_{2}^{\circ} $ when $m_{1}+m_{2}^{\prime}\leq M$ and $%
m_{1}+m_{2}= M$. Because $h$ is strictly convex, $m_{2}>
m_{2}^{\prime}>m_{2}^{\circ} $ implies that
\begin{equation}  \label{eq:nonbinding401}
\alpha m_{2}^{\prime}-h(m_{2}^{\prime})<\alpha m_{2}-h(m_{2}).
\end{equation}
Because $(m_{1}^{\ast}(t),m_{2}^{\ast}(t)$ is incentive compatible for type $%
t\in \left[t_{\ell},\overline{t}\right]$, we have that for each $t\in \left[
t_{\ell},\overline{t}\right]$,
\begin{equation}  \label{eq:nonbinding411}
\alpha m_{2}^{\ast}(t) +f\big(m_{1}^{\ast}(t), t\big)-c\big( %
m_{1}^{\ast}(t),t \big) -h(m_{2}^{\ast}(t))\geq \alpha m_{2}+f\big(m_{1},
t^{\ast}\big)-c\big( m_{1},t\big)-h(m_{2}).
\end{equation}
By combining equation (\ref{eq:nonbinding401}) and (\ref{eq:nonbinding411}),
we have that for each $t\in \left[t_{\ell},\overline{t}\right]$,
\begin{equation}  \label{eq:nonbinding412}
\alpha m_{2}^{\ast}(t) +f\big(m_{1}^{\ast}(t), t\big)-c\big( %
m_{1}^{\ast}(t),t \big)-h(m_{2}^{\ast}(t)) > w^{\prime}-c\big(m_{1},t\big) %
-h(m_{2}^{\prime})
\end{equation}
which shows that each sender type $t\in \left[t_{\ell},\overline{t}\right]$
is strictly worse off by deviating to off-path message pair$(m_{1},
m_{2}^{\prime})$ with $m_{1}\in \left[m_{1}^{\ast}(t_{\ell}),m_{1}^{\ast}(
\overline{t})\right]$, and $m_{2}^{\prime}>m_{2}^{\circ}$ such that $%
m_{1}+m_{2}^{\prime}\leq M$.

The proof is complete.
\end{proof}

\section*{Proofs of Theorem \protect\ref{lemma40} and Lemma \protect\ref%
{lemma4}}

We break the proofs in lemmas.

\begin{lemma}
\label{lemma400} Suppose that $Z(m_{1},m_{2})$ has a positive measure in
equilibrium and denote $t_{\circ}:=\text{max~} Z(m_{1},m_{2})$. Then at
least one of a sender's strategy pair $\hat{m_{1}}$, $\hat{m_{2}}$, is
discontinuous at $t_{\circ}$ such that $t_{\circ}$ $<\overline{t}$.
\end{lemma}

\begin{proof}
\textbf{of Lemma \ref{lemma400}.} Let $t_{\circ}<\overline{t}$. Suppose
towards a contradiction that both $\hat{m_{1}}$ and $\hat{m_{2}}$ are
continuous at $t_{\circ}$. Because supp $\mu(\hat{m_{1}}(t_{\circ}),\hat{
m_{2}}(t_{\circ}))=Z(m_{1},m_{2})$, and $\lim_{t\searrow t_{\circ}}\text{inf
supp}\mu(\hat{m_{1}}(t),\hat{m_{2}}(t))=t_{\circ}$, we have
\begin{equation}  \label{eq:jump1}
\mathbb{E}_{\mu(\hat{m_{1}}(t_{\circ}),\hat{m_{2}}(t_{\circ}))}f(\hat{m_{1}}
(t_{\circ}),z) <\lim_{t\searrow t_{\circ}}\mathbb{E}_{\mu(\hat{m_{1}}(t),
\hat{m_{2}}(t))}f(\hat{m_{1}}(t),z)
\end{equation}
for some generic type $z\in T$. Together with the continuity of $\hat{m_{1}}$
, $\hat{m_{2}}$, $c, f,$ and $h$, (\ref{eq:jump1}) implies
\begin{align}  \label{eq:jump2}
&\alpha \hat{m_{2}}(t_{\circ})+ \mathbb{E}_{\mu(\hat{m_{1}}(t_{\circ}),\hat{
m_{2}}(t_{\circ}))}f(\hat{m_{1}}(t_{\circ}),z)- c(\hat{m_{1}}
(t_{\circ}),t_{\circ})-h(\hat{m_{2}}(t_{\circ}))  \notag \\
<& \lim_{t\searrow t_{\circ}} \Big [ \alpha \hat{m_{2}}(t)+ \mathbb{E}_{\mu(
\hat{m_{1}}(t),\hat{m_{2}}(t))}f(\hat{m_{1}}(t),z)-c(\hat{m_{1}}(t),t)- h(
\hat{m_{2}}(t)) \Big ].
\end{align}
It follows from (\ref{eq:jump2}) that there exists some $\tau >0$ such that
deviation to the pair $(\hat{m_{1}}(t_{\circ}+\tau),\hat{m_{2}}
(t_{\circ}+\tau))$ makes type $t_{\circ}$ strictly better off which
contradicts the pair $(m_{1},m_{2})$ as the equilibrium choice for type $%
t_{\circ}$. We conclude that at least one of the strategy pairs, $\hat{m_{1}}
$ or $\hat{m_{2}}$ must be discontinuous at $t_{\circ}<\overline{t}$.
\end{proof}


\begin{proof}
\textbf{of Theorem \ref{lemma40}.} Denote $t_{\circ }:=\text{max~}%
Z(m_{1},m_{2})$. Suppose towards a contradiction that $m_{1}+m_{2}<M$. If $%
t_{\circ }=\overline{t}$, then $t_{\circ }$ has an incentive to slightly
increase the level of cognitive signal because the resource constraint is
slack and moreover, $t_{\circ }$ is the highest type in the entire type
space. That is, there is some $\varepsilon >0$ such that
\begin{align}
\alpha m_{2}+f(m_{1}+\varepsilon ,\overline{t})-c(m_{1}+\varepsilon
,t_{\circ })-h(m_{2})>&  \notag  \label{eq:profitabledeviation01} \\
& \alpha m_{2}+\mathbb{E}_{\mu (m_{1},m_{2})}f(m_{1},z)-c(m_{1},t_{\circ
})-h(m_{2}).
\end{align}

To establish inequality (\ref{eq:profitabledeviation01}), we first note
that, because type $t_{\circ}=\overline{t}$ is the highest in the type
interval that chooses the same signal pair $(m_{1},m_{2})$ in equilibrium,
we have
\begin{equation}  \label{eq:profitabledeviation02}
f(m_{1},\overline{t})>\mathbb{E}_{\mu(m_{1},m_{2})}f(m_{1},z).
\end{equation}
Because $f$, $c$ are continuous in the sender's actions, given (\ref%
{eq:profitabledeviation02}), we have that
\begin{align}  \label{eq:profitabledeviation03}
\lim_{\varepsilon\searrow 0}\Big[ \alpha m_{2}+ f(m_{1}+\varepsilon,
\overline{t})-c(m_{1}+\varepsilon,\overline{t})-h(m_ 2)\Big] &  \notag \\
> \alpha m_{2}+\mathbb{E}_{\mu(m_{1},m_{2})}f(m_{1},z)-c(m_{1},\overline{t}
)-h(m_{2}).
\end{align}
Inequality (\ref{eq:profitabledeviation03}) implies that there exists $%
\varepsilon>0$ such that (\ref{eq:profitabledeviation01}) holds. This is a
contradiction since $(m_{1},m_{2})$ is the equilibrium action pair for type $%
t_{\circ}=\overline{t}$.

Now consider the case where $t_{\circ}<\overline{t}$. Because of Lemma \ref%
{lemma400}, we have that at least one of the strategy pairs, $\hat{m_{1}}$, $%
\hat{m_{2}}$ must be discontinuous at $t_{\circ}<\overline{t}$. Suppose that
$\hat{m_{1}}$ is continuous at $t_{\circ}<\overline{t}$, but $\hat{m_{2}}$
is discontinuous at $t_{\circ}<\overline{t}$ when $m_{1}+m_{2}<M$. This
implies that $m_{1}=\lim_{t\searrow t_{\circ}} \hat{m_{1}}
(t)=\lim_{t\nearrow t_{\circ}} \hat{m_{1}}(t)$ and $m_{2}> \lim_{t\searrow
t_{\circ}} \hat{m_{2}}(t)$. Note that it is not possible for $%
\lim_{t\searrow t_{\circ}} \hat{m_{1}}(t)+ \lim_{t\searrow t_{\circ}} \hat{
m_{2}}(t)=M$. Otherwise, $m_{1}+m_{2}>\lim_{t\searrow t_{\circ}} \hat{m_{1}}
(t)+ \lim_{t\searrow t_{\circ}} \hat{m_{2}}(t)=M$ which contradicts the
resource constraint. Therefore, the only possibility here is $%
\lim_{t\searrow t_{\circ}} \hat{m_{1}}(t)+ \lim_{t\searrow t_{\circ}} \hat{
m_{2}}(t)<M$. Denote $m_{2}^{+}:= \lim_{t\searrow t_{\circ}} \hat{m_{2}}(t)$
. Given Assumption 3.(b) (strict convexity of $h$), Assumption 4, and $%
m_{2}^{+}<m_{2}$, only three subcases are possible; (a) $m_{2}=m_{2}^{\circ
} $ (b) $m_{2}^{+}=m_{2}^{\circ }$, and (c) $m_{2}<m_{2}^{\circ }$.

In subcase (a), where $m_{2}=m_{2}^{\circ }$, it follows that $%
m_{2}^{+}<m_{2}^{\circ }$. Because $m_{1} + m_{2}^{+}<M$, and $%
m_{2}^{+}<m_{2}^{\circ }$, there exists $m_{2}^{{\prime}+}$ for type $%
t\searrow t_{\circ}$ such that (i) $m_{2}^{+}<m_{2}^{{\prime}+}<m_{2}^{\circ
}$ and (ii) $r(m_{1}, m_{2}^{{\prime}+})=s$. Because $r(m_{1},
m_{2}^{+})=r(m_{1}, m_{2}^{{\prime}+})=s$, we have $\hat{\mu}(m_{1},
m_{2}^{+})=\hat{\mu}(m_{1}, m_{2}^{{\prime}+})$ given $\hat{\mu}\in \Psi $.
Accordingly, the utility difference for type $t\searrow t_{\circ}$ by
deviation to $(m_{1}, m_{2}^{{\prime}+})$ is
\begin{gather*}
\hat{w}(m_{1}, m_{2}^{{\prime}+})-h(m_{2}^{{\prime }+}) -c(m_{1}, t)-\left[
\hat{w}(m_{1}, m_{2}^{+})- h(m_{2}^{+})-c(m_{1}, t)\right] = \\
\alpha m_{2}^{{\prime }+}-h\left( m_{2}^{{\prime}+}\right) - [ \alpha
m_{2}^{+}-h\left( m_{2}^{+}\right)],
\end{gather*}
where the equality holds because $\hat{w}(m_{1}, m_{2}^{{\prime}+})-\hat{w}
(m_{1}, m_{2}^{+})=$ $\alpha m_{2}^{{\prime }+}- \alpha m_{2}^{+}$ given $%
r(m_{1}, m_{2}^{+})=r(m_{1}, m_{2}^{{\prime}+})$. Because $m_{2}^{+}<m_{2}^{{%
\ \ \prime}+}<m_{2}^{\circ }$, the strict convexity of $h$ implies that $%
\alpha m_{2}^{{\prime }+} -h\left( m_{2}^{{\prime}+}\right) > \alpha
m_{2}^{+}-h\left( m_{2}^{+}\right)$. Therefore, the utility difference is
strictly positive. This contradicts that $m_{2}^{+}$ is an equilibrium
choice of a non-cognitive signal for type $t\searrow t_{\circ}$.

In subcase (b), we have that $m_{2}^{+}=m_{2}^{\circ }$, which implies that $%
m_{2}^{\circ }<m_{2}$. The latter inequality means that the level of an
equilibrium choice of a non-cognitive signal for type $t_{\circ}$ exceeds
the optimal level of a non-cognitive signal. This is a contradiction, given
the strict convexity of $h$ and Assumption 4.

In subcase (c), we have that $m_{2}<m_{2}^{\circ }$. Because $m_{1}+m_{2}<M$
, and $m_{2}<m_{2}^{\circ }$, there exists $m_{2}^{\prime }$ for type $%
t_{\circ}<\overline{t}$ such that (i) $m_{2}<m_{2}^{\prime }<m_{2}^{\circ }$
and (ii) $r( m_{1},m_{2}^{\prime })=s.$ Because $r(m_{1}, m_{2})=r(m_{1},
m_{2}^{\prime})=s$, we have $\hat{\mu}(m_{1}, m_{2})=\hat{\mu}(m_{1},
m_{2}^{\prime})$ given $\hat{\mu}\in \Psi $. Accordingly, the utility
difference for type $t_{\circ}$ by deviation to $(m_{1}, m_{2}^{\prime})$ is
\begin{gather*}
\hat{w}(m_{1}, m_{2}^{\prime})-h(m_{2}^{\prime }) -c(m_{1}, t_{\circ})-\left[
\hat{w}(m_{1}, m_{2})- h(m_{2})-c(m_{1}, t_{\circ})\right] = \\
\alpha m_{2}^{\prime }-h\left( m_{2}^{\prime}\right) - [ \alpha
m_{2}-h\left( m_{2}\right)],
\end{gather*}
where the equality holds because $\hat{w}(m_{1}, m_{2}^{\prime})-\hat{w}
(m_{1}, m_{2})=$ $\alpha m_{2}^{\prime }- \alpha m_{2}$ given $r(m_{1},
m_{2})=r(m_{1}, m_{2}^{\prime})$. Because $m_{2}<m_{2}^{\prime
}<m_{2}^{\circ }$, the strict convexity of $h$ implies that $\alpha
m_{2}^{\prime } -h\left( m_{2}^{\prime}\right) > \alpha m_{2}-h\left(
m_{2}\right)$. Therefore, the utility difference is strictly positive. This
contradicts that $m_{2}$ is an equilibrium choice of a non-cognitive signal
for type $t_{\circ}<\overline{t}$. Therefore we conclude that in the case
where $\hat{m_{1}}$ is continuous at $t_{\circ}<\overline{t}$ and $\hat{
m_{2} }$ is discontinuous at $t_{\circ}<\overline{t}$, it must be the case
that $m_{1}+m_{2}=M$ when $Z(m_{1},m_{2})$ has a positive measure in
equilibrium.

Next, consider the case where $\hat{m_{1}}$ is discontinuous at $t_{\circ}<
\overline{t}$, but $\hat{m_{2}}$ is continuous at $t_{\circ}<\overline{t}$
when $m_{1}+m_{2}<M$. We follow the proof of inequality (\ref%
{eq:profitabledeviation01}) to similarly show that type $t_{\circ}<\overline{
t}$ has an incentive to slightly increase the level of cognitive signal in
this case. This contradicts the fact that $m_{1}$ is an equilibrium choice
of a cognitive signal for type $t_{\circ}<\overline{t}$. In conclusion, if $%
t_{\circ}<\overline{t}$, $m_{1}+m_{2}=M$ must hold when $\hat{m_{1}}$ is
discontinuous at $t_{\circ}<\overline{t}$ and $\hat{m_{2}}$ is continuous at
$t_{\circ}<\overline{t}$.

Finally, consider the case where both $\hat{m_{1}}$ and $\hat{m_{2}}$ are
discontinuous at $t_{\circ}<\overline{t}$ and $m_{1}+m_{2}<M$. This
discontinuity at $t_{\circ}<\overline{t}$ creates a product interval of
off-path message pairs call it $D$ defined as
\begin{equation*}
D=\Big\{(s_1,s_2)\in \mathcal{C}: (s_1,s_2)\in \displaystyle{\big( %
m_1,\lim_{t\searrow t_{\circ}} \hat{m_{1}}(t)\big)\times \big( %
\lim_{t\searrow t_{\circ}} \hat{m_{2}}(t), m_{2}\big)}\Big\}.
\end{equation*}

According to Lemma \ref{lemma_Off2}, Criterion D1 places all the weight on
type $t_{\circ}<\overline{t}$ for any deviation to $(m_{1}^{\prime},m_{2}^{
\prime})\in D$. That is, $\text{supp~}\mu(m^{\prime}_{1},m^{\prime}_{2})
=\{t_{\circ}\}$ for any $(m^{\prime}_{1},m^{\prime}_{2})\in D$, if the
belief $\mu(m^{\prime}_{1},m^{\prime}_{2})$ conditional on $%
(m^{\prime}_{1},m^{\prime}_{2})\notin \text{range}(m_{1},m_{2})$ passes
Criterion D1.
Therefore, we show that type $t_{\circ}<\overline{t}$ has a profitable
deviation to action pair $(m_{1}+\varepsilon,m_{2}-\varepsilon)\in D$, $%
\varepsilon>0$. That is, we show that that
\begin{align}  \label{eq:profit1}
\alpha m_{2}-\alpha \varepsilon +f(m_{1}+\varepsilon,t_{\circ})-c(m_{1}+
\varepsilon,t_{\circ})-h(m_ 2-\varepsilon)>&  \notag \\
\alpha m_{2}+\mathbb{E}_{\mu(m_{1},m_{2})}f(m_{1},t)-c(m_{1},t_{
\circ})-h(m_{2}).
\end{align}
Because type $t_{\circ}<\overline{t}$ is the highest in the type interval
that chooses the same signal pair $(m_{1},m_{2})$ in equilibrium, we have
that
\begin{equation}  \label{eq:profit2}
f(m_{1},t_{\circ})>\mathbb{E}_{\mu(m_{1},m_{2})}f(m_{1},t).
\end{equation}
Because $f$, $c$ and $h$ are continuous in the sender's actions, and
inequality (\ref{eq:profit2}), we have that
\begin{align}  \label{eq:profit3}
\lim_{\varepsilon\searrow 0}\Big[ \alpha m_{2}-\alpha \varepsilon +
f(m_{1}+\varepsilon,t_{\circ})-c(m_{1}+\varepsilon,t_{\circ})-h(m_
2-\varepsilon)\Big] >&  \notag \\
\alpha m_{2}+\mathbb{E}_{\mu(m_{1},m_{2})}f(m_{1},t)-c(m_{1},t_{
\circ})-h(m_{2}).
\end{align}
Inequality (\ref{eq:profit3}) implies that there exists $\varepsilon>0$ such
that (\ref{eq:profit1}) holds. This is a contradiction to $(m_{1},m_{2})$ as
the equilibrium action pair for type $t_{\circ}<\overline{t}$. This
concludes that if $t_{\circ}<\overline{t}$, $m_{1}+m_{2}=M$ must hold when
both $\hat{m_{1}}$ and $\hat{m_{2}}$ are discontinuous at $t_{\circ}<
\overline{t}$.

Therefore we conclude that, $m_{1}+m_{2}=M$ when $Z(m_{1},m_{2})$ has a
positive measure in equilibrium.
\end{proof}

\begin{proof}
\textbf{of Lemma \ref{lemma4}.} Let $Z(m_{1},m_{2})$ denote the set of types
of senders who choose the same signals $(m_{1},m_{2})$. Let $Z(m_{1},m_{2})$
be a set of positive measure in equilibrium, and denote $t_{\circ}:=\text{
max~} Z(m_{1},m_{2})$. Let $t_{\circ}$ $<$ $\overline{t}$. Suppose towards a
contradiction that $\displaystyle{\hat{m_{1}}(t_{\circ}) = \lim_{t\searrow
t_{\circ}} \hat{m_{1}}(t)}$ or $\displaystyle{\hat{m_{2}}(t_{\circ}) =
\lim_{t\searrow t_{\circ}} \hat{m_{2}}(t)}$. Because of Theorem \ref{lemma40}
, we have that $\displaystyle{\hat{m_{2}}(t_{\circ}) = \lim_{t\searrow
t_{\circ}} \hat{m_{2}}(t)}$ whenever $\displaystyle{\hat{m_{1}}(t_{\circ}) =
\lim_{t\searrow t_{\circ}} \hat{m_{1}}(t)}$, and vice versa.

Because $\text{supp}\mu(\hat{m_{1}}(t_{\circ}),\hat{m_{2}}(t_{\circ}))$$=$ $%
Z(m_{1},m_{2})$ and $\lim_{t\searrow t_{\circ}}$ $\text{inf supp}(\hat{m_{1}}
(t),\hat{m_{2}}(t))$=$t_{\circ}$, we have that
\begin{equation}  \label{eq:diff1}
\mathbb{E}_{\mu(\hat{m_{1}}(t_{\circ}),\hat{m_{2}}(t_{\circ}))}f(\hat{m_{1}}
(t_{\circ}),t_{\circ})< \lim_{t\searrow t_{\circ}}\mathbb{E}_{\mu(\hat{m_{1}}
(t),\hat{m_{2}}(t))}f(\hat{m_{1}}(t),t).
\end{equation}
Because $f$, $c$, and $h$ are continuous, inequality (\ref{eq:diff1})
together with $\displaystyle{\hat{m_{1}}(t_{\circ}) = \lim_{t\searrow
t_{\circ}} \hat{m_{1}}(t)}$, and $\displaystyle{\hat{m_{2}}(t_{\circ}) =
\lim_{t\searrow t_{\circ}} \hat{m_{2}}(t)}$ lead to
\begin{align}  \label{eq:indiff1}
&\alpha \hat{m_{2}}(t_{\circ})+ \mathbb{E}_{\mu(\hat{m_{1}}(t_{\circ}),\hat{
m_{2}}(t_{\circ}))}f(\hat{m_{1}}(t_{\circ}),t_{\circ})- c(\hat{m_{1}}
(t_{\circ}),t_{\circ})-h(\hat{m_{2}}(t_{\circ}))  \notag \\
<& \lim_{t\searrow t_{\circ}} \Big [ \alpha \hat{m_{2}}(t)+ \mathbb{E}_{\mu(
\hat{m_{1}}(t),\hat{m_{2}}(t))}f(\hat{m_{1}}(t),t)- c(\hat{m_{1}}(t),t)-h(
\hat{m_{2}}(t)) \Big ].
\end{align}
From equation (\ref{eq:indiff1}), there exists $\tau >0$ such that deviation
to $(\hat{m_{1}}(t_{\circ}+\tau),\hat{m_{2}}(t_{\circ}+\tau))$ makes type $%
t_{\circ}$ strictly better off which contradicts the equilibrium choice for
type $t_{\circ}$. Therefore, $\displaystyle{\hat{m_{1}}(t_{\circ})
<\lim_{t\searrow t_{\circ}} \hat{m_{1}}(t)}$ and $\displaystyle{\hat{m_{2}}
(t_{\circ}) >\lim_{t\searrow t_{\circ}} \hat{m_{2}}(t)}$.
\end{proof}

\section*{Proofs of Theorem \protect\ref{prop3} and Corollary \protect\ref%
{prop4}}

\begin{lemma}
\label{lemma2}If the measure of $Z(m_{1},m_{2})$ is positive in equilibrium,
$Z(m_{1},m_{2})$ is a connected interval.
\end{lemma}

\begin{proof}
\textbf{of Lemma \ref{lemma2}.} Let $Z(m_{1},m_{2})$ denote the set of types
of senders who choose the same signals $(m_{1},m_{2})\in \mathcal{C}$ and
let the measure of $Z(m_{1},m_{2})$ be positive in equilibrium. We show that
$Z(m_{1},m_{2})$ is an interval. That is, consider any $t_{1},t_{2}\in
Z(m_{1},m_{2})$. We show that for any $t^{\prime}\in T$, such that $%
t_{1}<t^{\prime}<t_{2}$, $t^{\prime}\in Z(m_{1},m_{2})$. Let $t_{1}:=\min
Z(m_{1},m_{2}) $ and $t_{2}:=\max Z(m_{1},m_{2})$ if the min and max exists,
(otherwise we replace min and max with inf and sup respectively). Because $%
Z(m_{1},m_{2})$ has a positive measure, we have that the set $Z(m_{1},m_{2})$
is nonempty and because of Lemma \ref{lemma40}, we have that $m_{1}+m_{2}=M$
. \newline
\textbf{case 1}: Let $(m_{1},m_{2})$ be the equilibrium action pair chosen
by both type $t_{1}$ and $t_{2}$. Consider any deviation $%
(m^{\prime}_{1},m^{\prime}_{2})\in \mathcal{C}$ such that $%
(m^{\prime}_{1},m^{\prime}_{2}) \neq (m_{1},m_{2})$ and $m^{\prime}_{1}\neq
m_{1}$. Because $t_{1},t_{2}\in Z(m_{1},m_{2})$, we have that $%
w-c(m_{1},t_{1})-h(m_{2})\geq w^{\prime}-c(m^{\prime}_{1},t_{1})
-h(m^{\prime}_{2})$ and $w-c(m_{1},t_{2})-h(m_{2})\geq
w^{\prime}-c(m^{\prime}_{1},t_{2}) -h(m^{\prime}_{2})$ where $w:=\alpha
m_{2}+\mathbb{E}_{\mu(m_{1},m_{2})}f(m_{1},z)$ and $w^{\prime}:=\alpha
m^{\prime}_{2} +\mathbb{E}_{\mu(m^{\prime}_{1},m^{\prime}_{2})}f(m^{
\prime}_{1},z)$. Rewriting these inequalities gives
\begin{equation}  \label{eq:lem1}
w-w^{\prime}+h(m^{\prime}_{2})-h(m_{2})\geq c(m_{1},t_{1})
-c(m^{\prime}_{1},t_{1})
\end{equation}
and
\begin{equation}  \label{eq:lem2}
w-w^{\prime}+h(m^{\prime}_{2})-h(m_{2})\geq c(m_{1},t_{2})-
c(m^{\prime}_{1},t_{2})
\end{equation}
for any message pair $(m^{\prime}_{1},m^{\prime}_{2})\in \mathcal{C}$ such
that $(m^{\prime}_{1},m^{\prime}_{2}) \neq (m_{1},m_{2})$ and $%
m^{\prime}_{1}\neq m_{1}$.

Given equation (\ref{eq:lem1}) and (\ref{eq:lem2}), first, we show that for
any $t^{\prime}\in T$ such that $t_{1}<t^{\prime}<t_{2}$, $%
w-c(m_{1},t^{\prime})-h(m_{2})> w^{\prime}
-c(m^{\prime}_{1},t^{\prime})-h(m^{\prime}_{2})$ for any $%
(m^{\prime}_{1},m^{\prime}_{2})\neq (m_{1},m_{2})$ and $m^{\prime}_{1}<m_{1}
$. Because $m^{\prime}_{1}<m_{1}$, we can apply the single crossing
condition for any $t_{1}< t^{\prime}$. That is,
\begin{equation}  \label{eq:lem3}
c(m_{1},t_{1})-c(m^{\prime}_{1},t_{1})> c(m_{1},t^{\prime})
-c(m^{\prime}_{1},t^{\prime}).
\end{equation}
Combining equation (\ref{eq:lem1}) and (\ref{eq:lem3}), we have
\begin{equation}  \label{eq:lem4}
w-w^{\prime}+h(m^{\prime}_{2})-h(m_{2})>
c(m_{1},t^{\prime})-c(m^{\prime}_{1},t^{\prime})
\end{equation}
which implies that $w-c(m_{1},t^{\prime})-h(m_{2})>
w^{\prime}-c(m^{\prime}_{1},t^{\prime})-h(m^{\prime}_{2})$ when $%
m^{\prime}_{1}<m_{1}$ for any $t^{\prime}$ with $t^{\prime}< t_{1}$.

Second, given equation (\ref{eq:lem1}) and (\ref{eq:lem2}), we show that for
any $t^{\prime}\in T$, such that $t_{1}<t^{\prime}<t_{2}$, we have $%
w-c(m_{1},t^{\prime})-h(m_{2})>
w^{\prime}-c(m^{\prime}_{1},t^{\prime})-h(m^{\prime}_{2})$ for any $%
(m^{\prime}_{1},m^{\prime}_{2})\neq (m_{1},m_{2})$ with $m^{
\prime}_{1}>m_{1} $. Because $m^{\prime}_{1}>m_{1}$, we can apply the single
crossing condition for any $t^{\prime}< t_{2}$. That is,
\begin{equation}  \label{eq:lem5}
c(m_{1},t_{2})-c(m^{\prime}_{1}, )>
c(m_{1},t^{\prime})-c(m^{\prime}_{1},t^{\prime}).
\end{equation}
Combining equation (\ref{eq:lem2}) and (\ref{eq:lem5}), we have
\begin{equation}  \label{eq:lem6}
w-w^{\prime}+h(m^{\prime}_{2})-h(m_{2})>
c(m_{1},t^{\prime})-c(m^{\prime}_{1},t^{\prime})
\end{equation}
which implies that $w-c(m_{1},t^{\prime})-h(m_{2})>
w^{\prime}-c(m^{\prime}_{1},t^{\prime})-h(m^{\prime}_{2})$ when $%
m^{\prime}_{1}>m_{1}$ for any $t^{\prime}$ with $t^{\prime}< t_{2}$.

\textbf{case 2}: Let $(m_{1},m_{2})\in \mathcal{C}$ be the equilibrium
action pair chosen by both type $t_{1}$ and $t_{2}$. Consider any deviation $%
(m^{\prime}_{1},m^{\prime}_{2})\in \mathcal{C}$ such that $%
m^{\prime}_{1}=m_{1}$ and $m^{\prime}_{2}\neq m_{2}$. Note that because $%
(m^{\prime}_{1},m^{\prime}_{2})\in \mathcal{C}$, it is not possible for $%
m^{\prime}_{2}>m_{2}$ when $m^{\prime}_{1}=m_{1}$. Therefore we only
consider the deviation in which $(m^{\prime}_{1},m^{\prime}_{2})\in \mathcal{%
\ \ C}$ such that $m^{\prime}_{1}=m_{1}$ and $m^{\prime}_{2}<m_{2}$.

Fix $t^{\ast}$ such that $t^{\ast}:=\inf \{t\in T: \hat{m_{1}}(t)=m_{1}\}$
so that $\mathbb{E}_{\mu(m_{1}, m_{2}^{\prime})}f(m_{1},z)=f(m_{1},t^{\ast})
$ for any $m^{\prime}_{2}<m_{2}$. It follows that $f(m_{1},t^{\ast})<\mathbb{%
\ \ E}_{\mu(m_{1},m_{2})}f(m_{1},z)$. Because $m^{\prime}_{2}<m_{2}\leq
m_{2}^{\circ}$, the strict convexity of $h$ implies that
\begin{equation}  \label{eq:lem7}
\alpha m_{2}^{\prime}-h(m_{2}^{\prime})< \alpha m_{2}-h(m_{2}),
\end{equation}
which leads to
\begin{equation}  \label{eq:lem40}
\alpha m_{2}+\mathbb{E}_{\mu(m_{1}, m_{2})}f(m_{1},z)
-c(m_{1},t)-h(m_{2}>\alpha m_{2}^{\prime}+\mathbb{E}_{\mu(m_{1},
m_{2})}f(m_{1},z)-c(m_{1},t)-h(m_{2}^{\prime})
\end{equation}
for any $t\neq t^{\ast}$ such that $t\in Z(m_{1},m_{2})$. Because $%
f(m_{1},t^{\ast})<\mathbb{E}_{\mu(m_{1},m_{2})}f(m_{1},z)$, we also have
that
\begin{equation}  \label{eq:lem41}
\alpha m_{2}^{\prime}+\mathbb{E}_{\mu(m_{1}, m_{2})}f(m_{1},z)-c(m_{1},t)-
h(m_{2}^{\prime}) > \alpha
m_{2}^{\prime}+f(m_{1},t^{\ast})-c(m_{1},t)-h(m_{2}^{\prime}).
\end{equation}
Combining (\ref{eq:lem40}) and (\ref{eq:lem41}), we have
\begin{equation}  \label{eq:lem42}
\alpha m_{2}+\mathbb{E}_{\mu(m_{1}, m_{2})}f(m_{1},z) -c(m_{1},t)-h(m_{2}>
\alpha m_{2}^{\prime}+f(m_{1},t^{\ast})-c(m_{1},t)-h(m_{2}^{\prime}).
\end{equation}
Because $\mathbb{E}_{\mu(m_{1}, m_{2}^{\prime})}f(m_{1},z)=f(m_{1},t^{\ast})$
, (\ref{eq:lem42}) implies that
\begin{equation}  \label{eq:lem43}
\alpha m_{2}+\mathbb{E}_{\mu(m_{1}, m_{2})}f(m_{1},z)
-c(m_{1},t)-h(m_{2}>\alpha m_{2}^{\prime}+ \mathbb{E}_{\mu(m_{1},
m_{2}^{\prime})}f(m_{1},z)-c(m_{1},t)-h(m_{2}^{\prime})
\end{equation}
which implies that no sender type $t\neq t^{\ast}$ such that $t\in
Z(m_{1},m_{2})$ has an incentive to deviate to $(m_{1},m^{\prime}_{2})\in
\mathcal{C}$ such that $m^{\prime}_{2}< m_{2}$.

Hence, we conclude from case 1 and case 2 that $t^{\prime}\in Z(m_{1},m_{2})$
when $t_{1},t_{2}\in Z(m_{1},m_{2})$ for any $t^{\prime}$ such that $%
t_{1}<t^{\prime}< t_{2}$. Thus $Z(m_{1},m_{2})$ is an interval.
\end{proof}


\begin{proof}
\textbf{of Theorem \ref{prop3}.} Let $Z(m_{1},m_{2})$ denote the set of
types of senders who choose the same signals $(m_{1},m_{2})$ and let $%
Z(m_{1},m_{2})$ be a set of positive measure in equilibrium. According to
Lemma \ref{lemma2}, $Z(m_{1},m_{2})$ is a connected interval. We now show
that $\text{max~} Z(m_{1},m_{2})=\overline{t}$. Suppose towards a
contradiction that bunching does not occur on the top, i.e., $t_{\circ}<
\overline{t}$. It follows immediately from Lemma \ref{lemma4} that $%
\displaystyle{\hat{m_{1}}(t_{\circ}) <\lim_{t\searrow t_{\circ}} \hat{m_{1}}
(t)}$ and $\displaystyle{\hat{m_{2}}(t_{\circ}) >\lim_{t\searrow t_{\circ}}
\hat{m_{2}}(t)}$. The discontinuity of $\hat{m_{1}}$ and $\hat{m_{2}}$ at $%
t_{\circ}<\overline{t}$ creates a product interval of off-path action pair.
Let us define the set of feasible off-path signal pairs by
\begin{equation*}
D=\Big\{(s_1,s_2)\in \mathcal{C}: (s_1,s_2)\in \displaystyle{\big(\hat{m_{1}}
(t_{\circ}),\lim_{t\searrow t_{\circ}} \hat{m_{1}}(t)\big)\times \big( %
\lim_{t\searrow t_{\circ}} \hat{m_{2}}(t),\hat{m_{2}}(t_{\circ})\big)}\Big\}.
\end{equation*}
According to Lemma \ref{lemma_Off2}, the only belief $\mu(m^{\prime}_{1},m^{
\prime}_{2})$ conditional on $(m^{\prime}_{1},m^{\prime}_{2})$ for any $%
(m^{\prime}_{1},m^{\prime}_{2})\in D$ that passes the Criterion D1 puts all
the weight on $t_{\circ}<\overline{t}$ hence, $supp~\mu(m^{\prime}_{1},m^{
\prime}_{2})=\{t_{\circ}\}$. 

Next, we show that type $t_{\circ}$ has a profitable deviation to action
pair $(m_{1}+\varepsilon,m_{2}-\varepsilon)\in D$, for some $\varepsilon>0$.
That is, we show that that
\begin{align}  \label{eq:profitabledeviation1}
\alpha m_{2}-\alpha \varepsilon +f(m_{1}+\varepsilon,t_{\circ})-c(m_{1}+
\varepsilon,t_{\circ})-h(m_ 2-\varepsilon)>&  \notag \\
\alpha m_{2}+\mathbb{E}_{\mu(m_{1},m_{2})}f(m_{1},t)-c(m_{1},t_{
\circ})-h(m_{2}).
\end{align}
Because type $t_{\circ}$ is the highest in the type interval that chooses
the same signal pair $(m_{1},m_{2})$ in equilibrium, we have that
\begin{equation}  \label{eq:profitabledeviation2}
f(m_{1},t_{\circ})>\mathbb{E}_{\mu(m_{1},m_{2})}f(m_{1},t).
\end{equation}
Because $f$, $c$ and $h$ are continuous in the sender's actions, and
inequality (\ref{eq:profitabledeviation2}), we have that
\begin{align}  \label{eq:profitabledeviation3}
\lim_{\varepsilon\searrow 0}\Big[ \alpha m_{2}-\alpha \varepsilon +
f(m_{1}+\varepsilon,t_{\circ})-c(m_{1}+\varepsilon,t_{\circ})-h(m_
2-\varepsilon)\Big] >&  \notag \\
\alpha m_{2}+\mathbb{E}_{\mu(m_{1},m_{2})}f(m_{1},t)-c(m_{1},t_{
\circ})-h(m_{2}).
\end{align}
Inequality (\ref{eq:profitabledeviation3}) implies that there exists $%
\varepsilon>0$ such that (\ref{eq:profitabledeviation1}) holds.

Therefore, there is no bunching in $Z(m_{1},m_{2})$ with $\big(\hat{m_{1}}
(t),\hat{m_{2}}(t)\big)=(m_{1},m_{2})$ for all $t\in Z(m_{1},m_{2})$ if $%
t_{\circ}<\overline{t}$. This implies that $t_{\circ}=\overline{t}$ whenever
there is bunching in $Z(m_{1},m_{2})$.

We now state the proof for item (b). First, note that because of Theorem \ref%
{lemma40}, $m_{1}=M$ is equivalent to $m_{2}=0$. Suppose towards a
contradiction that $m_{2}>0$. Because $m_{1}+m_{2}=M$ when $Z(m_{1},m_{2})$
has a positive measure in equilibrium (Theorem \ref{lemma40}), we must have $%
m_{1}<M$ when $m_{2}>0$. We follow the proof of (\ref%
{eq:profitabledeviation1}) to show that type $t_{\circ}=\overline{t}$ has an
incentive to profitably deviate to action pair $(M-\varepsilon, \varepsilon)$
for some $\varepsilon >0$ small enough. This leads to a contradiction. Thus $%
m_{1}=M$ when $Z(m_{1},m_{2})$ has a positive measure in equilibrium.
\end{proof}


\begin{proof}
\textbf{of Corollary \ref{prop4}.} Let $Z(m_{1},m_{2})$ denote the set of
types of senders who choose the same signals $(m_{1},m_{2})$ and suppose it
has a positive measure in equilibrium. According to Theorem \ref{prop3}, as
long as bunching occurs $Z(m_{1},m_{2})$ has a positive measure in
equilibrium, $\text{max~} Z(m_{1},m_{2}) =\overline{t}$ must hold.
Therefore, there is no bunching on $Z(m_{1},m_{2})$ such that $\text{max~}
Z(m_{1},m_{2})<\overline{t}$ when Criterion D1 is satisfied in equilibrium.
\end{proof}


\section*{Proof of Proposition \protect\ref{nonbinding53}}

\begin{proof}
\textbf{of Proposition \ref{nonbinding53}.} Because $m_{1}^{\ast}(t^{
\prime})=M$ for $t^{\prime}\in$ int($T$), we have that
\begin{align}  \label{eq:nonbinding55}
&f(M, t^{\prime})-c(M,t^{\prime}) > \lim_{t\nearrow t^{\prime}}\Big[ \alpha
m_{2}^{\ast}(t)
+f(m_{1}^{\ast}(t),t)-c(m_{1}^{\ast}(t),t^{\prime})-h(m_{2}^{\ast}(t))\Big].
\end{align}
Moreover, because $\mathbb{E}f(M,z|z \geq t^{\prime})>f(M, t^{\prime})$, ( %
\ref{eq:nonbinding55}) becomes
\begin{align}  \label{eq:nonbinding56}
&\mathbb{E}f(M,z|z\geq t^{\prime})-c(M,t^{\prime}) >\lim_{t\nearrow
t^{\prime}} \Big[ \alpha m_{2}^{\ast}(t)+f(m_{1}^{\ast}(t),t)-
c(m_{1}^{\ast}(t),t^{\prime})-h(m_{2}^{\ast}(t))\Big].
\end{align}
Because $f$, $h$, and $c$ are continuous functions, by applying the
intermediate value theorem, (\ref{eq:nonbinding54}) and (\ref%
{eq:nonbinding56}), implies that there exists some $t_{h}$ $\in (t_{\ell},
t^{\prime} ]$ such that
\begin{equation}  \label{eq:nonbinding57}
\alpha m_{2}^{\ast}(t_{h})+f(m_{1}^{\ast}(t_{h}),t_{h})-
c(m_{1}^{\ast}(t_{h}),t_{h})-h(m_{2}^{\ast}(t_{h}))= \mathbb{E}f(M,z|z\geq
t_{h})- c(M,t_{h}).
\end{equation}
We now show the uniqueness of such $t_{h}$. We show that no sender type $t$ $%
\in (t_{\ell}, t_{h})$ $\cup$ $(t_{h}, t^{\prime} )$ satisfies the
indifference condition in equation (\ref{eq:nonbinding52}), when (\ref%
{eq:nonbinding57}) holds.

Consider $t\in (t_{h},t^{\prime })$. Because of the incentive compatibility
constraint for $t_{h}$, we have that for $t\in (t_{h},t^{\prime })$,
\begin{align}
& \alpha m_{2}^{\ast }(t_{h})+f(m_{1}^{\ast }(t_{h}),t_{h})-c(m_{1}^{\ast
}(t_{h}),t_{h})-h(m_{2}^{\ast }(t_{h}))  \notag  \label{eq:nonbinding58} \\
>& \alpha m_{2}^{\ast }(t)+f(m_{1}^{\ast }(t),t)-c(m_{1}^{\ast
}(t),t_{h})-h(m_{2}^{\ast }(t)).
\end{align}%
Combining equation (\ref{eq:nonbinding57}) and inequality (\ref%
{eq:nonbinding58}), we obtain:
\begin{equation}
\mathbb{E}f(M,z|z\geq t_{h})-c(M,t_{h})>\alpha m_{2}^{\ast
}(t)+f(m_{1}^{\ast }(t),t)-c(m_{1}^{\ast }(t),t_{h})-h(m_{2}^{\ast }(t)).
\label{eq:nonbinding59}
\end{equation}%
For all $t\in (t_{h},t^{\prime })$, because $M>m_{1}^{\ast }(t)$, by using
the supermodularity property of $-c$ for $t>t_{h}$, we have that
\begin{equation}
c(M,t_{h})-c(m_{1}^{\ast }(t),t_{h})>c(M,t)-c(m_{1}^{\ast }(t),t).
\label{eq:nonbinding60}
\end{equation}%
Combining equation (\ref{eq:nonbinding59}) and inequality (\ref%
{eq:nonbinding60}), we obtain:
\begin{equation}
\mathbb{E}f(M,z|z\geq t_{h})-c(M,t)>\alpha m_{2}^{\ast }(t)+f(m_{1}^{\ast
}(t),t)-c(m_{1}^{\ast }(t),t)-h(m_{2}^{\ast }(t)).  \label{eq:nonbinding61}
\end{equation}%
for all $t\in (t_{h},t^{\prime })$. Because $t>t_{h}$, we have that $\mathbb{%
E}f(M,z|z\geq t)>\mathbb{E}f(M,z|z\geq t_{h})$. Therefore equation (\ref%
{eq:nonbinding61}) implies that for all $t\in (t_{h},t^{\prime })$,
\begin{equation}
\mathbb{E}f(M,z|z\geq t)-c(M,t)>\alpha m_{2}^{\ast }(t)+f(m_{1}^{\ast
}(t),t)-c(m_{1}^{\ast }(t),t)-h(m_{2}^{\ast }(t))  \label{eq:nonbinding62}
\end{equation}%
which violates (\ref{eq:nonbinding52}).

Now consider $t\in (t_{\ell },t_{h})$. Because $M>m_{1}^{\ast }(t_{h})$,
using the supermodularity property of $-c$ for $t<t_{h}$, we obtain:
\begin{equation}
c(M,t)-c(m_{1}^{\ast }(t_{h}),t)>c(M,t_{h})-c(m_{1}^{\ast }(t_{h}),t_{h}).
\label{eq:nonbinding63}
\end{equation}%
Combining inequality (\ref{eq:nonbinding57}) and (\ref{eq:nonbinding63}), we
obtain:
\begin{equation*}
\alpha m_{2}^{\ast }(t_{h})+f(m_{1}^{\ast }(t_{h}),t_{h})-c(m_{1}^{\ast
}(t_{h}),t)-h(m_{2}^{\ast }(t_{h}))>\mathbb{E}f(M,z|z\geq t_{h})-c(M,t).
\end{equation*}%
for all $t\in (t_{\ell },t_{h})$. Because of the incentive compatibility
constraint for $t\in (t_{\ell },t_{h})$, we have that
\begin{align}
& \alpha m_{2}^{\ast }(t)+f(m_{1}^{\ast }(t),t)-c(m_{1}^{\ast
}(t),t)-h(m_{2}^{\ast }(t))  \notag  \label{eq:nonbinding65} \\
>& \alpha m_{2}^{\ast }(t_{h})+f(m_{1}^{\ast }(t_{h}),t_{h})-c(m_{1}^{\ast
}(t_{h}),t)-h(m_{2}^{\ast }(t_{h})).
\end{align}%
Combining equation (\ref{eq:nonbinding64}) and inequality (\ref%
{eq:nonbinding65}), we obtain:
\begin{equation}
\alpha m_{2}^{\ast }(t)+f(m_{1}^{\ast }(t),t)-c(m_{1}^{\ast
}(t),t)-h(m_{2}^{\ast }(t))>\mathbb{E}f(M,z|z\geq t_{h})-c(M,t).
\label{eq:nonbinding66}
\end{equation}%
Because $\mathbb{E}f(M,z|z\geq t_{h})>\mathbb{E}f(M,z|z\geq t)$ for $t_{h}>t$%
, inequality (\ref{eq:nonbinding66}) implies that
\begin{equation}
\alpha m_{2}^{\ast }(t)+f(m_{1}^{\ast }(t),t)-c(m_{1}^{\ast
}(t),t)-h(m_{2}^{\ast }(t))>\mathbb{E}f(M,z|z\geq t)-c(M,t)
\label{eq:nonbinding67}
\end{equation}%
for all $t\in (t_{\ell },t_{h})$, which violates (\ref{eq:nonbinding52}).
Therefore, the only $t\in \lbrack t_{\ell },t^{\prime }]$ that solves
equation (\ref{eq:nonbinding52}) is type $t_{h}$.
\end{proof}

\section*{Proof of Theorem \protect\ref{theorem_2}}

\begin{proof}
\textbf{of Theorem \ref{theorem_2}.} The existence result comes from
Proposition \ref{prop_5} and Proposition \ref{nonbinding53}. In particular,
when $t_{h}=\overline{t}$, $\{\tilde{m_{1}},\tilde{m_{2}},\tilde{\mu},\tilde{%
w}\}$ follows the monotone D1 separating equilibrium $\{m_{1}^{\ast
},m_{2}^{\ast },\mu ^{\ast },w^{\ast }\}$ with the same $\underline{t}$ and $%
t_{\ell }$. Therefore, we will use some of the proof of Proposition \ref%
{prop_5}. Because there is a jump in $\left( \tilde{m_{1}},\tilde{m_{2}}%
\right) $ at $t_{h}<\overline{t}$, we have $\lim_{t\nearrow t_{h}}\tilde{m}%
_{1}(t)=m_{1}^{\ast }(t_{h})$ and $\lim_{t\nearrow t_{h}}\tilde{m}%
_{2}(t)=m_{2}^{\ast }(t_{h})$. Therefore, for every $t\in \left[ t_{h},%
\overline{t}\right] $, we have $\tilde{m_{1}}(t)=M$ and $\tilde{m_{2}}(t)=0$
according to Theorem \ref{prop3}.

It is straightforward to show that the beliefs in Condition 2 of Theorem \ref%
{theorem_2} satisfies the consistency. Note that Proposition \ref%
{nonbinding53} shows the uniqueness of $t_{h}$, and Lemma \ref{lemma_2}
shows the uniqueness of $\mu ^{\ast }$. We show below that no sender type
has an incentive to deviate from their equilibrium action pairs to any other
on-path or off-path equilibrium. Thanks to Lemma \ref{lemma_Off} and \ref%
{lemma_Off2}. \newline
\textbf{1. Off-path deviations:}\newline
There are two intervals of cognitive actions that are not observed in a
monotone D1 equilibrium (henceforth, equilibrium) with $\underline{t}%
<t_{\ell }\leq t_{h}<\overline{t}$: $(0,m_{1}^{\ast }(\underline{t}))$ and $%
[m_{1}^{\ast }(t_{h}),M)$. There are two intervals of non-cognitive actions
that are not observed in equilibrium: $(0,m_{2}^{\ast }(t_{h})]$, and $%
(m_{2}^{\circ },M]$.

(a) First, consider some cognitive action choice $m_{1}^{\prime }\in
(0,m_{1}^{\ast }(\underline{t}))$ and any corresponding $m_{2}^{\prime }$
such that $m_{1}^{\prime }+m_{2}^{\prime }\leq M$. We pick a belief
conditional on such a deviation to an off-path action pair $(m_{1}^{\prime
},m_{2}^{\prime })$ that puts all the probability weights on $\underline{t}$
. Accordingly, the market wage becomes $w^{\prime }=\alpha m_{2}^{\prime
}+f(m_{1}^{\prime },\underline{t}).$ We show that such a belief satisfies
Criterion D1.

Suppose that type $\underline{t}$ is weakly worse off by deviating to $%
(m_{1}^{\prime },m_{2}^{\prime })$ such that $m_{1}^{\prime }\in
(0,m_{1}^{\ast }(\underline{t}))$ and $m_{1}^{\prime }+m_{2}^{\prime }\leq M$
:
\begin{equation}
w^{\prime }-c\big(m_{1}^{\prime },\underline{t}\big)-h(m_{2}^{\prime })\leq
\alpha m_{2}^{\ast }(\underline{t})+f\big(m_{1}^{\ast }(\underline{t}),
\underline{t}\big)-c\big(m_{1}^{\ast }(\underline{t}),\underline{t}\big) %
-h(m_{2}^{\ast }(\underline{t})).  \label{eq:nonbinding1}
\end{equation}
Because $m_{1}^{\prime }<m_{1}^{\ast }(\underline{t})$, using the
supermodularity property of $-c$ for $t>\underline{t}$, we have that
\begin{equation}
c\big(m_{1}^{\ast }(\underline{t}),\underline{t}\big)-c\big(m_{1}^{\prime },
\underline{t}\big)>c\big(m_{1}^{\ast }(\underline{t}),t\big)-c\big( %
m_{1}^{\prime },t\big).  \label{eq:nonbinding2}
\end{equation}
By combining equation (\ref{eq:nonbinding100}) and (\ref{eq:nonbinding2}),
we have that
\begin{equation}
w^{\prime }-c\big(m_{1}^{\prime },t\big)-h(m_{2}^{\prime })<\alpha
m_{2}^{\ast }(\underline{t})+f\big(m_{1}^{\ast }(\underline{t}),\underline{t}
\big)-c\big(m_{1}^{\ast }(\underline{t}),t\big)-h(m_{2}^{\ast }(\underline{t}
)).  \label{eq:nonbinding3}
\end{equation}
Because $((m_{1}^{\ast }(t),m_{2}^{\ast }(t))$ is incentive compatible for
type $t>\underline{t}$, we have that
\begin{equation}
\alpha m_{2}^{\ast }(\underline{t})+f\big(m_{1}^{\ast }(\underline{t}),
\underline{t}\big)-c\big(m_{1}^{\ast }(\underline{t}),t\big)-h(m_{2}^{\ast
}( \underline{t}))<\alpha m_{2}^{\ast }(t)+f\big(m_{1}^{\ast }(t),t\big)-c %
\big( m_{1}^{\ast }(t),t\big)-h(m_{2}^{\ast }(t)).  \label{eq:nonbinding4}
\end{equation}
By combining equation (\ref{eq:nonbinding3}) and (\ref{eq:nonbinding4}), we
have that
\begin{equation}
w^{\prime }-c\big(m_{1}^{\prime },t\big)-h(m_{2}^{\prime })<\alpha
m_{2}^{\ast }(t)+f\big(m_{1}^{\ast }(t),t\big)-c\big(m_{1}^{\ast }(t),t\big) %
-h(m_{2}^{\ast }(t))  \label{eq:nonbinding5}
\end{equation}
which shows that type $t>\underline{t}$ is strictly worse off by deviating
to an off-path action pair $(m_{1}^{\prime },m_{2}^{\prime })$ with $%
m_{1}^{\prime }\in (0,m_{1}^{\ast }(\underline{t}))$ and $m_{1}^{\prime
}+m_{2}^{\prime }\leq M$.

Therefore the Criterion D1 is satisfied.

According to the implication of Criterion D1, it is sufficient to show that
( \ref{eq:nonbinding1}) is satisfied in order to show that no type has
incentive to deviate to an off-path action pair $(m_{1}^{\prime
},m_{2}^{\prime })$ with $m_{1}^{\prime }\in (0,m_{1}^{\ast }(\underline{t}%
)) $ and $m_{1}^{\prime }+m_{2}^{\prime }\leq M$. It is straightforward
given $w^{\prime }=\alpha m_{2}^{\prime }+f(m_{1}^{\prime },\underline{t})$
because $(m_{1}^{\ast }(\underline{t}),m_{2}^{\ast }(\underline{t}))$ is the
efficient action pair for type $\underline{t}.$

(b) Now consider a deviation to an off-path action pair $(m_{1}^{\prime
},m_{2}^{\prime })$ with $m_{1}^{\prime }\in \lbrack m_{1}^{\ast}(t_{h}), M)
$ such that $m_{1}^{\prime }+m_{2}^{\prime }\leq M$. We pick a belief
conditional on such an off-path action that put all the probability weights
on $t_{h}.$ Given such a belief, the market wage becomes $w^{\prime }=\alpha
m_{2}^{\prime }+f(m_{1}^{\prime },t_{h})$. We show that such a belief
satisfies Criterion D1. Suppose that type $t_{h}$ is weakly worse off by
deviating to an off-path action pair $(m_{1}^{\prime },m_{2}^{\prime })$
such that $m_{1}^{\prime }\in \lbrack m_{1}^{\ast }(t_{h}),M)$ and $%
m_{1}^{\prime }+m_{2}^{\prime }\leq M$:
\begin{equation}
w^{\prime }-c(m_{1}^{\prime },t_{h})-h(m_{2}^{\prime })\leq \mathbb{E}
f(M,z|z\geq t_{h})-c(M,t_{h}).  \label{eq:nonbinding106}
\end{equation}
Because $M>m_{1}^{\prime }$, the single crossing condition of $c$ for any
type $t>t_{h}$ implies that
\begin{equation}
c(M,t_{h})-c(m_{1}^{\prime },t_{h})>c(M,t)-c(m_{1}^{\prime },t).
\label{eq:nonbinding107}
\end{equation}
Combining equation (\ref{eq:nonbinding106}) and (\ref{eq:nonbinding107})
yields that for any $t>t_{h}$
\begin{equation}
w^{\prime }-c(m_{1}^{\prime },t)-h(m_{2}^{\prime })<\mathbb{E}f(M,z|z\geq
t_{h})-c(M,t).  \label{eq:nonbinding108}
\end{equation}
Because $\mathbb{E}f(M,z|z\geq t_{h})<\mathbb{E}f(M,z|z\geq t)$, inequality
( \ref{eq:nonbinding108}) implies that for any $t>t_{h}$
\begin{equation}
w^{\prime }-c(m_{1}^{\prime },t)-h(m_{2}^{\prime })<\mathbb{E}f(M,z|z\geq
t)-c(M,t).  \label{eq:nonbinding109}
\end{equation}
Therefore, any type $t>t_{h}$ is strictly worse off by the same deviation.

To see, whether any type $t<t_{h}$ is also strictly worse off by the same
deviation, note that (\ref{eq:nonbinding106}) is equivalent to the following
inequality because of the utility indifference condition for the threshold
type $t_{h}$:
\begin{equation}
w^{\prime }-c(m_{1}^{\prime },t_{h})-h(m_{2}^{\prime })\leq \alpha
m_{2}^{\ast }(t_{h})+f(m_{1}^{\ast }(t_{h}),t_{h})-c(m_{1}^{\ast
}(t_{h}),t)-h(m_{2}^{\ast }(t_{h})).  \label{eq:nonbinding109_B}
\end{equation}
Because $m_{1}^{\prime }\geq m_{1}^{\ast }(t_{h})$, the single crossing
condition of $c$ for type $t<t_{h}$ implies that
\begin{equation}
c(m_{1}^{\prime },t)-c(m_{2}^{\ast }(t_{h}),t)\geq c(m_{1}^{\prime
},t_{h})-c(m_{2}^{\ast }(t_{h}),t_{h}).  \label{eq:nonbinding110}
\end{equation}
Combining equation (\ref{eq:nonbinding109_B}) and (\ref{eq:nonbinding110})
yields
\begin{equation}
w^{\prime }-c(m_{1}^{\prime },t)-h(m_{2}^{\prime })\leq \alpha m_{2}^{\ast
}(t_{h})+f(m_{1}^{\ast }(t_{h}),t_{h})-c(m_{1}^{\ast
}(t_{h}),t)-h(m_{2}^{\ast }(t_{h})).  \label{eq:nonbinding111}
\end{equation}

Incentive compatibility constraint for $t<t_{h}$ and inequality (\ref%
{eq:nonbinding111}) implies that
\begin{equation}
w^{\prime }-c(m_{1}^{\prime },t)-h(m_{2}^{\prime })<\alpha m_{2}^{\ast
}(t)+f(m_{1}^{\ast }(t),t)-c(m_{1}^{\ast }(t),t)-h(m_{2}^{\ast }(t)),
\label{eq:nonbinding112}
\end{equation}
which implies that type $t<t_{h}$ is strictly worse off with the same
deviation. Therefore, Criterion D1 is satisfied.

The only thing we need to show is that type $t_{h}$ has no incentive to
deviate to an off-path action pair $(m_{1}^{\prime },m_{2}^{\prime })$ such
that $m_{1}^{\prime }+m_{2}^{\prime }\leq M$ and $m_{1}^{\prime }\in \lbrack
m_{1}^{\ast }(t_{h}),M)$.

First, consider the deviation to an off-path message pair $(m_{1}^{\prime
},m_{2}^{\prime })$ such that $m_{1}^{\prime }\in \lbrack m_{1}^{\ast
}(t_{h}),M)$ and the constraint is binding, i.e., $m_{1}^{\prime
}+m_{2}^{\prime }=M$. The utility for type $t_{h}$ upon such a deviation is $%
\alpha (M-m_{1}^{\prime })+f(m_{1}^{\prime },t_{h})-c(m_{1}^{\prime
},t_{h})-h(M-m_{1}^{\prime })$ given $w^{\prime }=\alpha m_{2}^{\prime
}+f(m_{1}^{\prime },t_{h})$.

From the first order condition in the separating equilibrium, we have that
\begin{align}
& f_{m_{1}}\left( m_{1}^{\ast }(t_{h}),\mu ^{\ast }(m_{1}^{\ast
}(t_{h}))\right) +f_{t}\big(m_{1}^{\ast }(t_{h}),\mu ^{\ast }(m_{1}^{\ast
}(t_{h}))\big)\mu ^{\ast \prime }(m_{1}^{\ast }(t_{h}))-c_{m_{1}}\big( %
m_{1}^{\ast }(t_{h}),t_{h}\big)-  \notag  \label{eq:nonbinding113} \\
& \alpha +h^{\prime }(M-m_{1}^{\ast }(t_{h}))=0
\end{align}
Because $\mu ^{\ast \prime }(\cdot )>0$, equation (\ref{eq:nonbinding113})
implies that
\begin{equation*}
f_{m_{1}}\left( m_{1}^{\ast }(t_{h}),t_{h})\right) -c_{m_{1}}\big( %
m_{1}^{\ast }(t_{h}),t_{h}\big)-\alpha +h^{\prime }(M-m_{1}^{\ast
}(t_{h}))<0.
\end{equation*}
Because $m_{1}^{\prime }>m_{1}^{\ast }(t_{h})$ and $f-c$ is strictly
concave, (\ref{eq:nonbinding113}) implies that we
\begin{align}
& \alpha (M-m_{1}^{\prime })+f(m_{1}^{\prime },t_{h})-c(m_{1}^{\prime
},t_{h})-h(M-m_{1}^{\prime })  \notag  \label{eq:nonbinding114} \\
& <\alpha (M-m_{1}^{\ast }(t_{h}))+f(m_{1}^{\ast
}(t_{h}),t_{h})-c(m_{1}^{\ast }(t_{h}),t_{h})-h(M-m_{1}^{\ast }(t_{h})),
\end{align}
which implies that type $t_{h}$ is worse off by the deviation to an off-path
message pair $(m_{1}^{\prime },m_{2}^{\prime })$ such that $m_{1}^{\prime
}\in \lbrack m_{1}^{\ast }(t_{h}),M)$ and $m_{1}^{\prime }+m_{2}^{\prime }=M$
.

Second, consider the deviation to an off-path action pair $(m_{1}^{\prime
},m_{2}^{\prime })$ such that $m_{1}^{\prime }+m_{2}^{\prime }<M$ and $%
m_{1}^{\prime }\in \lbrack m_{1}^{\ast }(t_{h}),M)$. Notice that $%
m_{1}^{\prime }+m_{2}^{\prime }<M=m_{1}^{\ast }(t_{h})+m_{2}^{\ast
}(t_{h})\leq m_{1}^{\prime }+m_{2}^{\ast }(t_{h})$ implies that $%
m_{2}^{\prime }<m_{2}^{\ast }(t_{h})<m_{2}^{\circ }$. Therefore by the
strict convexity of $h$, we have that
\begin{equation}
\alpha m_{2}^{\ast }(t_{h})-h(m_{2}^{\ast }(t_{h}))>\alpha m_{2}^{\prime
}-h(m_{2}^{\prime }).  \label{eq:nonbinding115}
\end{equation}
Also, by the strict concavity of $f-c$, because $m_{1}^{\ast }(t_{h})$ is
inefficiently high and $m_{1}^{\prime }>m_{1}^{\ast }(t_{h})$, we have that
\begin{equation}
f(m_{1}^{\ast }(t_{h}),t_{h})-c(m_{1}^{\ast }(t_{h}),t_{h})>f(m_{1}^{\prime
},t_{h})-c(m_{1}^{\prime },t_{h}).  \label{eq:nonbinding116}
\end{equation}
Combining (\ref{eq:nonbinding115}) and (\ref{eq:nonbinding116}), we have
that
\begin{align}
& \alpha m_{2}^{\prime }+f(m_{1}^{\prime },t_{h})-c(m_{1}^{\prime
},t_{h})-h(m_{2}^{\prime })  \notag  \label{eq:nonbinding117} \\
<& \alpha m_{2}^{\ast }(t_{h})+f(m_{1}^{\ast }(t_{h}),t_{h})-c(m_{1}^{\ast
}(t_{h}),t_{h})-h(m_{2}^{\ast }(t_{h})).
\end{align}

Note that $\alpha m_{2}^{\prime }+f(m_{1}^{\prime },t_{h})$ is the market
wage conditional on the deviation to an off-path action pair $(m_{1}^{\prime
},m_{2}^{\prime })$. Therefore no sender type has an incentive to deviate to
off-path message pair $(m_{1}^{\prime },m_{2}^{\prime })$ such that $%
m_{1}^{\prime }\in \lbrack m_{1}^{\ast }(t_{h}),M)$ and $m_{1}^{\prime
}+m_{2}^{\prime }<M$.

(c) Consider a deviation to the action pair $(m_{1},m_{2}^{\prime})$ with $%
m_{1}\in \big[m_{1}^{\ast}(\underline{t}),m_{1}^{\ast}(t_{\ell})\big)$ and $%
m_{2}^{\prime}\in\left(0,m_{2}^{\ast}(t_{h})\right]$ such that $m_{1}+
m_{2}^{\prime}< M$.

Define $t^{\ast}$ as the type such that $m_{1}^{\ast}(t^{\ast})=m_{1}$ for $%
m_{1}\in \big[m_{1}^{\ast}(\underline{t}),m_{1}^{\ast}(t_{\ell})\big)$. Then
$\mathbb{E}_{\mu(m_{1}, m_{2}^{\prime})}f(m_{1},z)=f(m_{1},t^{\ast})$ for
some generic type $z$. Let $w^{\prime}:=\alpha m_{2}^{\prime}+f\big(m_{1},
t^{\ast}\big)$ denote the wage a sender receives upon choosing the action
pair $(m_{1}, m_{2}^{\prime})$ such that $m_{1}+ m_{2}^{\prime}<M$. For each
$m_{1}\in \left[m_{1}^{\ast}(\underline{t}),m_{1}^{\ast}(t_{\ell})\right)$,
let $m_{2}$ be the non-cognitive signal chosen by a sender along with $m_{1}$
on the equilibrium path. Because $h$ is strictly convex, $m_{2}^{\prime}\leq
m_{2}^{\ast}(t_{h})< m_{2}=m_{2}^{\circ}$ implies that
\begin{equation}  \label{eq:nonbinding118}
\alpha m_{2}^{\prime}-h(m_{2}^{\prime})<\alpha m_{2}-h(m_{2}).
\end{equation}
Because $((m_{1}^{\ast}(t),m_{2}^{\ast}(t))$ is incentive compatible for
type $t\in \left[\underline{t},t_{\ell}\right)$, we have that for each $t\in %
\left[\underline{t},t_{\ell}\right)$,
\begin{equation}  \label{eq:nonbinding119}
\alpha m_{2}^{\ast}(t) +f\big(m_{1}^{\ast}(t), t\big)-c\big( %
m_{1}^{\ast}(t),t \big) -h(m_{2}^{\ast}(t))> \alpha m_{2}+f\big(m_{1},
t^{\ast}\big)-c\big( m_{1},t\big)-h(m_{2}).
\end{equation}
By combining equation (\ref{eq:nonbinding118}) and (\ref{eq:nonbinding119}),
we have that for each $t\in \left[\underline{t},t_{\ell}\right)$,
\begin{equation}  \label{eq:nonbinding120}
\alpha m_{2}^{\prime}+f\big(m_{1}, t^{\ast}\big)- c\big(m_{1},t\big) %
-h(m_{2}^{\prime}) < \alpha m_{2}^{\ast}(t) +f\big(m_{1}^{\ast}(t), t\big)-c %
\big(m_{1}^{\ast}(t),t\big)-h(m_{2}^{\ast}(t)),
\end{equation}
which shows that each sender type $t\in \left[\underline{t},t_{\ell}\right)$
is strictly worse off by deviating to off-path message pair $(m_{1},
m_{2}^{\prime})$ such that $m_{1}\in \left[m_{1}^{\ast}(\underline{t}
),m_{1}^{\ast}(t_{\ell})\right)$, $m_{2}^{\prime}\in (0,m_{2}^{\ast}(t_{h})]
$ and $m_{1}+m_{2}^{\prime}< M$.

(d) Consider a deviation to the action pair $(m_{1},m_{2}^{\prime})$ with $%
m_{1}\in \big[m_{1}^{\ast}(t_{\ell}),m_{1}^{\ast}(t_{h})\big)$, $%
m_{2}^{\prime}\in \left(0,m_{2}^{\ast}(t_{h})\right]$ such that $m_{1}+
m_{2}^{\prime}<M$.

Define $t^{\ast}$ as the type such that $m_{1}^{\ast}(t^{\ast})=m_{1}$ for $%
m_{1}\in \big[m_{1}^{\ast}(t_{\ell}),m_{1}^{\ast}(t_{h})\big)$. Then $%
\mathbb{E}_{\mu(m_{1}, m_{2}^{\prime})}f(m_{1},z)=f(m_{1},t^{\ast})$ for
some generic type $z$. Let $w^{\prime}:=\alpha m_{2}^{\prime}+f\big(m_{1},
t^{\ast}\big)$ denote the wage a sender receives upon choosing the action
pair $(m_{1}, m_{2}^{\prime})$ such that $m_{1}+ m_{2}^{\prime}<M$. For each
$m_{1}\in \big[m_{1}^{\ast}(t_{\ell}),m_{1}^{\ast}(t_{h})\big)$, let $m_{2}$
be the non-cognitive signal chosen by a sender along with $m_{1}$ on the
equilibrium path. Because $m_{1}+ m_{2}=M$, it is clear that $%
m_{2}^{\prime}<m_{2}$. Because $h$ is strictly convex, $m_{2}^{
\prime}<m_{2}<m_{2}^{\circ}$ implies that
\begin{equation}  \label{eq:nonbinding124}
\alpha m_{2}^{\prime}-h(m_{2}^{\prime})<\alpha m_{2}-h(m_{2}).
\end{equation}
Because $((m_{1}^{\ast}(t),m_{2}^{\ast}(t))$ is incentive compatible for
type $t\in [t_{\ell},t_{h})$, we have that for each $t\in [t_{\ell},t_{h})$,
\begin{equation}  \label{eq:nonbinding125}
\alpha m_{2}^{\ast}(t) +f\big(m_{1}^{\ast}(t), t\big)-c\big( %
m_{1}^{\ast}(t),t \big) -h(m_{2}^{\ast}(t))\geq \alpha m_{2}+f\big(m_{1},
t^{\ast}\big)-c\big( m_{1},t\big)-h(m_{2}).
\end{equation}
By combining equation (\ref{eq:nonbinding124}) and (\ref{eq:nonbinding125}),
we have that for each $t\in [t_{\ell},t_{h})$,
\begin{equation}  \label{eq:nonbinding126}
\alpha m_{2}^{\ast}(t) +f\big(m_{1}^{\ast}(t), t\big)-c\big( %
m_{1}^{\ast}(t),t \big)-h(m_{2}^{\ast}(t)) > w^{\prime}-c\big(m_{1},t\big) %
-h(m_{2}^{\prime})
\end{equation}
which shows that each sender type $t\in [t_{\ell},t_{h})$ is strictly worse
off by deviating to $(m_{1},m_{2}^{\prime})$ if $m_{1}\in \big[
m_{1}^{\ast}(t_{\ell}),m_{1}^{\ast}(t_{h})\big)$ and some corresponding $%
m_{2}^{\prime}\in \left(0,m_{2}^{\ast}(t_{h})\right]$ such that $m_{1}+
m_{2}^{\prime}<M$.

e) Consider a deviation to the action pair $(m_{1},m_{2}^{\prime})$ with $%
m_{1}\in \big[m_{1}^{\ast}(\underline{t}),m_{1}^{\ast}(t_{\ell})\big)$, and $%
m_{2}^{\prime}\in \left(m^{\circ}_{2},M\right)$ such that $m_{1}+
m_{2}^{\prime}\leq M$.

Define $t^{\ast}$ as the type such that $m_{1}^{\ast}(t^{\ast})=m_{1}$ for $%
m_{1}\in \big[m_{1}^{\ast}(\underline{t}),m_{1}^{\ast}(t_{\ell})\big)$. Then
$\mathbb{E}_{\mu(m_{1}, m_{2}^{\prime})}f(m_{1},z)=f(m_{1},t^{\ast})$ for
some generic type $z$. Let $w^{\prime}:=\alpha m_{2}^{\prime}+f\big(m_{1},
t^{\ast}\big)$ denote the wage a sender receives upon choosing the action
pair $(m_{1}, m_{2}^{\prime})$ such that $m_{1}+ m_{2}^{\prime}<M$. For each
$m_{1}\in \big[m_{1}^{\ast}(\underline{t}),m_{1}^{\ast}(t_{\ell})\big)$, let
$m_{2}$ be the non-cognitive signal chosen by a sender along with $m_{1}$ on
the equilibrium path. Because $h$ is strictly convex, $m_{2}=m_{2}^{
\circ}<m_{2}^{\prime}$ implies that
\begin{equation}  \label{eq:nonbinding121}
\alpha m_{2}^{\prime}-h(m_{2}^{\prime})<\alpha m_{2}-h(m_{2}).
\end{equation}
Because $((m_{1}^{\ast}(t),m_{2}^{\ast}(t))$ is incentive compatible for
type $t\in \left[\underline{t},t_{\ell}\right)$, we have that for each $t\in %
\left[\underline{t},t_{\ell}\right)$,
\begin{equation}  \label{eq:nonbinding122}
\alpha m_{2}^{\ast}(t) +f\big(m_{1}^{\ast}(t), t\big)-c\big( %
m_{1}^{\ast}(t),t \big) -h(m_{2}^{\ast}(t))\geq \alpha m_{2}+f\big(m_{1},
t^{\ast}\big)-c\big( m_{1},t\big)-h(m_{2}).
\end{equation}
By combining equation (\ref{eq:nonbinding121}) and (\ref{eq:nonbinding122}),
we have that for each $t\in \left[\underline{t},t_{\ell}\right)$,
\begin{equation}  \label{eq:nonbinding123}
\alpha m_{2}^{\ast}(t) +f\big(m_{1}^{\ast}(t), t\big)-c\big( %
m_{1}^{\ast}(t),t \big)-h(m_{2}^{\ast}(t)) > w^{\prime}-c\big(m_{1},t\big) %
-h(m_{2}^{\prime})
\end{equation}
which shows that each sender type $t\in \left[\underline{t},t_{\ell}\right)$
is strictly worse off by deviating to off-path message pair $%
(m_{1},m_{2}^{\prime})$ for $m_{1}\in \big[m_{1}^{\ast}(\underline{t}
),m_{1}^{\ast}(t_{\ell})\big)$ and some corresponding $m_{2}^{\prime}\in
\left(m^{\circ}_{2},M\right)$ such that $m_{1}+ m_{2}^{\prime}\leq M$.

f) Consider a deviation to the action pair $(m_{1},m_{2}^{\prime})$ with $%
m_{1}\in \big[m_{1}^{\ast}(t_{\ell}),m_{1}^{\ast}(t_{h})\big]$, and $%
m_{2}^{\prime}\in \left(m^{\circ}_{2},M\right)$ such that $m_{1}+
m_{2}^{\prime}\leq M$.

Define $t^{\ast}$ as the type such that $m_{1}^{\ast}(t^{\ast})=m_{1}$. Then
$\mathbb{E}_{\mu(m_{1}, m_{2}^{\prime})}f(m_{1},z)=f(m_{1},t^{\ast})$ for
some generic type $z$. Let $w^{\prime}:=\alpha m_{2}^{\prime}+f\big(m_{1},
t^{\ast}\big)$ denote the wage a sender receives upon choosing the action
pair $(m_{1}, m_{2}^{\prime})$ with $m_{1}\in \left[m_{1}^{\ast}(t_{
\ell}),m_{1}^{\ast}(t_{h})\right]$, and $m_{2}^{\prime}>m_{2}^{\circ}$ such
that $m_{1}+m_{2}^{\prime}\leq M$. For each $m_{1}\in \left[
m_{1}^{\ast}(t_{\ell}),m_{1}^{\ast}(t_{h})\right]$, let $m_{2}$ be the
non-cognitive signal chosen by a sender along with $m_{1} $ on the
equilibrium path. Because $m_{2}^{\prime}>m_{2}^{\circ}$, it is clear that $%
m_{2}> m_{2}^{\prime}>m_{2}^{\circ} $ when $m_{1}+m_{2}^{\prime}\leq M$ and $%
m_{1}+m_{2}= M$. Because $h$ is strictly convex, $m_{2}>
m_{2}^{\prime}>m_{2}^{\circ} $ implies that
\begin{equation}  \label{eq:nonbinding501}
\alpha m_{2}^{\prime}-h(m_{2}^{\prime})<\alpha m_{2}-h(m_{2}).
\end{equation}
Because $(m_{1}^{\ast}(t),m_{2}^{\ast}(t)$ is incentive compatible for type $%
t\in \left[t_{\ell},t_{h}\right]$, we have that for each $t\in \left[
t_{\ell},t_{h}\right]$,
\begin{equation}  \label{eq:nonbinding511}
\alpha m_{2}^{\ast}(t) +f\big(m_{1}^{\ast}(t), t\big)-c\big( %
m_{1}^{\ast}(t),t \big) -h(m_{2}^{\ast}(t))\geq \alpha m_{2}+f\big(m_{1},
t^{\ast}\big)-c\big( m_{1},t\big)-h(m_{2}).
\end{equation}
By combining equation (\ref{eq:nonbinding501}) and (\ref{eq:nonbinding511}),
we have that for each $t\in \left[t_{\ell},t_{h}\right]$,
\begin{equation}  \label{eq:nonbinding512}
\alpha m_{2}^{\ast}(t) +f\big(m_{1}^{\ast}(t), t\big)-c\big( %
m_{1}^{\ast}(t),t \big)-h(m_{2}^{\ast}(t)) > w^{\prime}-c\big(m_{1},t\big) %
-h(m_{2}^{\prime})
\end{equation}
which shows that each sender type $t\in \left[t_{\ell},t_{h}\right]$ is
strictly worse off by deviating to off-path message pair$(m_{1},
m_{2}^{\prime})$ with $m_{1}\in \left[m_{1}^{\ast}(t_{\ell}),m_{1}^{
\ast}(t_{h})\right]$, and $m_{2}^{\prime}\in \left(m^{\circ}_{2},M\right)$
such that $m_{1}+m_{2}^{\prime}\leq M$.

\textbf{2. On-path deviations:}\newline
Note that $[m_{1}^{\ast }(\underline{t}),m_{1}^{\ast }(t_{\ell }))\cup
\lbrack m_{1}^{\ast }(t_{\ell }),m_{1}^{\ast }(t_{h}))\cup \{M\}$ is the set
of on-equilibrium-path cognitive message in the partially separating
equilibrium. We show that no sender type has an incentive to deviate from
their equilibrium action pair to any other $(m_{1},m_{2})$ if $m_{1}\in
\lbrack m_{1}^{\ast }(\underline{t}),m_{1}^{\ast }(t_{\ell }))\cup \lbrack
m_{1}^{\ast }(t_{\ell }),m_{1}^{\ast }(t_{h}))\cup \{M\}$ and some
corresponding $m_{2}$ such that $m_{1}+m_{2}\leq M$.

\textbf{case a:} No sender type $t\in [\underline{t},t_{\ell})$ chooses
on-path action pairs $(m_{1},m_{2})$ with $m_{1}\in [m_{1}^{\ast}(
\underline{t}),m_{1}^{\ast}(t_{\ell})) \cup [m_{1}^{\ast}(
t_{\ell}),m_{1}^{\ast}(t_{h})) \cup \{M\}$ and some corresponding $m_{2}$
such that $m_{1}+ m_{2}\leq M$:

First, notice that for $t\in \lbrack \underline{t},t_{\ell })$, because $%
(m_{1}^{\ast }(t),m_{2}^{\ast }(t))$ is incentive compatible as shown in the
fully separating equilibrium, we have that for $m_{1}\in \lbrack m_{1}^{\ast
}(\underline{t}),m_{1}^{\ast }(t_{\ell }))$ with $m_{1}\neq m_{1}^{\ast }(t)$%
,
\begin{align}
& \alpha m_{2}^{\ast }(t)+f(m_{1}^{\ast }(t),t)-c(m_{1}^{\ast
}(t),t)-h(m_{2}^{\ast }(t))  \notag  \label{eq:nonbinding70} \\
>& \alpha m_{2}+f(m_{1},\mu ^{\ast }(m_{1}))-c(m_{1},t)-h(m_{2})
\end{align}%
which shows that no sender type $t\in \lbrack \underline{t},t_{\ell })$
chooses action pairs $(m_{1},m_{2})$ with $m_{1}\in \lbrack m_{1}^{\ast }(%
\underline{t}),m_{1}^{\ast }(t_{\ell }))$ and some corresponding $m_{2}$
such that $m_{1}+m_{2}<M$ and $m_{1}\neq m_{1}^{\ast }(t)$.

Second, as shown in the proof of incentive compatibility (part c) of
Proposition (\ref{prop_5}), we have that
\begin{align}
& \alpha m_{2}^{\ast }(t)+f(m_{1}^{\ast }(t),t)-c(m_{1}^{\ast
}(t),t)-h(m_{2}^{\ast }(t))  \notag  \label{eq:nonbinding70_A} \\
>& \alpha m_{2}^{\ast }(t^{\prime })+f(m_{1}^{\ast }(t^{\prime }),t^{\prime
})-c(m_{1}^{\ast }(t^{\prime }),t)-h(m_{2}^{\ast }(t^{\prime }))
\end{align}%
for all $t\in \lbrack \underline{t},t_{\ell })$ and for all $t^{\prime }\in %
\left[ t_{\ell },t_{h}\right] $ which shows that $t\in \left[ \underline{t}%
,t_{\ell }\right) $ do not have incentive to deviate to on-path equilibrium
action pair $(m_{1},m_{2})$ with $m_{1}\in \lbrack m_{1}^{\ast }(t_{\ell
}),m_{1}^{\ast }(t_{h}))$, and a corresponding equilibrium $m_{2}$ such that
$m_{1}+m_{2}=M$.

Third, consider $m_{1}=M$. Recall that for sender type $t_{h}$, we have that
\begin{equation}  \label{eq:nonbinding72}
\alpha m_{2}^{\ast}(t_{h})+f(m_{1}^{\ast}(t_{h}),t_{h})-
c(m_{1}^{\ast}(t_{h}),t_{h})-h(m_{2}^{\ast}(t_{h}))= \mathbb{E}f(M,z|z\geq
t_{h})- c(M,t_{h}).
\end{equation}
Because $M>m_{1}^{\ast}(t_{h})$, using the supermodularity property of $-c$
for $M>m_{1}^{\ast}(t_{h})$ and $t<t_{h}$, we obtain:
\begin{equation}  \label{eq:nonbinding73}
c(M,t)- c(m_{1}^{\ast}(t_{h}),t)>c(M,t_{h})- c(m_{1}^{\ast}(t_{h}),t_{h}).
\end{equation}
Combining inequality (\ref{eq:nonbinding72}) and (\ref{eq:nonbinding73} ),
we obtain:
\begin{align}  \label{eq:nonbinding74}
\alpha m_{2}^{\ast}(t_{h})+f(m_{1}^{\ast}(t_{h}),t_{h})-
c(m_{1}^{\ast}(t_{h}),t)-h(m_{2}^{\ast}(t_{h})) > \mathbb{E}f(M,z|z\geq
t_{h})- c(M,t).
\end{align}
Because of the incentive compatibility constraint for $t\in [\underline{t},
t_{\ell})$, we have that
\begin{align}  \label{eq:nonbinding75}
&\alpha m_{2}^{\ast}(t)+f(m_{1}^{\ast}(t),t)-
c(m_{1}^{\ast}(t),t)-h(m_{2}^{\ast}(t))  \notag \\
>&\alpha m_{2}^{\ast}(t_{h})+f(m_{1}^{\ast}(t_{h}),t_{h})-
c(m_{1}^{\ast}(t_{h}),t)-h(m_{2}^{\ast}(t_{h})).
\end{align}
Combining equation (\ref{eq:nonbinding74}) and inequality (\ref%
{eq:nonbinding75}), we obtain:
\begin{equation}  \label{eq:nonbinding76}
\alpha m_{2}^{\ast}(t)+f(m_{1}^{\ast}(t),t)-
c(m_{1}^{\ast}(t),t)-h(m_{2}^{\ast}(t))> \mathbb{E}f(M,z|z\geq t_{h})-
c(M,t).
\end{equation}
Because $\mathbb{E}f(M,z|z\geq t_{h})>\mathbb{E}f(M,z|z\geq t)$ for $t_{h}>t
$, inequality (\ref{eq:nonbinding76}) implies that for $t\in (\underline{t}
, t_{\ell})$,
\begin{equation}  \label{eq:nonbinding77}
\alpha m_{2}^{\ast}(t)+f(m_{1}^{\ast}(t),t)-
c(m_{1}^{\ast}(t),t)-h(m_{2}^{\ast}(t))> \mathbb{E}f(M,z|z\geq t)- c(M,t)
\end{equation}
which shows that the sender of type $t\in [\underline{t},t_{\ell})$ has no
incentive to increase his cognitive action to $M$.

\textbf{case b:} No sender type $t\in [t_{\ell},t_{h})$ chooses on-path
action pairs $(m_{1},m_{2})$ with $m_{1}\in [m_{1}^{\ast}( \underline{t}
),m_{1}^{\ast}(t_{\ell})) \cup [m_{1}^{\ast}( t_{\ell}),m_{1}^{\ast}(t_{h}))
\cup \{M\}$ and some corresponding $m_{2}$ such that $m_{1}+ m_{2}\leq M$:

First, notice that because $m_{2}^{\ast}(t^{\prime})=m_{2}^{\circ}$ for all $%
t^{\prime}\in \left(\underline{t},t_{\ell}\right)$ is optimal, every sender
type $t\in \left(t_{\ell},\overline{t}\right)$ have incentive to deviate to
this level of non-cognitive signal. However for $m_{1}\in
\left
		[m_{1}^{\ast}(\underline{t}),m_{1}^{\ast}(t_{\ell})\right)$,
because $m_{1}+m_{2}^{\circ}< M$, the corresponding $m_{1}\in
\left
		[m_{1}^{\ast}( \underline{t}),m_{1}^{\ast}(t_{\ell})\right)$ and $%
m_{2}^{\circ}$ are independently chosen in equilibrium. We show that sender
types $t\in \left(t_{\ell},\overline{t}\right)$ do not have incentive to
deviate to the equilibrium level of a cognitive signal chosen by a sender
type $t^{\prime}\in \left(\underline{t},t_{\ell}\right)$.

We follow the proof of incentive compatibility (part d) of Proposition (\ref%
{prop_5}) to show that
\begin{align}
& \alpha m_{2}^{\ast }(t)+f(m_{1}^{\ast }(t),t)-c(m_{1}^{\ast
}(t),t)-h(m_{2}^{\ast }(t))  \notag  \label{eq:nonbinding77_A} \\
>& \alpha m_{2}^{\ast }(t^{\prime })+f(m_{1}^{\ast }(t^{\prime }),t^{\prime
})-c(m_{1}^{\ast }(t^{\prime }),t)-h(m_{2}^{\ast }(t^{\prime }))
\end{align}%
for all $t\in \left[ t_{\ell },t_{h}\right) $ and for all $t^{\prime }\in
\left( \underline{t},t_{\ell }\right) $ which shows that no $t\in \left[
t_{\ell },t_{h}\right) $ has incentive to deviate to on-path message pair $%
(m_{1},m_{2})$ with $m_{1}\in \lbrack m_{1}^{\ast }(\underline{t}%
),m_{1}^{\ast }(t_{\ell }))$, and a corresponding equilibrium $m_{2}$ such
that $m_{1}+m_{2}<M$.

Second, notice that for $t\in \lbrack t_{\ell },t_{h})$, because $%
(m_{1}^{\ast }(t),m_{2}^{\ast }(t))$, for $t\in \lbrack t_{\ell },t_{h})$ is
incentive compatible as shown in the fully separating equilibrium, we have
that for $m_{1}\in \lbrack m_{1}^{\ast }(t_{\ell }),m_{1}^{\ast }(t_{h}))$
with $m_{1}\neq m_{1}^{\ast }(t)$,
\begin{align}
& \alpha m_{2}^{\ast }(t)+f(m_{1}^{\ast }(t),t)-c(m_{1}^{\ast
}(t),t)-h(m_{2}^{\ast }(t))  \notag  \label{eq:nonbinding8} \\
>& \alpha m_{2}+f(m_{1},\mu ^{\ast }(m_{1}))-c(m_{1},t)-h(m_{2})
\end{align}%
for $t\in \lbrack t_{\ell },t_{h})$, which shows that no sender type $t\in
\lbrack t_{\ell },t_{h})$ chooses action pairs $(m_{1},m_{2})$ with $%
m_{1}\in \lbrack m_{1}^{\ast }(t_{\ell }),m_{1}^{\ast }(t_{h}))$ and some
corresponding $m_{2}$ such that $m_{1}+m_{2}=M$ and $m_{1}\neq m_{1}^{\ast
}(t)$.

Third, consider $m_{1}=M$. Recall that for sender type $t_{h}$, we have that
\begin{equation}
\alpha m_{2}^{\ast }(t_{h})+f(m_{1}^{\ast }(t_{h}),t_{h})-c(m_{1}^{\ast
}(t_{h}),t_{h})-h(m_{2}^{\ast }(t_{h}))=\mathbb{E}f(M,z|z\geq
t_{h})-c(M,t_{h}).  \label{eq:nonbinding82}
\end{equation}%
Because $M>m_{1}^{\ast }(t_{h})$, using the supermodularity property of $-c$
for $M>m_{1}^{\ast }(t_{h})$ and $t<t_{h}$, we obtain:
\begin{equation}
c(M,t)-c(m_{1}^{\ast }(t_{h}),t)>c(M,t_{h})-c(m_{1}^{\ast }(t_{h}),t_{h}).
\label{eq:nonbinding83}
\end{equation}%
Combining inequality (\ref{eq:nonbinding82}) and (\ref{eq:nonbinding83} ),
we obtain:
\begin{equation*}
\alpha m_{2}^{\ast }(t_{h})+f(m_{1}^{\ast }(t_{h}),t_{h})-c(m_{1}^{\ast
}(t_{h}),t)-h(m_{2}^{\ast }(t_{h}))>\mathbb{E}f(M,z|z\geq t_{h})-c(M,t).
\end{equation*}%
Because of the incentive compatibility constraint for $t\in \lbrack t_{\ell
},t_{h})$, we have that
\begin{align}
& \alpha m_{2}^{\ast }(t)+f(m_{1}^{\ast }(t),t)-c(m_{1}^{\ast
}(t),t)-h(m_{2}^{\ast }(t))  \notag  \label{eq:nonbinding85} \\
>& \alpha m_{2}^{\ast }(t_{h})+f(m_{1}^{\ast }(t_{h}),t_{h})-c(m_{1}^{\ast
}(t_{h}),t)-h(m_{2}^{\ast }(t_{h})).
\end{align}%
Combining equation (\ref{eq:nonbinding84}) and inequality (\ref%
{eq:nonbinding85}), we obtain:
\begin{equation}
\alpha m_{2}^{\ast }(t)+f(m_{1}^{\ast }(t),t)-c(m_{1}^{\ast
}(t),t)-h(m_{2}^{\ast }(t))>\mathbb{E}f(M,z|z\geq t_{h})-c(M,t).
\label{eq:nonbinding86}
\end{equation}%
Because $\mathbb{E}f(M,z|z\geq t_{h})>\mathbb{E}f(M,z|z\geq t)$ for $t_{h}>t$%
, inequality (\ref{eq:nonbinding86}) implies that for $t\in \lbrack t_{\ell
},t_{h})$,
\begin{equation}
\alpha m_{2}^{\ast }(t)+f(m_{1}^{\ast }(t),t)-c(m_{1}^{\ast
}(t),t)-h(m_{2}^{\ast }(t))>\mathbb{E}f(M,z|z\geq t)-c(M,t)
\label{eq:nonbinding87}
\end{equation}%
which shows that the sender of type $t\in \lbrack t_{\ell },t_{h})$ has no
incentive to increase his cognitive action to $M$.

\textbf{case c:} Type $t_{h}$ does not have an incentive to deviate from his
equilibrium action pair $(M,0)$ to any other action pairs $(m_{1},m_{2})$
with $m_{1}\in [m_{1}^{\ast}( \underline{t}),m_{1}^{\ast}(t_{\ell})) \cup
[m_{1}^{\ast}( t_{\ell}),m_{1}^{\ast}(t_{h})) \cup \{M\}$ and some
corresponding $m_{2}$ such that $m_{1}+ m_{2}\leq M$:

The equilibrium action pair for type $t_{h}$ is $(M,0)$. Recall that we
showed in equation (\ref{eq:nonbinding57}) that type $t_{h}$ is indifferent
between the action pairs $(M,0)$ and $(m_{1}^{\ast}(t_{h}),m_{2}^{
\ast}(t_{h}))$. Therefore the equilibrium action pair for type $t_{h}$ is
unique according to equation (\ref{eq:nonbinding57}).

\textbf{case d:} No sender type $t\in \left(t_{h},\overline{t}\right]$
chooses action pairs $(m_{1},m_{2})$ with $m_{1}\in [m_{1}^{\ast}(
\underline{t}),m_{1}^{\ast}(t_{\ell})) \cup [m_{1}^{\ast}(
t_{\ell}),m_{1}^{\ast}(t_{h})) \cup \{M\}$ and some corresponding $m_{2}$
such that $m_{1}+ m_{2}\leq M$:

Because of the incentive compatibility constraint for $t_{h}$, we have that
\begin{align}  \label{eq:nonbinding90}
& \alpha m_{2}^{\ast}(t_{h})+f(m_{1}^{\ast}(t_{h}),t_{h})-
c(m_{1}^{\ast}(t_{h}),t_{h})-h(m_{2}^{\ast}(t_{h}))  \notag \\
>&\alpha m_{2}^{\ast}(t)+f(m_{1}^{\ast}(t),t)-
c(m_{1}^{\ast}(t),t_{h})-h(m_{2}^{\ast}(t)),
\end{align}
for all $t\in \left(t_{h},\overline{t}\right]$. Combining equation (\ref%
{eq:nonbinding57}) and inequality (\ref{eq:nonbinding90}), we obtain:
\begin{equation}  \label{eq:nonbinding91}
\mathbb{E}f(M,z|z\geq t_{h})- c(M,t_{h})>\alpha
m_{2}^{\ast}(t)+f(m_{1}^{\ast}(t),t)-
c(m_{1}^{\ast}(t),t_{h})-h(m_{2}^{\ast}(t)).
\end{equation}

Because $M>m_{1}^{\ast}(t)$, using the supermodularity property of $-c$ for $%
t>t_{h}$, we have that
\begin{equation}  \label{eq:nonbinding92}
c(M,t_{h})- c(m_{1}^{\ast}(t),t_{h})>c(M,t)- c(m_{1}^{\ast}(t),t).
\end{equation}
Combining equation (\ref{eq:nonbinding91}) and inequality (\ref%
{eq:nonbinding92}), we obtain:
\begin{equation}  \label{eq:nonbinding93}
\mathbb{E}f(M,z|z\geq t_{h})- c(M,t)>\alpha
m_{2}^{\ast}(t)+f(m_{1}^{\ast}(t),t)-
c(m_{1}^{\ast}(t),t)-h(m_{2}^{\ast}(t)),
\end{equation}
for all $t\in \left(t_{h},\overline{t}\right]$. Also note that $\mathbb{E}
f(M,z|z\geq t)>\mathbb{E} f(M,z|z\geq t_{h})$ for $t>t_{h}$. Therefore
equation (\ref{eq:nonbinding93}) implies that for $t\in (t_{h},t^{\prime})$
(where $t^{\prime}$ is as given under condition 2),
\begin{equation}  \label{eq:nonbinding94}
\mathbb{E}f(M,z|z\geq t)- c(M,t)>\alpha
m_{2}^{\ast}(t)+f(m_{1}^{\ast}(t),t)-
c(m_{1}^{\ast}(t),t)-h(m_{2}^{\ast}(t)).
\end{equation}
Incentive compatibility for type $t\in \left(t_{h},\overline{t}\right]$
together with (\ref{eq:nonbinding94}) gives
\begin{equation}  \label{eq:nonbinding94_A}
\mathbb{E}f(M,z|z\geq t)- c(M,t)>\alpha
m_{2}^{\ast}(t^{\prime\prime})+f(m_{1}^{\ast}(t^{\prime\prime}),t^{\prime
\prime})-
c(m_{1}^{\ast}(t^{\prime\prime}),t)-h(m_{2}^{\ast}(t^{\prime\prime}))
\end{equation}
for all $t\in \left(t_{h},\overline{t}\right]$ and for all $%
t^{\prime\prime}\in \left[\underline{t},t_{h}\right)$ which shows that the
sender of type $t\in \left(t_{h},\overline{t}\right]$ has no incentive to
change his action to any other action in $(m_{1},m_{2})$ with $m_{1}\in
[m_{1}^{\ast}( \underline{t}),m_{1}^{\ast}(t_{\ell})) \cup [m_{1}^{\ast}(
t_{\ell}),m_{1}^{\ast}(t_{h})) \cup \{M\}$ and some corresponding $m_{2}$
such that $m_{1}+ m_{2}\leq M$.
\end{proof}

\end{document}